\documentclass[usenatbib]{mnras}

\usepackage[T1]{fontenc}

\DeclareRobustCommand{\VAN}[3]{#2}
\let\VANthebibliography\thebibliography
\def\thebibliography{\DeclareRobustCommand{\VAN}[3]{##3}\VANthebibliography}

\usepackage{graphicx}	
\usepackage{amsmath}	
\usepackage{amssymb}	
\usepackage{color}
\usepackage{xcolor}

\usepackage{newtxtext,newtxmath}

\newcommand{\hinv}{\ensuremath{\, h^{-1}}}%

\newcommand{\mpc}{\ensuremath{\, {\rm Mpc}}}         
\newcommand{\gpc}{\ensuremath{\, {\rm Gpc}}}

\newcommand{\eg}{{\sl e.g.}, }

\newcommand{\thetasmooth}{\theta}

\newcommand{\erf}{{\rm erf}}

\newcommand{\CDF}{{\rm CDF}}
\newcommand{\PDF}{{\rm PDF}}
\newcommand{\nhat}{\mathbf{\hat{n}}}

\definecolor{purple}{RGB}{128, 0, 128}

\newcommand{\citepinprep}[1]{(\textcolor{blue}{#1 et al., in prep.})}

\newcommand*{\vcenteredhbox}[1]{\begingroup
\setbox0=\hbox{#1}\parbox{\wd0}{\box0}\endgroup}

\newcommand{\OrcidID}[1]{ \href[urlcolor = red]{https://orcid.org/#1}{\textcolor{lightgray}{\faOrcid}}}
\newcommand{\OrcidIDName}[2]{\href{https://orcid.org/#1}{#2}}

\title[Beyond the 3rd Moment with CDFs]{Beyond the 3rd moment: A practical study of using lensing convergence CDFs for cosmology with DES Y3}

\usepackage{eso-pic}
\AddToShipoutPictureBG*{%
  \AtPageUpperLeft{%
    \hspace{0.75\paperwidth}%
    \raisebox{-3.5\baselineskip}{%
      \makebox[0pt][l]{\textnormal{DES-2023-0768}}
}}}%

\AddToShipoutPictureBG*{%
  \AtPageUpperLeft{%
    \hspace{0.75\paperwidth}%
    \raisebox{-4.5\baselineskip}{%
      \makebox[0pt][l]{\textnormal{FERMILAB-PUB-23-415-PPD}}
}}}%

\author[DES Collaboration]{
\parbox{\textwidth}{
\OrcidIDName{0000-0003-3312-909X}{D.~Anbajagane}\thanks{Corresponding author email: dhayaa@uchicago.edu}(\vcenteredhbox{\includegraphics[height=1.2\fontcharht\font`\B]{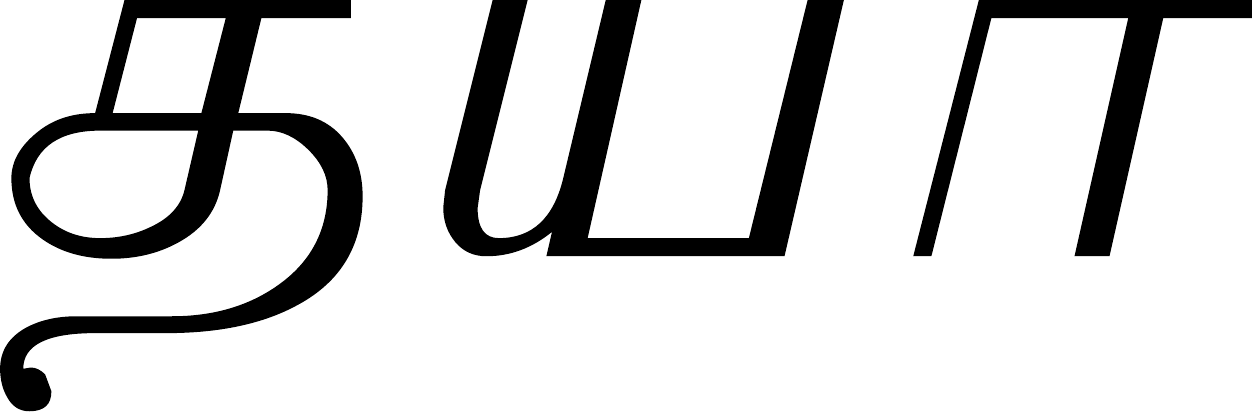}}),$^{1, 2}$ 
\OrcidIDName{0000-0002-7887-0896}{C.~Chang},$^{1,2}$
A.~Banerjee,$^{3}$
T.~Abel,$^{4,5,6}$
M.~Gatti,$^{7}$
V.~Ajani,$^{8}$
A.~Alarcon,$^{9}$
A.~Amon,$^{10,11}$
E.~J.~Baxter,$^{12}$
K.~Bechtol,$^{13}$
M.~R.~Becker,$^{9}$
G.~M.~Bernstein,$^{7}$
A.~Campos,$^{14}$
A.~Carnero~Rosell,$^{15,16,17}$
M.~Carrasco~Kind,$^{18,19}$
R.~Chen,$^{20}$
A.~Choi,$^{21}$
C.~Davis,$^{5}$
J.~DeRose,$^{22}$
H.~T.~Diehl,$^{23}$
S.~Dodelson,$^{14,24}$
C.~Doux,$^{7,25}$
A.~Drlica-Wagner,$^{1,23,2}$
K.~Eckert,$^{7}$
J.~Elvin-Poole,$^{26}$
S.~Everett,$^{27}$
A.~Fert\'e,$^{6}$
D.~Gruen,$^{28}$
R.~A.~Gruendl,$^{18,19}$
I.~Harrison,$^{29}$
W.~G.~Hartley,$^{30}$
E.~M.~Huff,$^{27}$
B.~Jain,$^{7}$
M.~Jarvis,$^{7}$
N.~Jeffrey,$^{31}$
T.~Kacprzak,$^{8}$
N.~Kokron,$^{4,5}$
N.~Kuropatkin,$^{23}$
P.-F.~Leget,$^{5}$
N.~MacCrann,$^{32}$
J.~McCullough,$^{5}$
J.~Myles,$^{4,5,6}$
A. Navarro-Alsina,$^{33}$
S.~Pandey,$^{7}$
J.~Prat,$^{1,2}$
M.~Raveri,$^{34}$
R.~P.~Rollins,$^{35}$
A.~Roodman,$^{5,6}$
E.~S.~Rykoff,$^{5,6}$
C.~S{\'a}nchez,$^{7}$
L.~F.~Secco,$^{2}$
I.~Sevilla-Noarbe,$^{36}$
E.~Sheldon,$^{37}$
T.~Shin,$^{38}$
M.~A.~Troxel,$^{20}$
I.~Tutusaus,$^{39,40,41}$
L.~Whiteway,$^{31}$
B.~Yanny,$^{23}$
B.~Yin,$^{14}$
Y.~Zhang,$^{42,43}$
T.~M.~C.~Abbott,$^{42}$
S.~Allam,$^{23}$
M.~Aguena,$^{16}$
O.~Alves,$^{44}$
F.~Andrade-Oliveira,$^{44}$
J.~Annis,$^{23}$
D.~Bacon,$^{45}$
J.~Blazek,$^{46}$
D.~Brooks,$^{31}$
R.~Cawthon,$^{47}$
L.~N.~da Costa,$^{16}$
M.~E.~S.~Pereira,$^{48}$
T.~M.~Davis,$^{49}$
S.~Desai,$^{50}$
P.~Doel,$^{31}$
I.~Ferrero,$^{51}$
J.~Frieman,$^{23,2}$
G.~Giannini,$^{52}$
G.~Gutierrez,$^{23}$
S.~R.~Hinton,$^{49}$
D.~L.~Hollowood,$^{53}$
K.~Honscheid,$^{54,55}$
D.~J.~James,$^{56}$
K.~Kuehn,$^{57,58}$
O.~Lahav,$^{31}$
J.~L.~Marshall,$^{59}$
J. Mena-Fern{\'a}ndez,$^{36}$
F.~Menanteau,$^{18,19}$
R.~Miquel,$^{60,52}$
A.~Palmese,$^{14}$
A.~Pieres,$^{16,61}$
A.~A.~Plazas~Malag\'on,$^{5,6}$
K.~Reil,$^{6}$
E.~Sanchez,$^{36}$
M.~Smith,$^{62}$
M.~E.~C.~Swanson,$^{63}$
G.~Tarle,$^{44}$
and P.~Wiseman$^{62}$
\begin{center} (DES Collaboration) \end{center}
}
}

\begin{document}
\label{firstpage}
\pagerange{\pageref{firstpage}--\pageref{lastpage}}
\maketitle

\begin{abstract}
Widefield surveys of the sky probe many clustered scalar fields --- such as galaxy counts, lensing potential, gas pressure, etc. --- that are sensitive to different cosmological and astrophysical processes. Our ability to constrain such processes from these fields depends crucially on the statistics chosen to summarize the field. In this work, we explore the cumulative distribution function (CDF) at multiple scales as a summary of the galaxy lensing convergence field. Using a suite of N-body lightcone simulations, we show the CDFs' constraining power is modestly better than the 2nd and 3rd moments of the field, as they approximately capture the information from all moments of the field in a concise data vector. We then study the practical aspects of applying the CDFs to observational data, using the first three years of the Dark Energy Survey (DES Y3) data as an example, and compute the impact of different systematics on the CDFs. The contributions from the point spread function are 2-3 orders of magnitude below the cosmological signal, while those from reduced shear approximation contribute $\lesssim 1\%$ to the signal. Source clustering effects and baryon imprints contribute 1-10\%. Enforcing scale cuts to limit systematics-driven biases in parameter constraints degrades these constraints a noticeable amount, and this degradation is similar for the CDFs and the moments. We also detect correlations between the observed convergence field and the shape noise field at $13\sigma$. We find that the non-Gaussian correlations in the noise field must be modeled accurately to use the CDFs, or other statistics sensitive to all moments, as a rigorous cosmology tool.
\end{abstract}





\section{Introduction}

The structure in the Universe --- namely the distribution of matter --- contains significant information on all kinds of physical processes; from the largest cosmological scales which probe the initial conditions of the Universe, to the galaxy and halo scales which probe both non-linear, gravitational evolution as well as baryonic imprints due to astrophysical processes, to the intra-galaxy scales where the gas and stellar phase space exhibit distinct structures from the rich physics of magneto-hydrodynamics. It is clear that the observed fields are abundant with information on both cosmology and astrophysics. It is then pertinent to question how best to extract the information from these fields, i.e. how best to maximize the constraints we can place on physical phenomena through measurements of these fields.

In the scenario where the field is a mean-zero Gaussian random field that is isotropic and homogeneous, the only degree of freedom for the field is the covariance between the pixels/voxels in real-space (or alternatively, the power spectra in Fourier-space). In such a scenario, it is clear that the maximal constraining power is obtained by measuring the power spectra, i.e. the only degree of freedom. For cosmological fields, the initial conditions seeding structure formation are Gaussian to a very good approximation, as has been verified by the cosmic microwave background (CMB) observations \citep[][]{Planck2016GaussianityTest, Planck2020GaussianityTest}, and a large part of the cosmological information in the resulting late time density field is still Gaussian, i.e. encoded in the variance of the field. Thus, the power spectra are a good way to extract information from the late-time fields as well.

However, there still remains significant, additional information beyond the power spectra. Even in the fiducial $\Lambda$CDM case --- where $\Lambda$CDM is the cosmological model with cold dark matter (CDM) and the cosmological constant $\Lambda$ --- and the initial conditions contain no primordial non-Gaussianities, the presence of nonlinear, gravitational evolution generates signatures beyond the power spectra. This is commonly called ``higher-order information''\footnote{Power spectra are referred to as ``2-point statistics'' and they capture up to second-order information as they are fundamentally a variance measure and contain two orders of the field. ``Higher-order'' here refers to higher than second-order information, which needs to be captured by beyond 2-point statistics, or sometimes referred to as ``higher-order statistics''.} and represents information in the field that is not captured by the power spectra. Such information still encodes signatures from cosmological and astrophysical processes, and is often highly complementary to the 2-point constraints; as a result, the combination of power spectra with higher-order information leads to constraints that are better than the trivial sum of the individual parts \citep[\eg][]{Fluri2018DeepLearning,  Fluri2019DeepLearningKIDS, Gatti2020Moments, Zurcher2021WLForecast, Gatti2022MomentsDESY3, Fluri2022wCDMKIDS, Lanzieri2023HOS}

There exists a rich body of literature on different, complementary ways to extract this non-Gaussian information from continuous scalar fields like the density field or the weak lensing convergence field. The N-point correlation functions (or their Fourier equivalents, the poly-spectra) are the most well-known and widely used statistic, and measure the correlation of $N$ points in space, where the points are separated by some distances. For $N = 3$, these statistics are computationally expensive to compute, and for $N = 4$ they are mostly prohibitive unless measured in specific limiting cases. Given this, many alternative methods have been explored to capture some/all of this information in a computationally inexpensive way. Some of the most commonly known/used methods include moments \citep{Petri2015MomentsMinkowski, Peel2018Moments, Chang2018MassMap, Gatti2020Moments, Gatti2022MomentsDESY3}, Minkowski Functionals \citep{Mecke1994Minkowski, Blake2014Minkowski, Petri2015MomentsMinkowski, Parroni2020Minkowski}, density-split statistics \citep{Gruen2018DensitySplitY1, Friedrich2018DensitySplit} and more. Similar statistics exist for the discrete fields, such as counts-in-cells \citep{Baugh1995CIC, Adelberger1998CIC} and the k-Nearest Neighbor (kNN) distributions \citep{Banerjee2021kNN, Banerjee2021CrossCorr}. For the weak lensing field, the 3-point information has been pursued either through the direct measurement or approximate summaries like the density-split statistics \citep{Gruen2018DensitySplitY1, Friedrich2018DensitySplit}, mass aperture moments \citep{Secco2022MassAp}, field moments \citep{Petri2015MomentsMinkowski, Gatti2020Moments, Gatti2022MomentsDESY3}, and integrated shear functions \citep{Halder2021Integrated3ptShear}. Weak lensing peaks \citep{Kratochvil2010WLPeaks, Shan2018HOSKiDS, Martinet2018HOSKiDS,  Zurcher2022WLPeaks} probe a specific, fixed combination of N-point functions, as is the case with other statistics like cosmic void distribution functions \citep{Davies2021WLVoids} and persistent homology \citep{Heydenreich2021BettiNumbersWL, Heydenreich2022Y1BettiNumberWL}. Field-level inference tools are also employed \citep{Fluri2018DeepLearning, Jeffrey2020DLReconstruction, Fluri2019DeepLearningKIDS, Fluri2022wCDMKIDS}, while others explore machine learning-informed, but still interpretable, statistics such as scattering transforms \citep{Cheng2021WeakLensingST} and wavelet phase harmonics \citep{Allys2020WPHandLSS}.

An outstanding question is identifying the ``maximally'' informative statistic for summarizing, and extracting constraints from, the fully nonlinear late-time density/convergence field. This is an unsolved problem given we do not a priori know the exact cosmological information contained in the different non-Gaussian signatures (including those beyond the 3-point function) across both linear and non-linear scales. Thus, to ensure we use all the available cosmological information in the field, it is desirable to consider statistics that capture all orders of statistical information (rather than just one order, or a specific combination of orders). The kNN distributions have been formally shown to be such a statistic for discrete tracers \citep{Banerjee2021kNN} as they capture volume integrals of all N-point auto/cross-correlation functions of the field. While these kNN distributions are constructed for discrete tracer fields, \citet{Banerjee2023TracerFieldkNN} demonstrated that the analogous statistic for continuous fields are the CDFs of the field smoothed on different length scales.

The CDFs --- or the probability distribution functions (PDFs), which are interchangeable ideas given they are connected by a linear integral transform --- are the main statistic of focus in this work and have been theoretically known as a good non-Gaussian statistic for lensing fields since more than two decades ago \citep{Jain1998LensingPDF, Kruse2000NGTail}. The CDF is also an intuitive, visually informative statistic for non-Gaussian features and is often used to check and validate reconstructed lensing fields \citep{White2000WLsimulations, Chang2018MassMap, Niall2021MassMap}. Previous works have also shown that the  lensing PDF significantly improves constraints in $w\rm CDM$ compared to the standard 2-point functions \citep{Giblin2023LensingPDF}, while more works have shown the utility of the \textit{3D matter density} PDF in probing both $w\rm CDM$ and other extended cosmologies \citep{Uhlemann2020PDFNeutrino, Friedrich2020PDFBulkfNL,  Boyle2021MatterPDF, Gough2022MatterPDFMG, Cataneo2022MatterPDFMG}.

While the benefits of using the CDF --- namely the level of cosmological non-Gaussianity it can capture --- have been explored in the past, this has mostly been in the more idealistic regime where some key observational factors were not included in the analysis. Thus, while we have had a prior understanding of the benefits of using PDFs/CDFs of the lensing field, we currently have an incomplete picture of the practical challenges in using this statistic to infer cosmological constraints. 

In this work, we measure the CDFs of the lensing field from the first three years (Y3) of the Dark Energy Survey (DES) data and validate that the common lensing systematics --- such as point spread function (PSF) contributions, reduced shear approximation, source clustering, and baryon imprints --- have an impact on this statistic that is either negligible or can be adequately mitigated. Many of these tests have been extensively performed for 2-point statistics \citep*{Gatti2021ShearCatalog} and have also been done for some 3-point statistics \citep{Secco2022MassAp, Gatti2022MomentsDESY3}. The CDFs are sensitive to information at all orders, and validating the impact of these observational/modelling systematics on the CDFs also provides validation for higher-order information beyond the 3-point.

This work is organized as follows: first, we introduce the formalism for the CDFs in \S\ref{sec:Formalism}. In \S\ref{sec:data} we describe the datasets and simulations used in this work, as well as the procedures used to forward-model the simulations to match the DES Y3 data. In \S\ref{sec:AnalysisFisher} we define the data vector used for the rest of this work, and also demonstrate the Fisher constraining power of the CDFs for DES Y3-like data. In \S\ref{sec:ResultsDESY3} we measure the CDFs on the DES Y3 weak lensing maps, and quantify the signal-to-noise of the measurements. We then validate the impact of different effects --- PSF contributions, source clustering, reduced shear approximation, and baryonic imprints --- on this statistic and discuss any scale cuts required to mitigate these effects. Finally, we conclude in \S\ref{sec:Conclusion}.

\section{CDF Formalism} \label{sec:Formalism}

We begin in \S \ref{sec:CDFs} by describing the formalism of the CDF statistics used in this work, including the exact measurement procedure. In \S \ref{sec:kNNReview} we briefly review the k-Nearest Neighbor distributions, which are a recently introduced statistic for discrete tracers that summarize all higher-order information, and we discuss how the analogous, continuous-field statistic is the CDF. Finally, in \S \ref{sec:GaussianConsistency} we validate the CDFs using Gaussian fields. Note that the CDFs are closely related to other statistics in the literature and we will describe these later on in \S \ref{sec:Conclusion}.

\begin{figure*}
    \centering
    \includegraphics[width = 1.8\columnwidth]{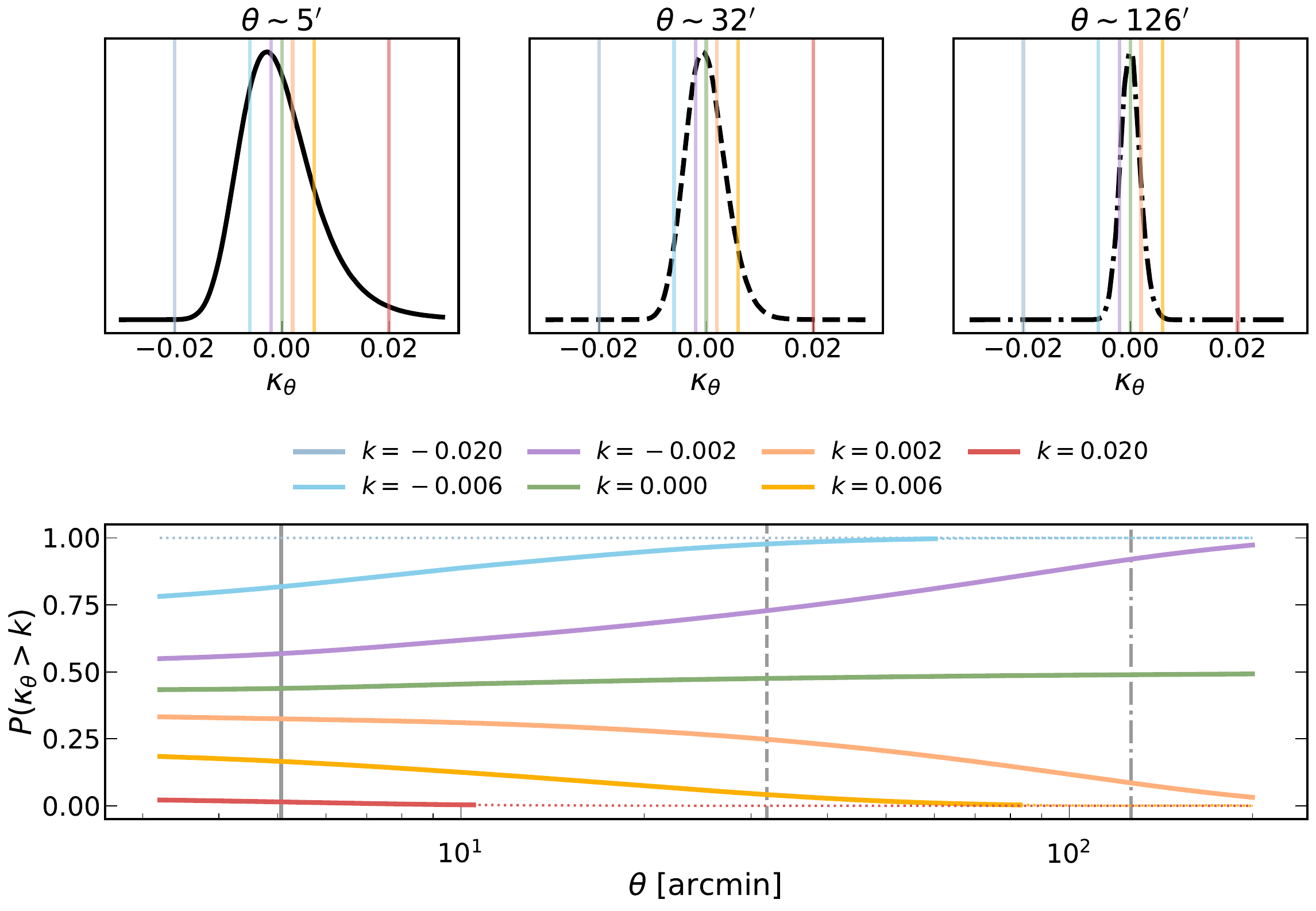}
    \caption{\textbf{Bottom:} The probability that $\kappa_\theta$, the average convergence within circles of apertures $\theta$, exceeds a chosen threshold $k$. We use seven thresholds and show measurements for a noiseless convergence field corresponding to the fourth tomographic redshift bin in DES Y3. The solid lines are converted to dotted ones when the CDFs fall into the $99.7\%$ ($3\sigma$) tail. The gray-blue line is always in the tail for this particular measurement. \textbf{Top:} The PDFs of $\kappa_\theta$ for different choices of aperture, $\theta$. The three aperture scales that we show PDFs for are indicated by the vertical gray lines in the bottom panel. The PDFs are estimated from noiseless convergence fields and are smoothed with a Gaussian for visualization purposes. The vertical lines in these top three panels are the thresholds we use. The probability to exceed is the integral from each threshold up to $P(\kappa = \infty)$. For high thresholds, we have a lower probability to exceed and vice-versa for low thresholds.}
    \label{fig:Illustration}
\end{figure*}

\subsection{Cumulative Distribution Functions} \label{sec:CDFs}

The CDFs\footnote{The entire formalism could also be done using PDFs instead of CDFs. The latter is simply a more natural/convenient choice when connecting to the kNN formalism, as we describe in Section \ref{sec:kNNReview}.} used in this work are defined as follows. Given a set of uniform/random points in a field, with spheres of radius $r$ around each point, the CDFs summarize the fraction of spheres that have an enclosed density --- i.e. the mean density within radius $r$ --- that exceeds a chosen threshold. In 2D, the density becomes a surface density, $\Sigma$, and the radius is a projected aperture, $\theta$. The calculation of the fraction of points whose enclosed surface densities exceed a threshold can be formally written down using the following expression,
\begin{equation}
    \CDF(\theta, k) = P(\kappa_\theta > k),
\label{eqn:threshold_condition}
\end{equation}
where $\kappa_\theta \equiv \kappa(<\thetasmooth)$ is the average surface overdensity within an aperture $\thetasmooth$. This measurement can also be trivially modified to use the surface density, rather than overdensity, just switching $\kappa \rightarrow \bar{\Sigma}(1 + \kappa)$, where $\bar{\Sigma}$ is the mean surface density field. It can also be done with the surface \textit{mass}, by simply multiplying the surface density with the aperture area associated with scale $\theta$.

For a given map, the CDF measurement is performed as follows: 

\textbf{First}, we fill the map with a grid of points.
Without loss of generality, we take these points to be located at the center of the HEALPix pixels (with $\texttt{NSIDE} = 1024$), as this greatly simplifies the calculations. Increasing the number of points in the grid (i.e. the number of pixels) will improve the precision of the measurement, as is the case with the traditional 2-point correlations.

\textbf{Second}, we pick a certain aperture scale, $\thetasmooth$, and for each point we compute $\kappa_\theta$, the convergence smoothed on scale $\theta$. The smoothing is done in harmonic space using a harmonic tophat filter
\begin{equation}
    B(\ell) = 2\frac{J_1(\ell\thetasmooth)}{\ell \thetasmooth},
\end{equation}
where $J_1(x)$ is the Bessel function of the first order. The choice of tophat over a Gaussian filter is because the former allows for an easy interpretation of an enclosed quantity within a given physical scale. Our computing procedure is the same for any other choice of filter as well.

\textbf{Third}, we measure what fraction of the grid points satisfy the inequality in equation \ref{eqn:threshold_condition}, which is the probability, $P(\kappa_\theta > k)$. The choice of thresholds is a degree of freedom in the measurement, and we describe our choices in Section \ref{sec:DataVectorDef}.

\textbf{Fourth and finally,} steps 2 and 3 are repeated for a range of scales and thresholds to extract the distribution, $P(\kappa_\theta > k)$, for different choices of $\theta$. The exact choice of scales and thresholds used in this work is described in Section \ref{sec:DataVectorDef}. 

Figure \ref{fig:Illustration} illustrates how the CDFs are constructed in a given field, and highlights some generic features of the CDFs. In the limit where the variance $\sigma^2 \rightarrow \infty$, we expect $P(\kappa_\theta > k) \rightarrow 0.5$, and where $\sigma^2 \rightarrow 0$, then we expect $P(\kappa_\theta > k) \rightarrow 0$ if $k > 0$, and $P(\kappa_\theta > k) \rightarrow 1$ if $k < 0$. In Figure \ref{fig:Illustration} we see that all curves are closer to $P = 0.5$ on small scales where the field's variance is high compared to the threshold values, and move towards $P = 0$ or $P = 1$ on large scales where the large smoothing scale suppresses the field's variance to values lower than the thresholds. Additionally, we see $P(\kappa_\theta > 0) \approx 0.4$ at small-scales, where the distribution is log-normal (see top panels of Figure \ref{fig:Illustration}) and so the median of the distribution is not the same as the mean, $\langle \kappa \rangle = 0$. At large-scales, we find $P(\kappa_\theta > 0) \approx 0.5$ as the distribution becomes more Gaussian.

Thinking in 3D space, the CDFs extract $P(>\rho \,|\, R)$, the conditional distribution of the enclosed mean density given radius, as well as $P(R \,|\, >\rho)$, the conditional distribution of radii or volumes given a density threshold. These two distributions can be related using Bayes' theorem,
\begin{equation}\label{eqn:HaloCollapseEqnRho}
    P(>\rho\,|\,R)  = P(R\,|\,>\rho)\frac{P(>\rho)}{P(R)}.
\end{equation}
Note that given the enclosed density $\rho$ and spherical radius $R$, we can easily obtain a mass $M \equiv \frac{4}{3}\pi R^3\rho$. So the above can be rewritten as
\begin{equation}\label{eqn:HaloCollapseEqnMass}
    P(>M\,|\,R)  = P(R\,|\,>M)\frac{P(>M)}{P(R)}.
\end{equation}
Equation \eqref{eqn:HaloCollapseEqnRho} better elucidates the connection between the CDFs and the ideas from halo collapse. The quantity $P(>200\rho_c\,|\,R)$ is simply the fraction of volumes that contain a halo, where the halos are identified/defined as overdensities of at least $\rho >200\rho_c$, with $\rho_c$ being the critical density of the Universe.

We can also generalize the CDF formalism to multi-field probes by computing the joint CDFs of multiple fields; this is simply,
\begin{equation}
    P(\kappa_{\theta, 1} > k_1, \kappa_{\theta, 2} > k_2\,|\,\thetasmooth),
\end{equation}
where $\kappa_{\theta, 1}$ and $\kappa_{\theta, 2}$ are two different fields (eg. different tomographic bins of a single type of field, or different types of fields). While we are allowed to choose different values for the thresholds $k_1$ and $k_2$, we will enforce $k = k_1 = k_2$ henceforth for simplicity in the data vector. In this work, we will consider the cross-correlation between tomographic bins as part of our measurement. Note that the 2-field version of the CDFs formally contains all the 1-field information as well. This connection is identical to how 2D PDFs contain the marginal 1D distributions within them. \footnote{A simple example is the 2D CDF, $P(\kappa_{\theta, 1} > k_1, \kappa_{\theta, 2} > k_2\,|\,\thetasmooth)$ taken in the limit $k_2 = -\infty$. In this case, $\kappa_{\theta, 2}$ is always above the threshold $k_2$ and so the 2D CDF reduces to a 1D CDF, $P(\kappa_{\theta, 1} > k_1, \kappa_{\theta, 2} > k_2\,|\,\thetasmooth) \rightarrow P(\kappa_{\theta, 1} > k_1\,|\,\thetasmooth)$} We will use both 1-field and 2-field CDFs as part of our main data vector. The 3-field and 4-field CDFs will formally have additional information beyond the 1-field and 2-field CDFs, though our tests have shown there is only marginal improvement in cosmological constraints for the analysis choices described here (\eg tomographic bin, angular scales, and thresholds).

For some tests, we will also postprocess the 2-field CDFs to isolate just the cross-covariance/correlation. This is done by performing the redefinitions described in \citet{Banerjee2021CrossCorr},
\begin{equation}\label{eqn:CrossCorr}
    \psi_{1, 2}(k) = \CDF_{1, 2}(k) - \CDF_1(k)\CDF_2(k),
\end{equation}
which takes the joint probability to exceed in two different fields and removes the product of the individual probability to exceed for each field. The quantity $\psi_{1, 2}(k)$ is 0 if the fields are completely uncorrelated, and non-zero otherwise. The sign of $\psi_{1, 2}(k)$, for any threshold $k$, indicates the sign of the correlation between the two fields at that threshold.

We can also extend this formalism to more than 2 fields (eg. an $\mathcal{ABC}$ correlation of three fields). While we do not consider such measurements in our analysis here, we note their potential utility both for cosmological information, but also as further compressions of the data vector. Note that there is no benefit to repeating a field twice (eg. $\mathcal{AAB}$) \textit{if} we also fix the threshold $k$ for all the fields. The joint probability $P(\kappa_1 > k, \kappa_1 > k, \kappa_2 > k)$ is exactly similar to $P(\kappa_1 > k, \kappa_2 > k)$.

While we have discussed the CDFs in terms of lensing convergence, it is not necessary to be limited to this quantity. For example, one could consider the kinetic or thermal Sunyaev-Zeldovich fields \citep{Sunyaev1972SZEffect, Carlstrom2002SZReview}, which are generated by baryons in halos and thus inherit the non-Gaussian features of the structure traced by these halos.

\subsection{Connection to kNN distributions for discrete fields} \label{sec:kNNReview}

The kNN distributions \citep{Banerjee2021kNN, Banerjee2021CrossCorr} are a novel way to summarize the clustering in a field of discrete tracers, such as galaxies or halos. They have been formally shown to capture volume integrals of all N-point functions of the tracer field, but can be computed in $\mathcal{O}(N\log N)$ time, where $N$ is the number of tracers. Thus, they have the same computational efficiency as a 2-point correlation function, but capture integrals of all the information held in the N-point functions (2-point, 3-point, 4-point, etc.). This statistic has already been measured in observational data, particularly to quantify the signal-to-noise of all correlations (both Gaussian and non-Gaussian) in a clustered field \citep{Wang2022kNNSDSS}.

The kNNs are computed by taking a field of tracers with a known number density $n_{\rm tr}$, and then generating a large set of random points in this field as one would for computing an N-point clustering function (although a set of uniform points would be a sufficient choice as well). For each point, one computes the distance to the nearest tracer neighbor. The distribution of distances to the $k^{\rm th}$ nearest neighbor forms a kNN distribution. This statistic is probing the distribution $P(V\,|\,>k_{\rm tr})$, i.e. the distribution of volumes that contain at least $k_{\rm tr}$ tracers, where $k_{\rm tr}$ takes integer values. Assuming spherical volumes, this can be reformulated as the distribution $P(R\,|\,>k_{\rm tr})$. Given kNNs depend on the counts of tracers enclosed within a volume, it is sensitive to volume integrals of all the correlation functions. However, the fact that the sensitivity is to a \textit{volume integral} of the functions means signals from specific configurations of the N-point functions will be mixed together. \footnote{For the 2-point function, there is no configuration information as the correlations depend on just distance, $r$. For N-point correlations of N > 3, the geometry connecting the N points will contain additional information, though the exact information contained in this geometry remains an open question.}

In the limit of $n_{\rm tr} \rightarrow \infty$, the number counts threshold $>k_{\rm tr}$ becomes a density threshold $>\rho$, and the conditional distribution becomes $P(R\,|\,>\rho)$ which can be related, using Bayes' theorem, to the distribution probed by the CDFs, $P(>\rho\,| R)$. A detailed discussion on this connection between kNNs and CDFs be found in \citet[see their Section 2.1]{Banerjee2023TracerFieldkNN}. The analytic connection between the two statistics directly confirms that the CDFs can be \textit{formally} shown to contain all volume integrals of higher-order functions, and this makes them better suited for summarizing a field, where we do not apriori know the exact cosmological information contained in all the non-Gaussian signatures of the field. In addition, this connection means the CDFs are the natural statistic to cross-correlate discrete and continuous fields while using the kNN formalism for the former \citep{Banerjee2023TracerFieldkNN}.

\subsection{Consistency relations for Gaussian fields} \label{sec:GaussianConsistency}

In the Gaussian limit of $P(\kappa_\theta) = \mathcal{N}(\kappa_\theta;\mu,\sigma)$ --- where $\mathcal{N}$ is a normal distribution with mean $\mu$ and variance $\sigma^2$ --- there are three degrees of freedom for the CDF: the mean and variance of the map at each aperture scale, and the threshold $k$. The threshold is an input parameter, and the mean of the map is taken to be $\mu = 0$ given $\kappa$ is derived from the overdensity field and so is defined as a perturbation field with the mean background subtracted. Thus, the variance is the only unconstrained parameter, and this variance can also be measured directly on the map. Formally, a Gaussian CDF is parameterized as,
\begin{equation}
    \CDF(k) = 1 - \int_{k}^\infty \mathcal{N}(x - \mu, \sigma)dx = \frac{1}{2}\bigg[1 + \erf\bigg(\frac{{k} - \mu}{\sigma\sqrt{2}}\bigg)\bigg].
\end{equation}

We can thus use the variance measured from the map smoothed on a given scale, $\theta$, to predict the CDFs at that scale. For a purely Gaussian field, the measurements and predictions must agree. The same exercise is trivially extended for the 2-field CDFs. In the Gaussian limit, the joint PDF of any set of fields is given by a multivariate normal distribution,
\begin{equation}\label{eqn:multivariate_normal}
    \PDF = \frac{1}{\sqrt{(2\pi)^n\det{\Sigma}}}\exp\bigg[-\frac{1}{2}(\vec{\kappa}- \mu)^T\Sigma^{-1}(\vec{\kappa}- \mu)\bigg],
\end{equation}
where the column vector $\vec{\kappa} = \{\kappa_1, \kappa_2, \ldots \kappa_n\}$ are the kappa value in each field, and denote the point in multi-dimensional space where we evaluate the probability. The PDF in Equation \eqref{eqn:multivariate_normal} can be integrated, assuming some set of thresholds for each field, to obtain the CDF. Recall that in this work we set all thresholds to the same value $k$. We also use $\mu = 0$. The unknown degrees of freedom for the distribution are then entirely in the covariance matrix. Thus if we know this covariance matrix, we can always predict the CDFs exactly.

\begin{figure}
    \centering
    \includegraphics[width = \columnwidth]{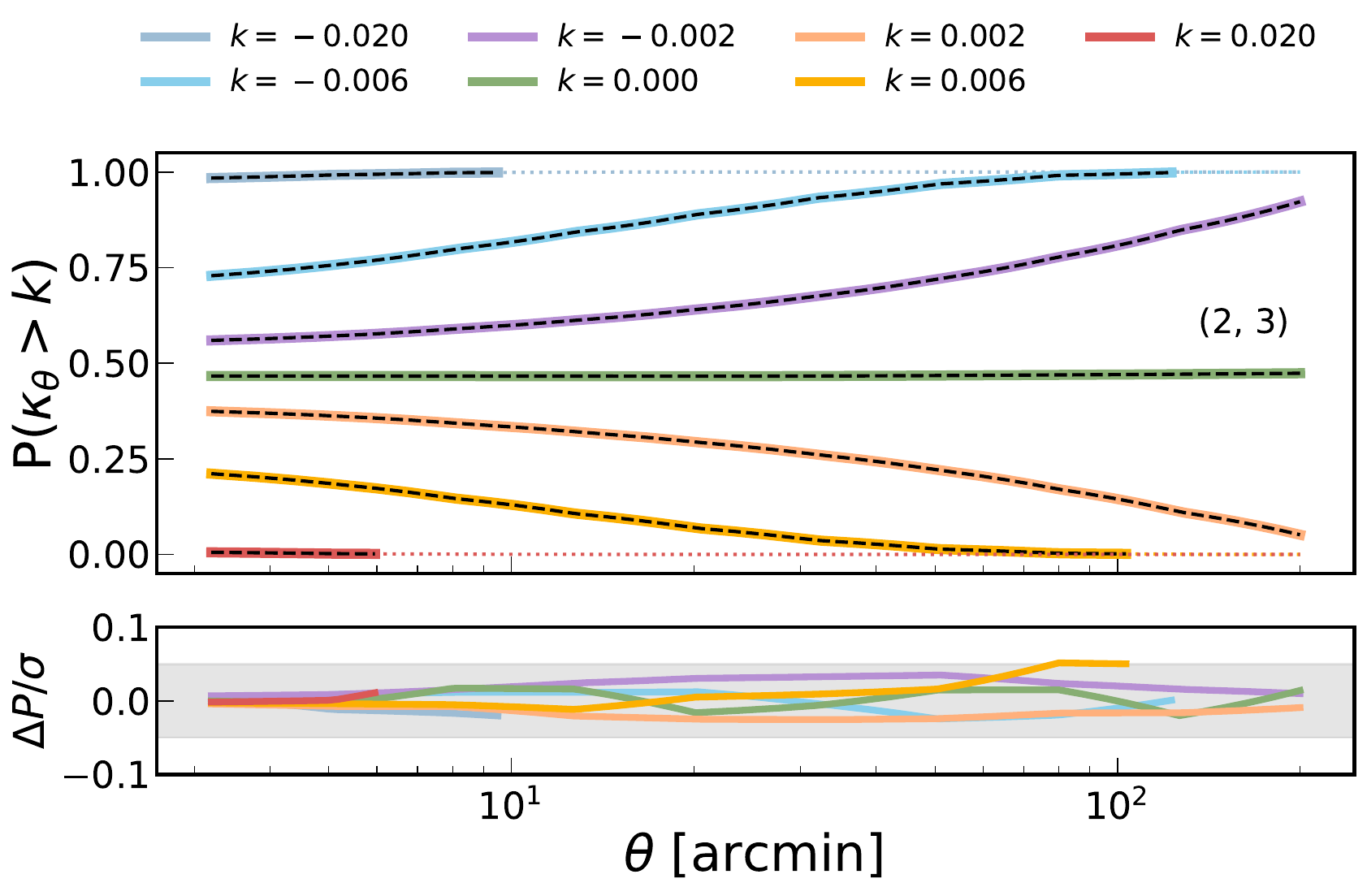}
    \caption{\textbf{Top:} The 2-field CDFs averaged over an ensemble of 1000 \textit{noiseless, full-sky, Gaussian} convergence maps. The $n(z)$ for the two fields corresponds to the third and fourth tomographic bins of DES Y3. The solid lines are converted to dotted styles when the CDF is outside the range $[0.003, 0.997]$ (approximately corresponding to the 3$\sigma$ bounds). The black dashed lines show the predictions for the CDFs given the covariance of the two fields at a given smoothing scale, $\theta$; under the assumption the fields are Gaussian, the predictions must match the measurement. \textbf{Bottom:} The difference between the CDF measurement and Gaussian-field predictions, $\Delta P = P_{\rm meas} - P_{\rm theory}$ normalized by the uncertainty in the CDFs --- where the uncertainty is cosmic variance and is the observational limit for measurement uncertainty --- estimated from the 1000 realizations. The gray band shows $\Delta/\sigma < 0.1$. In all cases the difference, $\Delta P$, is within this region and is completely negligible.}
    \label{fig:ConsistencyRelation}
\end{figure}

We verify this in Figure \ref{fig:ConsistencyRelation} for our analysis setup. The top panel shows the 2-field CDF measured on noiseless, simulated maps whose signal mimics the DES Y3 data used in this work (see Section \ref{sec:DESY3Data} for more details). In particular, the convergence map has the same redshift distribution as the 3rd and 4th tomographic bins. These are all Gaussian maps made by postprocessing N-body products, as detailed below in Section \ref{sec:GaussianMaps}. The dashed lines (prediction) are consistent with the solid ones (measurement). The bottom panel shows the Gaussian model predictions are within $0.05 \sigma$ of the measurements, where the $\sigma$ of the data vector is just cosmic variance and thus represents the observational limit in precision.

\section{Data}\label{sec:data}

We first describe in Section \ref{sec:simulations} the different simulations used in our analysis. We then detail the DES Y3 data in Section \ref{sec:DESY3Data} and in Section \ref{sec:ForwardModelProcess} we describe how the simulated maps are forward modelled to imitate the DES Y3 data.

All maps used in this work are made with the \textsc{Healpix} convention of $\texttt{NSIDE}=1024$. This corresponds to a pixel scale of 3.2 arcmin. The one exception are the products used from the \textsc{Cosmogrid} suite, described in Section \ref{sec:Cosmogrid}, which are \texttt{NSIDE} = 512.

\subsection{Simulations} \label{sec:simulations}

While the CDF is a statistic that can be used to summarize any scalar field, in this work we are specifically interested in the lensing convergence, $\kappa$, which is a line-of-sight integral of the density field,
\begin{equation}\label{eqn:convergence_definition}
    \kappa(\nhat, z_s) = \frac{3}{2}\frac{H_0^2\Omega_{\rm m}}{c^2}\int_0^{z_s}\!\!\!\delta(\nhat, z_j) \frac{\chi_j(\chi_s - \chi_j)}{a(z_j)\chi_s}dz_j\frac{d\chi}{dz}\bigg|_{z_j},
\end{equation}
where $z_s$ is the redshift of the ``source'' plane/galaxies being lensed, $\nhat$ is the pointing direction on the sky, $\delta$ is the overdensity field, $\chi$ is the comoving distance from an observer to a given redshift, $a$ is the scale factor, $H_0$ is the Hubble constant, $\Omega_{\rm m}$ is the matter energy density fraction at $z = 0$, and $c$ is the speed of light. We use the shorthand $\chi(z_s) \equiv \chi_s$ and $\chi(z_j) \equiv \chi_j$.

We model this convergence using full-sky density maps from different N-body simulations, with each simulation serving a different purpose in this work. We detail these different simulations below. Such simulations are uniquely suited for modeling these fields in the non-linear regime. For quasi-linear and linear regimes, analytic models can also be utilized \citep[\eg][]{Barthelemy2023PDF}

\subsubsection{Anbajagane23 simulations (A23)}\label{sec:A23Sims}

In this work, we use a suite of N-body simulations run with the \textsc{Pkdgrav3} solver \citep{Potter2017Pkdgrav3}, where the suite has been specialized for performing Fisher forecasts for widefield surveys. This simulation suite will be formally introduced as part of a future work \citepinprep{Anbajagane}, and we describe here just the essential features of the runs and the relevant data products used in this work. The A23 simulations are run in $1 \hinv\gpc$ boxes, starting at $z = 127$, with $N = 512^3$ dark matter particles. The initial conditions for all the A23 simulations are obtained from the \textsc{Quijote} suite \citep{Navarro2020Quijote}, and so the simulations are essentially lightcone runs of the \textsc{Quijote} simulations specialized for widefield survey analyses. The original \textsc{Quijote} suite was designed for studying the Fisher information of the nonlinear structure, as well as building emulators sampling different cosmological parameters, but the data products are inadequate for producing mock lightcones of the lensing/density field. These products include snapshots and halo catalogs at only five redshifts, which is too coarse a redshift resolution for building lightcones. Hence we have rerun a subset of these simulations to create accurate full-sky lensing and density maps.

The A23 suite contains 100 simulations for computing the derivatives of the lensing/density field with respect to multiple cosmological parameters, of which three are of interest to us --- $\Omega_m$, $\sigma_8$, and $w$. The suite runs 100 simulations with parameter values slightly higher than the fiducial, and 100 simulations with values slight lower than the fiducial, and these two sets are used compute these derivatives. The fiducial cosmology is from  \citet{Planck2016CosmoParams}, and the derivatives are computed over differences of $\Delta \Omega_m = 0.02$, $\Delta \sigma_8 = 0.03$ and $\Delta w = 0.05$, which are all the same settings as the \textsc{Quijote} suite. The A23 suite also has $2000$ simulations at the fiducial cosmology which are used to compute the covariance matrix for our data vector. Since each all-sky map can have 4 completely independent DES footprints, we have a total of $8000$ realizations to use for the covariance, and $400$ realizations to compute the derivatives for the Fisher forecast.

While the original \textsc{Quijote} suite was run using \textsc{Gadget3} \citep[last described in][]{Springel2005Gagdet2}, we use \textsc{Pkdgrav3} which has already been employed extensively to perform both theoretical studies of the lensing field as well as simulation-based analyses of data from different weak lensing surveys \citep{Fluri2019DeepLearningKIDS, Gatti2022MomentsDESY3, Zurcher2022WLPeaks}. The \textsc{Pkdgrav3} solver automatically builds lightcones as it solves the gravitational dynamics of the system forward in time, and so our final outputs are the lightcone shells --- i.e. \textsc{Healpix} maps --- of the density field at different redshifts. The simulation box is tiled/repeated as needed to construct large enough volumes to then build full-sky lightcones to a given redshift. This repetition will bias any large-scale correlations in the lightcone, but in this work we only consider scales much smaller than the box size. 

The simulations have a total of 100 timesteps/shells, with 95 shells between $0 < z < 10$. This gives us a high redshift resolution of between $\Delta z \approx 0.01 - 0.05$ in that redshift range, with the exact value depending on the shell. The timesteps in this redshift range are spaced uniformly in proper time, $t$, and this corresponds to different $z$ and comoving distances depending on the cosmology. These density shells are then post-processed via Equation \eqref{eqn:convergence_definition}, with the integral over $z_j$ replaced by a simple discrete sum, to create a lensing convergence field at each source plane redshift, $z_s$. This technique uses the Born approximation, which computes the total effective deflection due to lensing but along an undeflected ray path. A more precise calculation uses full rayracing, which calculates these deflections while constantly deflecting/updating the ray path. \citet{Petri2017Born} found the Born approximation leads to differences of $\lesssim 5\%$ for the third moments statistic we will use in Section \ref{sec:MainFisher}, but this is subdominant to the current uncertainties of $\approx 15\%$.

Note that we have not performed any resolution-convergence tests. The numerical requirements for this work are less stringent as we do not use the simulations for cosmological inference, but rather for (i) performing a Fisher analysis (Section \ref{sec:MainFisher}), where the relevant quantities are relative and not absolute differences in the simulations as we vary cosmological parameters, and for (ii) computing covariance matrices for our systematic checks (Section \ref{sec:ResultsDESY3}).

\subsubsection{Takahashi17 simulations (T17)}\label{sec:Takahashi}

The Takahashi17 simulations \citep{Takahashi2017Sims} are a suite of N-body simulations run at a WMAP9 cosmology \citep{Hinshaw2013WMAP9}, and have a higher particle resolution than the A23 suite, with $2048^3$ particles. They, however, have lower redshift resolution than the A23 suite with $38$ shells between $0 < z < 5$. The shells are spaced equally in comoving distance, with widths of $150 \mpc \hinv$, and this leads to redshift spacing of roughly $\delta z \sim 0.05-0.2$. The T17 simulations have been used to model/test higher-order statistics in many works \citep{Gatti2020Moments, Secco2022MassAp, Munshi2023PosDepCorr, Gong2023Integrated3pt, Heydenreich2023ThirdOrder} for modelling, validation etc. and so we measure our statistics on these simulations for completeness. There are 108 independent full-sky maps, and that gives us a total of 432 DES Y3 cutouts.

\subsubsection{Cosmogrid}\label{sec:Cosmogrid}

\textsc{Cosmogrid} is a large suite of simulations that span the $w\rm CDM$ parameter space, including the sum of the neutrino masses, and are designed for simulation-based modelling of widefield survey data \citep{Kacprzak2023Cosmogrid}. They were run using \textsc{Pkdgrav3}, similar to the A23 simulations, and have a $900 \mpc/h$ box size with $832^3$ particles. The simulations are run at 2500 points spanning the parameter space, with 7 realizations at each point. They have 140 timesteps, with 70 equally spaced steps in proper time between $4 < z < 99$, and another 70 equally spaced steps in proper time between $0 < z < 4$. The spacing is different in each of the two regimes.

In this work, we use \textsc{Cosmogrid} to test the impact of baryons on the lensing CDF statistic. For this purpose, we use the fiducial runs which are 200 simulations run at fixed cosmology \citep[][see their Table 2]{Kacprzak2023Cosmogrid}. We use both the default N-body run as well as the run postprocessed using the method of \citet{Schneider2019Baryonification} so the density field looks like that of a hydrodynamic simulation with baryons. We discuss this more in Section \ref{sec:Baryons}. While the raw maps are available at \texttt{NSIDE} = 2048, the maps we used to study baryons have \texttt{NSIDE} = 512 --- which is lower than the fiducial resolution of \texttt{NSIDE} = 1024 used in thus work --- and we discuss the impact of this in Section \ref{sec:Baryons} as well.

\subsubsection{Gaussian maps}\label{sec:GaussianMaps}

For the purpose of validating non-Gaussian statistics, it is useful to have maps that are purely Gaussian --- i.e. are represented entirely by a power spectrum --- rather than ones that contain a realistic level of non-linearity/non-Gaussianity. We use the power spectrum measured on the N-body maps, which contain the relevant non-linearities, to then create consistent Gaussian maps. These maps will by construction have the same non-linear power spectra as the original maps. The method employed for doing this is the same as \citet[][see their Appendix A]{Giannantonio2008Gaussianize}. It involves computing all auto- and cross-spectra between the relevant fields on the simulated maps, and then using these spectra with random phases to generate spherical harmonic modes $a_{\ell m}$ that are appropriately correlated to reproduce the input auto- and cross-power spectra. The $a_{\ell m}$ can then be transformed to obtain a real-space map. By definition, such maps will have no higher-order information and be described entirely by their power spectra.

If we have two maps $X$ and $Y$, and want to generate Gaussian maps that have the same auto and cross-power spectrum as $X$ and $Y$, we obtain the $a_{\ell m}$ via

\begin{align} \label{eqn:GaussianizeMaps}
    a_{\ell m}^X & = \eta_{\ell m}^X T^{XX} = \eta_{\ell m}^X \sqrt{C_\ell ^{XX}}, \nonumber\\[10pt]
    a_{\ell m}^Y & = \eta_{\ell m}^X T^{XY} + \eta_{\ell m}^Y T^{YY} \nonumber\\
    & = \eta_{\ell m}^X \frac{C_\ell ^{XY}}{\sqrt{C_\ell ^{XX}}} + \eta_{\ell m}^Y \sqrt{C_\ell ^{YY} - \frac{(C_\ell ^{XY})^2}{C_\ell ^{XX}}},
\end{align}
where $\eta_{\ell m}$ is a complex random normal variable with zero mean and unit variance, and $T_{ij}$ are coefficients derived from the power spectra, with a general form given by,

\begin{equation} \label{eqn:GaussianizeTij}
T^{ij}=
\begin{cases}
		 \sqrt{C^{ji}-\sum_{k=1}^{j-1}(T^{ik})^{2}}	, & \text{if $i=j$;}\\[10pt]
            \frac{1}{T^{jj}}\bigg(C^{ji}-\sum_{k=1}^{j-1}T^{ik}T^{jk} \bigg), & \text{if $i>j$.}
		 \end{cases}
\end{equation}
and Equations \eqref{eqn:GaussianizeMaps} and \eqref{eqn:GaussianizeTij} above have been reproduced from \citet[][see Appendix C]{Omori2022Agora}. 

For producing real maps, the $m = 0$ coefficients must be handled separately as they should have no imaginary component \citep[see Appendix B in][for an example]{Sellentin2023Almanac}. Thus, we explicitly remove their imaginary component, by setting ${\rm Im}(a_{\ell m=0}) = 0$, and then rescale the coefficients as $a_{\ell m=0} \rightarrow \sqrt{2}a_{\ell m=0}$.\footnote{Formally, our complex variable satisfies $\langle \eta \rangle = 0$ and $\langle \eta \eta^*\rangle = 1$. Thus, the real and imaginary components of $\eta$ have variance $0.5$ each. For the $a_{\ell m=0}$ coefficients, we remove their imaginary component, and so their real component must be rescaled for the coefficients to have the intended unit variance.} From these final $a_{\ell m}$ values we generate the Gaussian maps using the \textsc{healpy} routine, \texttt{alm2map}.

Note that when we post-process the Gaussian maps to mimic the DES year 3 observations (see Section \ref{sec:ForwardModelProcess}), only the true convergence field is Gaussian. The procedures applied to the field to post-process it --- such as non-Gaussian noise, and survey masks of complicated geometries --- will still induce a non-zero non-Gaussianity in the final simulated convergence field, but these non-Gaussianities will not be cosmological in origin.

\subsection{Dark Energy Survey Year 3 (DES Y3)}\label{sec:DESY3Data}

The Dark Energy Survey \citep{DES2005} is an optical imaging survey of 5000 deg$^2$ of the southern sky, and is currently the largest precision photometric dataset for cosmology. We use the data from the Year 3 data release \citep{Sevilla2021Y3Gold}, and in particular the galaxy shape catalogs. This is the same dataset used for the fiducial 2-point correlation function shear results \citep*{Secco2022Shear, Amon2022Shear} and harmonic power spectrum results \citep{Doux2022HarmomicShearY3}, as well as the higher-order statistics such as the moments \citep{Gatti2022MomentsDESY3}, mass aperture \citep{Secco2022MassAp}, and peaks \citep{Zurcher2022WLPeaks}. In this work, the Y3 \textsc{Metacalibration} galaxy shape catalog \citep*{Gatti2021ShearCatalog} is used to make a map of the ellipticities, which is then converted into a convergence map via the Kaiser Squires method \citep{Kaiser1993KS}. This is the same technique used in previous works on the mass map \citep*{Chang2018MassMap, Niall2021MassMap}. We perform all our measurements and tests on these maps.

We also use the DES Y3 PSF and reserved star shape catalogs from \citet{Jarvis2021PIFF} to estimate the impact of PSF contributions to the signal observed by our statistic. The shape catalogs are used to make a PSF ``mass map'' the same way the galaxy ellipticities are used, and this mass map is used to test the PSF contributions (see Section \ref{sec:RoweStats} for more detail). The same star shape catalog was used to test PSF contributions for both the shear 2-point function \citep*{Gatti2021ShearCatalog}  and the 3-point function \citep{Secco2022MassAp, Gatti2022MomentsDESY3}.

\subsection{Making simulated DES Y3-like Mass maps}\label{sec:ForwardModelProcess}

We modify the simulated convergence/mass maps described in the above sections to include all the relevant observational effects of the DES Y3 data. Note that the main purpose of the maps is both to perform realistic forecasts of the cosmological constraints (Section \ref{sec:AnalysisFisher}), and to validate the contribution of different systematics to the CDFs data vector (Section \ref{sec:ResultsDESY3}). In this work, we do not use these simulations to get cosmology constraints from the DES Y3 data vector.

To make the mock maps, we start from the true convergence field, $\kappa$, and use an inverse Kaiser-Squires (KS) transform \citep{Kaiser1993KS} to obtain the two shear components, $\gamma_{1,2}$. The shear is the true observable of a weak lensing survey given we measure galaxy shapes. The KS transform can be quickly performed in harmonic space as

\begin{equation}\label{eqn:Kappa2Shear}
    \gamma^{\ell m}_E + i\gamma^{\ell m}_B = -\sqrt{\frac{(\ell + 2)(\ell - 1)}{\ell(\ell + 1)}} \bigg(\kappa^{\ell m}_E + i\kappa^{\ell m}_B \bigg),
\end{equation}
where the subscripts denote the E-mode and B-mode shear/convergence maps respectively. In the full-sky limit, where we have no survey masks, this is an exact expression. The technique has been validated for realistic data and found to have adequate accuracy \citep*{Chang2018MassMap, Niall2021MassMap}.

\textbf{Redshift distribution/bins:} We use four tomographic redshift bins with source galaxy $n(z)$ distributions matching DES Y3 \citep*{Myles2021PhotoZ}; the mean redshifts of these bins are $z_{\rm mean} \in \{0.336, 0.521, 0.741, 0.935\}$. The true shear maps corresponding to each bin are obtained via a weighted sum of the shear maps in each redshift, where the weights are the $n(z)$ distributions.

\textbf{Noise realization:} The noise is obtained using the DES Y3 \textsc{Metacalibration} shape catalog from \citet*{Gatti2021ShearCatalog}, using the same technique as \citet{Gatti2022MomentsDESY3}. The galaxy shapes are randomly rotated to remove all spatial correlations of the galaxy ellipticities, thus removing any cosmological signal. We then place galaxies in pixels of a $\texttt{NSIDE}=1024$ map, and compute the weighted average of the shear components in each pixel of the map, $\gamma_{1, 2}^{\rm noise}(\nhat)$, using the weights provided in the catalog. We add this noise to the true shear maps, $\gamma_{1, 2}$, separately for each tomographic redshift bin. This ensures the Y3 data and the simulated noise maps have the exact same variations in source/survey depth, and as we will show later, these variations create a strong non-Gaussian feature in the map (Section \ref{sec:NonGaussSignal}).

\textbf{Multiplicative bias:} The measured galaxy shapes have a bias of the order $1\%$ that has been calibrated using large suites of image simulations of the DES Y3 survey \citep{MacCrann2022ImsimsY3}. We include these bias terms, $m$, in the maps by simply multiplying the shears as $\gamma_{1, 2} \rightarrow (1 + m)\gamma_{1, 2}$.

\textbf{Mask:} We only use map pixels that have at least one DES Y3 galaxy in each of the four redshift bins. All pixels that do not fall into this category are discarded, and this defines the survey mask which is used in all further analyses, both for the simulations and for the DES Y3 data.

\textbf{Kaiser-Squires reconstruction:} Following the steps above, we obtain a spin-2 shear field, $\gamma_{1, 2}$, per DES Y3 tomographic redshift bin, that has noise, multiplicative bias, and a mask applied to it. We then convert this back to a convergence field using equation \eqref{eqn:Kappa2Shear} to obtain a noisy convergence map for each redshift bin. We only use the E-mode convergence map in our analysis. This map is then used as our final DES Y3-like map. Other, more sophisticated map-making techniques have been explored in the Y3 data as a replacement to KS reconstruction. A detailed description can be found in \citet*{Niall2021MassMap}. The KS method remains the simplest method that is also quick and accurate. The simplicity in compute time is a particularly attractive feature here as we make $\mathcal{O}(10^4)$ mock DES Y3 maps in this work. Note that the mass maps we generate from DES Y3 data in Section \ref{sec:DESY3Data} are also created by making the shear maps $\gamma_{1, 2}$ and using the KS transformation to obtain the convergence field.

In Section \ref{sec:ResultsDESY3}, we will add other effects to the mock maps --- such as PSFs, higher-order shear effects, and so on --- to test their impact on the measured signal and quantify which effects can be safely ignored and which effects may require scale cuts on the data vector. We do not address the impact of intrinsic alignments in this work, as we treat it as a systematic that can be properly modelled, and thus marginalized over, in a full cosmological analysis as opposed to an effect that contaminates the data vector and requires scale cuts.

\section{CDF analysis setup and Fisher constraints}\label{sec:AnalysisFisher}

We define the CDF data vector for DES Y3 in \S\ref{sec:DataVectorDef} and show the Fisher information in this CDFs data vector, as well as data vectors of other closely-related statistics, in \S\ref{sec:MainFisher}.

\subsection{Defining CDFs data vector}\label{sec:DataVectorDef}

 \begin{figure*}
    \centering
    \includegraphics[width = 2\columnwidth]{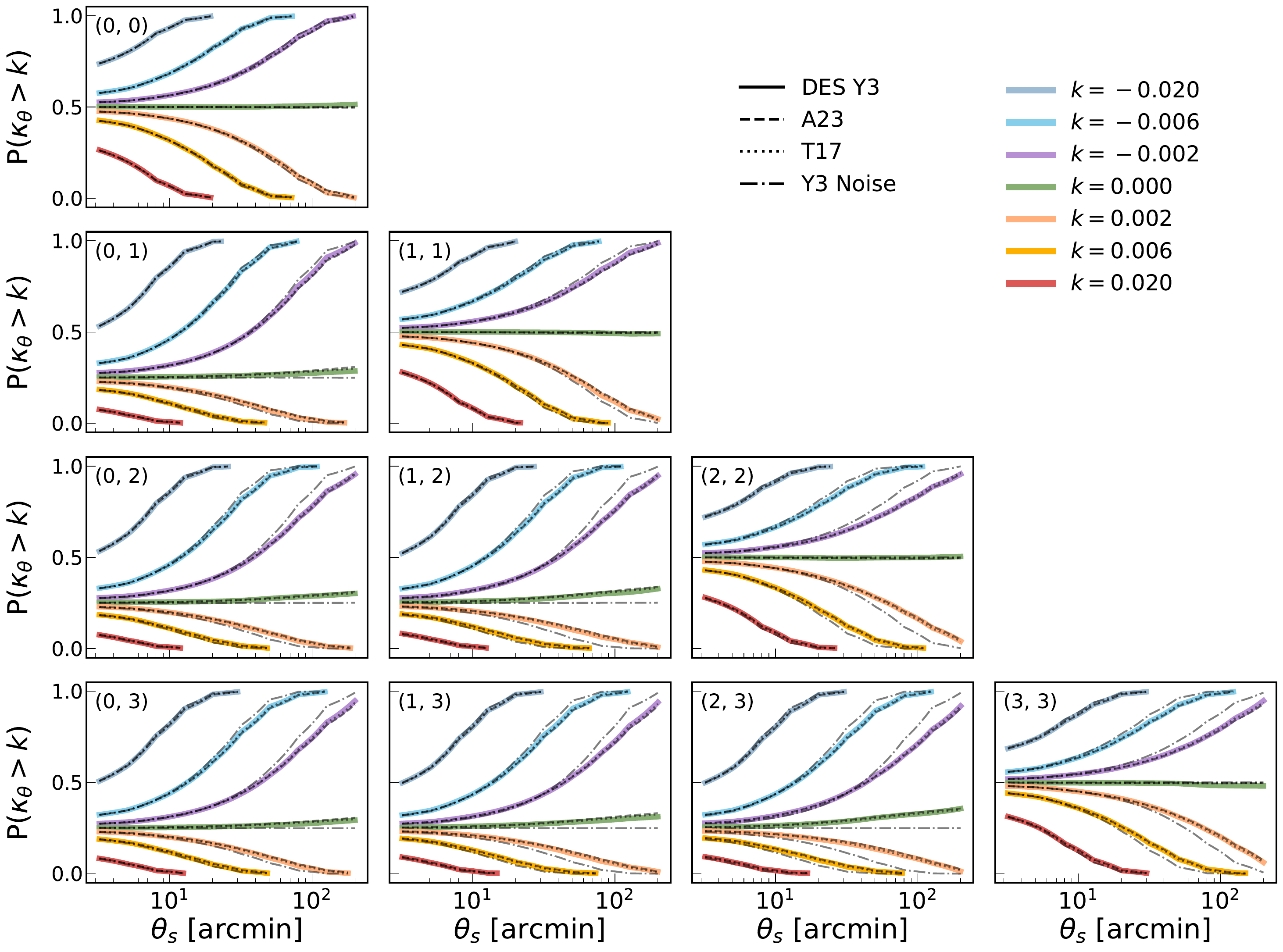}
    \caption{The fiducial datavector used in this work. Colored solid lines are measurements of the CDFs on DES Y3 mass maps, dark dashed lines are from the A23 suite, dotted lines are from T17, and the dashed-dotted are from just shape noise maps with no cosmological signal. All simulated maps have the same DES Y3 shape noise field, survey mask, n(z) distribution, and are put through the same convergence reconstruction method. The panels show 1-field or 2-field CDFs for different bin combinations, with the specific combination denoted in the corner of each panel. There are clear differences between the noise-only CDFs and the DES Y3 data CDFs, particularly on larger scales and in higher redshift bins, which are the expected imprints for a cosmological signal in the lensing convergence maps. The A23 and T17 simulation predictions are a decent match to the Y3 data.}
    \label{fig:DataVector}
\end{figure*}

In this work, we measure all possible 1-field and 2-field CDFs for the four tomographic bins of DES Y3. This results in four 1-field ``auto'' CDFs, and six 2-field ``cross'' CDFs. We measure the CDFs across 10 smoothing scales, spaced logarithmically between $3.2 \arcmin$ and $200 \arcmin$. The choice of scales matches the moments-based DES Y3 analysis of \citet{Gatti2022MomentsDESY3}. For each scale, we use 7 thresholds $k \in \{-20 , -6, -2,  0  ,  20,  6,  20\} \times 10^{-3}$. These were chosen by looking at the variance of the field at the smallest and largest smoothing scale, and ensuring at least two thresholds did not asymptote to 0 or 1 at each scale. Using the Fisher forecast below we have checked that these thresholds probe most of the relevant information while being practical to implement, and we do not perform a more methodic study of the optimal threshold choices. We have, however, verified that removing any one of the seven thresholds leads to a fractional change in the constraints of 5\% to 10\%. We did not test adding more thresholds as the longer data vector leads to poorer numerical convergence, which then makes it difficult to robustly identify the increase in constraining power provided by the additional thresholds. 

For all CDF measurements, we only focus on the range of scales where $0.05 < \CDF(k, \theta) < 0.95$, which excludes the $\sim 2\sigma$ region of the distribution for each threshold $k$ and smoothing scale $\theta$. This removes measurements of the tails of the distribution where noise can cause spurious signals, and it also helps remove regions where the CDF has asymptoted to constant values of 0 or 1. We have confirmed that using different choices, such as $3\sigma$ or $4\sigma$ cuts, leads to a fractional difference of $<5\%$ in the Fisher constraints. While the tails of the distribution are a sensitive probe of the non-Gaussian information, they are also much noisier and so the actual constraining power from this region of the distribution is not significant. The  ``bulk'' of the distribution --- for example, the 1 to 2$\sigma$ region --- is still quite sensitive to non-Gaussian features while being less susceptible to noise \citep[\eg][]{Friedrich2020PDFBulkfNL, Uhlemann2020PDFNeutrino}.

Our initial data vector has size $N = 10 \text{ z-bins} \times 10 \text{ scales} \times 7 \text{ thresholds} = 700$ data points. The procedure above of focusing only on $0.05 < \CDF < 0.95$ removes more datapoints as multiple thresholds reach asymptotic behavior of $\CDF = 0$ and $\CDF = 1$ at large smoothing scales, especially for the lower redshift bins where the variance of the convergence field is lower.\footnote{The density field has a higher variance at lower redshifts, but the lensing kernel has a lower amplitude for low-redshift sources and so the variance of the convergence field increases with redshift.} In practice, the data vector for DES Y3-like maps has $N = 460$ points. Note that different thresholds reach these asymptotic values at different scales. Figure \ref{fig:Illustration} illustrates this behavior.

Figure \ref{fig:DataVector} presents the data vector measured on the DES Y3 data as well as different simulations described in Section \ref{sec:simulations}. The 1-field (2-field) CDFs are shown in the diagonal (off-diagonal) panels. The colored lines show $P(\kappa_\theta > k)$, the fraction of the map that exceeds a given threshold at a given smoothing scale, where each color is a different threshold. At a fixed threshold, the probability is driven to 0 or 1 with larger $\theta$, and this behavior is discussed in Section \ref{sec:CDFs}.

The threshold $k = 0$ is special as it is the mean of the 1D marginal distributions, and so its probability for the 1-field CDFs is $P \approx 0.5$ across all scales.\footnote{Figure \ref{fig:Illustration} shows the true convergence field is log-normal on small-scale, and thus has $P(\kappa_\theta > 0) \neq 0.5$. However, for \textit{noisy} convergence fields, the noise dominates the cosmological signal on small scales and this noise is a symmetric distribution (the odd moments are zero, as discussed in Section \ref{sec:NonGaussSignal}). This restores the measurements to $P(\kappa_\theta > 0) \approx 0.5$ as mentioned.} In the 2-field case the probability for $k = 0$ is $P(\kappa_{\theta, 1} > 0, \kappa_{\theta, 2} > 0) \approx 0.25$ but has scale-dependent deviations. This is because the correlation between the two fields alters this probability, and this correlation has a scale dependence, meaning the deviations from $P \approx 0.25$ will also be scale-dependent as expected.

We can also see a clear visual difference between the CDFs of the shape noise field (dashed-dotted) and those of the observed convergence field. In particular, the 1-field CDFs of the (3, 3) bin show the clearest difference at larger scales. The shape noise field has a notably smaller variance than the observed convergence field, and this causes the CDFs to asymptote to 0 or 1 more quickly compared to the CDFs of the data. We also find that the T17 predictions are quite similar to those of A23, and that the simulations are generally a decent match to the data.

\subsection{Fisher Information}\label{sec:MainFisher}

\begin{figure}
    \centering
    \includegraphics[width=1\columnwidth]{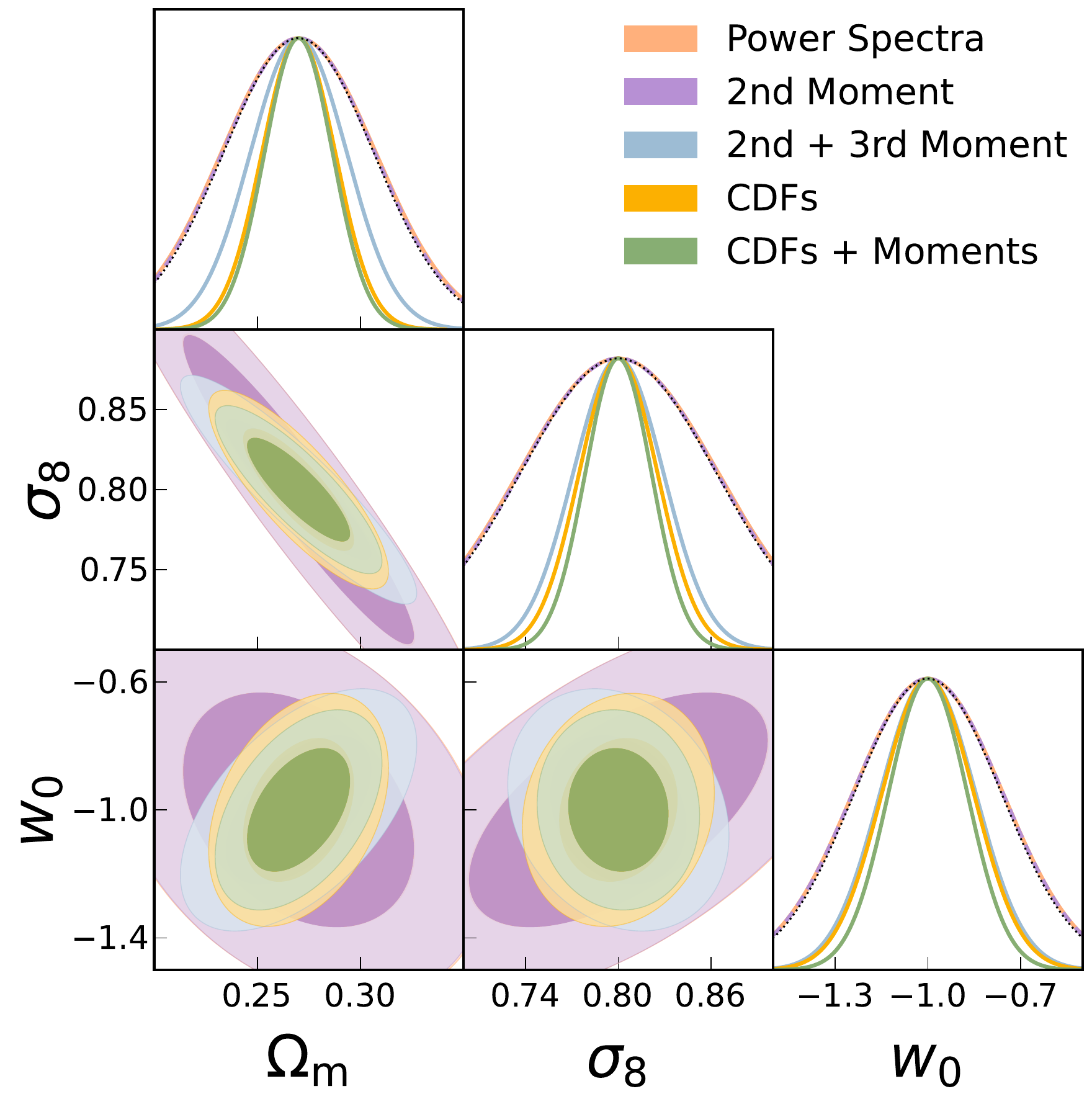}
    \caption{The Fisher information of different statistics for $\sigma_8$, $\Omega_{\rm m}$ and $w_0$ when using  DES Y3-like data. The power spectra and 2nd moment probe only the Gaussian information and their contours overlap completely (the peach contour is hidden underneath the purple). Adding the 3rd moment significantly improves the constraints, and the CDF, which approximately contains all moments, improves upon that a non-negligible but diminishing amount. The degeneracy direction of 2+3 moments and the CDFs is also visibly different, and combining them leads to a further 20-30\% improvement in constraints. The black dashed lines in the diagonal panels show the 1D constraints from CDFs measured on a purely Gaussian field, and these are consistent with those from the other Gaussian statistics. The constraints are tabulated in Table \ref{tab:FisherConstraints}.}
    \label{fig:Fisher}
\end{figure}

We use the data vectors and covariance matrices constructed from the A23 simulations to perform a Fisher forecast for three $w\rm CDM$ parameters that are the target of current and future lensing surveys --- $\Omega_m$, $\sigma_8$, and $w_0$. We measure three broad types of summary statistics for this forecast:

\textbf{Gaussian Statistics}, such as angular power spectra and the 2nd moments of the field are well-known for being sensitive to only the variance of the field, and the variance is often denoted the Gaussian part of the distribution. These statistics provide a good baseline for cosmological constraints obtained from current fiducial analyses, which primarily use such Gaussian statistics. The angular power spectra are measured in 20 bins in the range $10 < \ell < 2048$. The 2nd moments are measured on the maps smoothed with a tophat across 10 scales that are logarithmically spaced in the range $3.2 \arcmin < \theta < 200 \arcmin$.

\textbf{Higher-order moments} are a natural extension to the 2nd moments where one averages higher powers of the fields, $\langle \kappa ^N \rangle$. The most common one is the 3rd moment (or skewness), though the 4th moment (or kurtosis) has also been measured in lensing data before across a smaller range of angular scales $2\arcmin < \theta < 8\arcmin$ \citep{VanWaerbeke2013CFHTLens}. In this work we measure the 2nd and 3rd moments in the range $3.2 \arcmin < \theta < 200 \arcmin$.

Finally, the \textbf{CDF} is the non-Gaussian statistic that is the focus of this work. The data vector definition is described in Section \ref{sec:DataVectorDef}, and the measurement on DES Y3 data and some simulated mock maps is shown in Figure \ref{fig:DataVector}.

Note that the datavectors of these higher-order statistics tend to be long, and this is particularly an issue when computing the covariance numerically, as the number of realizations needed for the covariance increases with the data vector size. However, the A23 simulation suite contains 8000 DES Y3-like maps, and this number is far larger than the length of any data vector computed in this work.

We can now estimate the Fisher information with the standard approach,
\begin{equation}\label{eqn:Fisher}
    \textbf{F}_{ij} = \sum_{m,n}\frac{d\widetilde{X}_m}{d\theta_i}\big(\mathcal{C}^{-1}\big)_{mn}\frac{d\widetilde{X}_n}{d\theta_j},
\end{equation}
where $\frac{d\widetilde{X}_m}{d\theta_i}$ is the mean derivative of point $m$ in data vector $X$ with respect to parameter $\theta_i$, where the mean is computed using 400 DES Y3 realizations. $\mathcal{C}^{-1}$ is the inverse of the numerically estimated covariance matrix and this is computed while accounting for the Hartlap factor \citep{Hartlap2007},
\begin{equation}\label{eqn:invertcov}
    \mathcal{C}^{-1} \rightarrow \frac{N_{\rm sims} - N_{\rm data} - 2}{N_{\rm sims} - 1} \,\mathcal{C}^{-1}.
\end{equation}
The Hartlap factor for all data vectors in this work is $\gtrsim 0.9$. We have verified that the Fisher information --- for all the statistics we present --- changes by $<1\%$ even if we halve the number of realizations used to compute the covariance matrix, from $N = 8000 \rightarrow 4000$. Similarly, halving the number of realizations used in computing the derivatives, $N = 400 \rightarrow 200$, changes the Fisher information by $<1\%$ for most statistics; the one exception is the CDFs, where the change in Fisher information is still at the 5-10\% level. However, a numerical uncertainty of this level does not change our qualitative interpretations below.

Figure \ref{fig:Fisher} shows the Fisher information of each statistic. The parameter constraints are obtained by inverting the Fisher matrix of Equation \eqref{eqn:Fisher}. First, we see that the angular power spectra and the 2nd moments have indistinguishable constraints, and this is the expected behavior as one is simply a transformation of the other; given the $\mathcal{C}_\ell$, one can predict the 2nd moments exactly via an integral, and vice versa. \footnote{This assumes we measure both harmonic-space and real-space over a wide enough range of scales to perform the transform. The agreement between $\mathcal{C}_\ell$ and 2nd moments in Figure \ref{fig:Fisher} then implies we chose an appropriately wide range of scales.} We also see that the CDFs measured on a Gaussian version of the simulated Y3-like fields, shown by the gray dotted line in the diagonal panels, have constraints very consistent with those of the power spectrum and 2nd-moment. We show in Appendix \ref{appx:Cov_Matrix} that the statistics used in this figure all follow a Gaussian likelihood even when measured on fully non-linear, non-Gaussian fields --- which is not always the case for higher-order statistics as has been found in previous works \citep{Park2022NGCov, Euclid2023NGCov}. 

Including the 3rd moment alongside the 2nd moment improves the constraints significantly for all parameters. This is primarily because of the different degeneracy directions for the different moments \citep{Gatti2020Moments, Gatti2022MomentsDESY3}. 

The CDFs improve the constraints compared to the combination of 2nd and 3rd moments. This confirms that there is still usable information beyond the 3rd moment in the convergence field, particularly in constraining $\Omega_m$. However, the modest improvement in going from the 2nd + 3rd moments to the CDFs (when compared to the increase from 2nd moments to 2nd + 3rd moments) shows that there is less information from the 4th moment and beyond. We explicitly check the information content of the 4th and 5th moments later in Figure \ref{fig:NonGaussSNR}. We have separately verified that the constraining power of the moments approach agrees better with that of the CDFs if we include the 4th and 5th moments in the former.

In general, we find that the CDFs do better than the combination of the 2nd and 3rd moments by around $\approx 20\%$ in the three parameters we focus on (Table \ref{tab:FisherConstraints}). They are also more compact, meaning the data vector for the CDFs (N = 460) is notably smaller than the data vectors for the higher-order moments --- from progressively including the 4th Moment (N = 650) or 5th Moment (N = 1210) --- while still providing constraints that are better than using up to the 5th moment. Combining the CDFs with the 2nd and 3rd moments leads to constraints that are 20-30\% better than using just the 2nd and 3rd Moments. We have verified in Appendix \ref{appx:Cov_Matrix} that the combined data vector also follows a Gaussian likelihood.

We also use the Figure of Merit (FoM), which is defined as the inverse of the area/volume of the ellipsoid formed by the parameter constraints,
\begin{equation}\label{eqn:FoM}
    {\rm FoM}_\theta = \sqrt{\frac{1}{\det (F^{-1})_\theta}},
\end{equation}
where $\theta$ is the subset of parameters used to define the FoM and in our case is $\theta \in \{\Omega_{\rm m}, \sigma_8, w_0\}$. The FoM metric provides a concise way to summarize the constraining power in a multidimensional parameter space. We list the FoM values of our data vectors in Table \ref{tab:FisherConstraints}. Including the 3rd moments improves the FoM, relative to the 2nd moments, by a factor of 3. Including the CDFs improves it by 15\%, relative to the FoM of the combination of the 2nd and 3rd moments. Combining the CDFs with the 2nd and 3rd moments improves the latter's FoM by 65\% and the former's FoM by 40\%.

\begin{table}
    \centering
    \begin{tabular}{c|c|c|c|c|c}
       \hline
       \hline
       Analysis & $\sigma(\Omega_{\rm m})$ & $\sigma(\sigma_8)$ & $\sigma(w_0)$ & $\rm FoM$ & $N_{\rm dof}$ \\
       \hline
       \multicolumn{6}{c}{\textit{Fisher information (This work)}}\\
       \hline
       Power Spectra  & 0.037 & 0.064 & 0.24 & 1.00 & 200\\
       2nd Moment & 0.037 &  0.064 & 0.24 & 1.02 & 100\\
       2nd + 3rd Moments & 0.023 & 0.029 & 0.15 & 2.95 & 300\\
       CDFs & \textbf{0.018} & \textbf{0.025} & \textbf{0.15} & \textbf{3.47} &  \textbf{460}\\
       CDFs + Moments & \textbf{0.016} & \textbf{0.021} & \textbf{0.12} &  \textbf{5.01} & \textbf{760}\\
       \hline
       \multicolumn{5}{c}{\textit{DES Y3}}\\ 
       \hline
       Cosmic Shear & 0.051 & 0.083  & --\\
       2+3 Moments & 0.030 & 0.050  & --\\
       Peaks & 0.060 & 0.099 & -- \\
       \hline
       \multicolumn{5}{c}{\textit{KiDS-1000}}\\
       \hline
       Cosmic Shear & 0.050 & 0.080 & -- \\
       Field Level ML & 0.096 & 0.206 & 0.29\\
       \hline
    \end{tabular}
    \caption{The Fisher information constraints for a joint analysis of $\Omega_{\rm m}$, $\sigma_8$, and $w_0$, the Figure of Merit (FoM), and the size of the data vectors. The FoM, defined in equation \eqref{eqn:FoM}, is a metric for the constraining power in multi-dimensional parameter space. All presented FoM values are normalized by the FoM of the Power Spectra. We compare with the shear results from different Gaussian and non-Gaussian statistics used on the DES Y3 or KiDS 1000 data. For DES Y3 we show constraints from Cosmic Shear \citep*{Amon2022Shear, Secco2022Shear}, 2nd and 3rd Moments \citep{Gatti2022MomentsDESY3}, and Peaks \citep{Zurcher2022WLPeaks}. For KiDS 1000, we show results from cosmic shear \citep{Asgari2021ShearKIDS} and a field-level analysis \citep{Fluri2022wCDMKIDS}. The DES analysis of the 2nd + 3rd Moments uses different, more conservative analysis choices (scale cuts, nuisance parameters etc.) compared to the Fisher forecast done here with the 2nd + 3rd Moments, resulting in its relatively looser constraints.}
    \label{tab:FisherConstraints}
\end{table}

\section{Lensing CDFs in DES Y3 data}\label{sec:ResultsDESY3}

We now discuss measurements of the CDF on the DES Y3 data in \S \ref{sec:MeasurementSNR}, including the non-Gaussian aspect of the noise field in \S \ref{sec:NonGaussSignal}, and then detail the contributions from different effects that can impact the inference process: PSFs in \S\ref{sec:RoweStats}, source clustering in \S\ref{sec:SourceClustering}, higher-order shear effects in \S\ref{sec:HigherOrderShear} and baryonic effects in \S\ref{sec:Baryons}. Finally, we discuss scale cuts in \S\ref{sec:ScaleCuts}.

\subsection{CDF measurement \& signal-to-noise}\label{sec:MeasurementSNR}

\begin{figure*}
    \centering
    \includegraphics[width = 2\columnwidth]{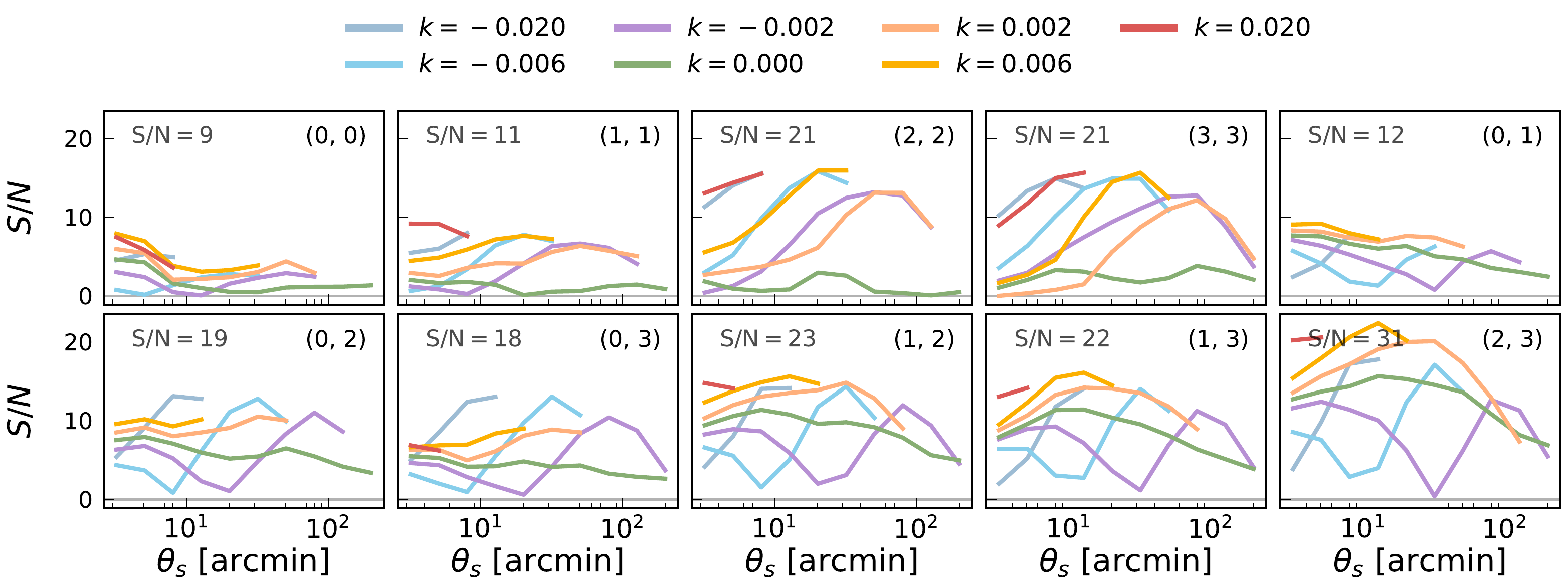}
    \caption{The S/N of the DES Y3 data vector. There is a clear signal observed in the CDFs with ${\rm  S/N} = 45.3$ which is slightly higher than, but generally consistent with, the S/N of the 2-point analyses in DES \citep[${\rm  S/N} = 40.2$, see Section IV of][]{Secco2022MassAp}. We show the ${\rm  S/N}$ from individual bin combinations as text in the upper left panels. The upper right text in a panel denotes the bin combinations used in a certain CDF measurement. Note that the measurements are significantly correlated so one cannot trivially add the S/N of different bins together.}
    \label{fig:SNR}
\end{figure*}

In Figure \ref{fig:DataVector}, we have already shown the DES Y3 measurements in solid lines, with the noise-only data vector in dotted gray lines and the A23 version of DES Y3-like map in the gray dashed lines. There is a clear cosmological signal as evidenced by the difference between the noise-only and DES Y3 measurements. Figure \ref{fig:SNR} now shows the signal-to-noise of the cosmological component for each datapoint in the data vector. This is computed as the residuals normalized by the uncertainty, ${\rm S/N} = |\CDF_{\rm Y3} - \CDF_{\rm N}|/\sigma(\CDF_{\rm A23})$. We then also combine the statistical significance of the individual points, accounting for the covariance between them, and find a total signal-to-noise of $\rm S/N = 45.3$.

If the difference between the signal+noise and noise-only fields is a difference in only their even moments (eg. variance and kurtosis) then for the 1-field CDFs (the ``auto-correlation'' part) in Figure \ref{fig:SNR}, the S/N of a positive threshold should be similar to that of a negative threshold of the same amplitude. We see some indication of this via visual inspection of the 1-field CDF of the 3rd and 4th tomographic bin. We also see an asymmetry in the S/N, and this is a sign of an additional skewness caused by the signal field --- for example, in the (0, 0) bin the amplitude of the yellow line ($k = 0.006$) is higher than the light blue one ($k = -0.006$). Thus, we can also visually see indications that this statistic captures non-Gaussian signatures.

Note that while we quote a signal-to-noise for the full set of residuals, we do not use it as a robust estimate of the amount of information. This is because the CDFs respond to noise and signal nonlinearly, \footnote{Even in the Gaussian case, the CDF heuristically goes as $\int \exp[1/\sigma^2]dx$, so changes in $\sigma$ lead to highly nonlinear responses in the CDF.} so a $\chi^2$ statistic is not the ideal way to quantify deviations \textit{if the deviations are large}, which is the case between measurements of the noise-only maps and the noisy convergence maps. The interpretation of a $\chi^2$ in the large-deviation regime is unclear. Note that this is not a problem for our Fisher forecast as the residuals are small given the shifts in the cosmology parameters, as needed for the derivatives, are also small.

Given the results of Figure \ref{fig:Fisher}, where we find the CDFs are a useful and complementary statistic for constraining cosmology, and Figure \ref{fig:SNR}, where we find the CDFs in DES Y3 have a clear cosmological signal with signs of both the Gaussian and non-Gaussian part, we would like to now test the robustness of this statistic to the relevant observational effects in the Y3 weak lensing data. We will explore exactly this in the following subsections:

\begin{itemize}
    \item Naturally we would want to know how much of the cosmological information seen in Figure \ref{fig:SNR} is non-Gaussian -- this requires a more precise understanding of the non-Gaussianity in the noise field (\S\ref{sec:NonGaussSignal}).
    \item The measured shape of galaxies will have some contributions from the PSF, which can then lead to non-cosmological spatial correlations of the galaxy ellipticities -- we find this is negligible (\S \ref{sec:RoweStats}).
    \item Source galaxies, which trace the density field, will be clustered and this can impact the observed convergence field -- this has a noticeable impact (\S \ref{sec:SourceClustering}).
    \item The source clustering also leads to correlations between the shape noise field and the convergence field, as seen in the CDFs -- we can model this correlation effectively (\S\ref{sec:SourceClustering}).
    \item The impact of ignoring higher-order shear effects when modelling the data vector -- this is also negligible (\S\ref{sec:HigherOrderShear}).
    \item The effect of baryonic physics on our statistics  -- as expected from previous works, this is important (\S\ref{sec:Baryons}).
    \item Given the tests above, we detail the analysis choices one would need to make --- under our current modelling ability --- to robustly infer cosmology using the CDFs (\S\ref{sec:ScaleCuts}).
\end{itemize}

The impact from other common systematic factors, such as $n(z)$ uncertainties, multiplicative bias uncertainties, and intrinsic alignments, is not considered here. These effects can all be marginalized in the inference and modeling process when obtaining cosmological constraints via the CDFs data vector. Such marginalization has already been performed for multiple different analyses of higher-order statistics \citep[\eg][]{Gatti2022MomentsDESY3, Zurcher2022WLPeaks}.

\subsection{Non-Gaussianity of shape noise fields}\label{sec:NonGaussSignal}

\begin{figure}
    \centering
    \includegraphics[width = \columnwidth]{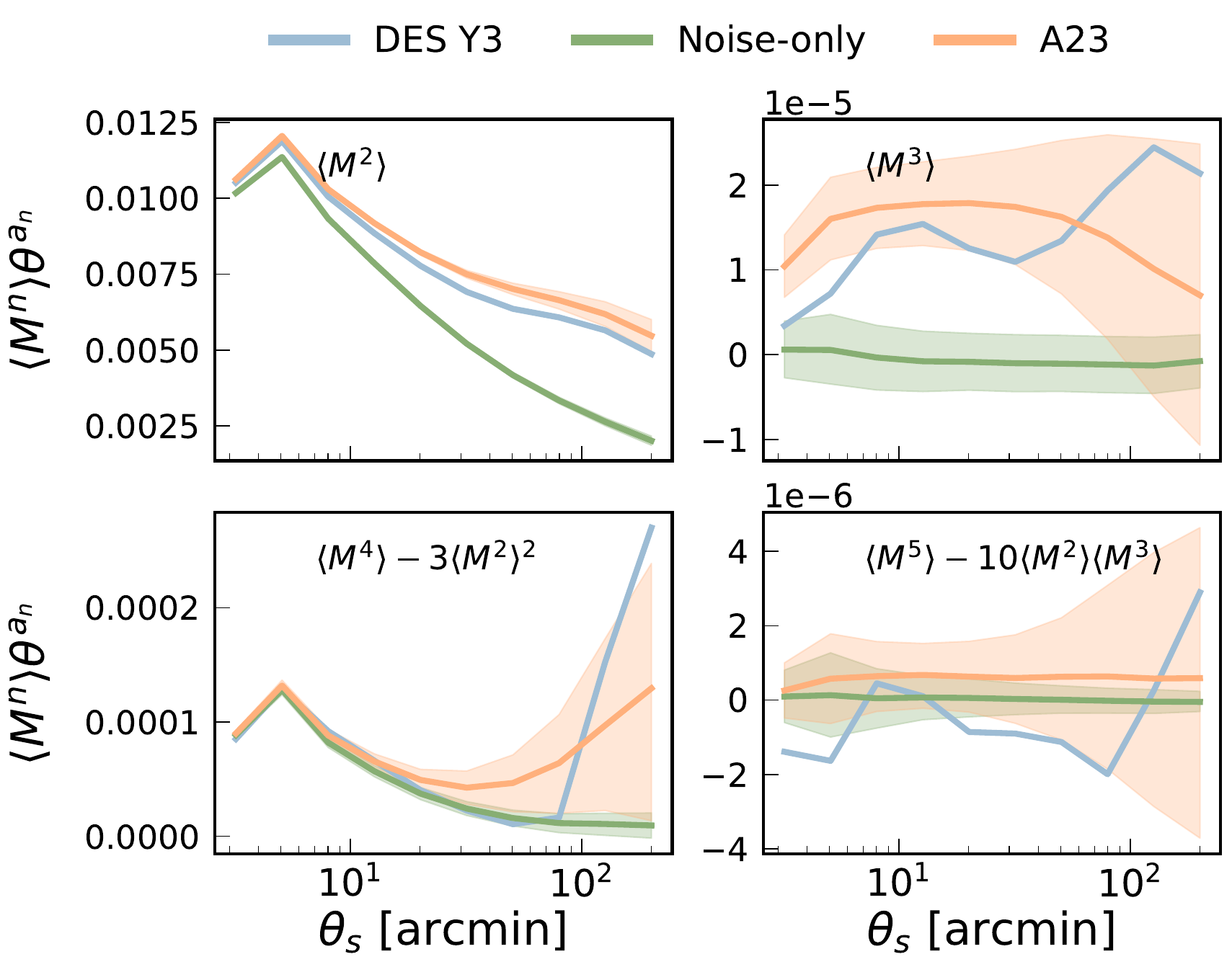}
    \caption{The moments of the 4th tomographic bin, as a function of smoothing scales, for the DES Y3 map, the noise-only maps, and the A23 maps. The 4th and 5th moments (bottom panels) have their disconnected components subtracted out. The bands show $1\sigma$ uncertainties for the Noise-only and A23 maps from the $\mathcal{O}(10^4)$ realizations used in this work. The moments are re-scaled by $\theta^{a_{\rm n}}$ as a visualization choice, where $a_{\rm n} = n/2$ and $n \in \{2, 3, 4, 5\}$ is the moment order. The 2nd and 3rd moments have significant information beyond the noise. The 4th moment is significant on the smallest scales, but this contribution is entirely from the noise field since the blue/orange and green lines are almost perfectly overlaid. On larger scales, there is a weak, cosmological signal. The 5th moment is fully consistent with no signal across all scales.}
    \label{fig:NonGaussSNR}
\end{figure}

To quantify the level of cosmological non-Gaussianity observed by the CDFs, one first needs to understand the non-Gaussianity in the noise field. This is particularly relevant for us as the CDFs are sensitive to all moments of the field, meaning all moments of the cosmological signal but also all moments of the noise field. For this particular investigation, we will switch to using the fields' moments to summarize the noise field and cosmological field at different orders. We do this as the moments can easily isolate the signal from different orders, which helps disentangle the information contained in the CDFs.

Figure \ref{fig:NonGaussSNR} shows the 2nd to 5th moments of DES Y3 mass map, as well as the shape noise map, for the 4th tomographic bin. We find that there is a significant non-Gaussianity in the noise, particularly in the 4th moment and on small scales. Such a feature is naturally expected if the field of source galaxy number counts is not uniform. In the limit that the galaxy counts are uniform across the whole DES Y3 footprint, then every pixel in the map has the same number of galaxies, and thus would have the same shape noise per pixel. In reality, the number of source galaxies per pixel varies across the footprint, either from survey observing conditions or from the intrinsic clustering of sources due to structure formation (see Section \ref{sec:SourceClustering} or Figure \ref{fig:NoiseSignalCorr}). In this case, the noise variance per pixel varies across the footprint, and summing the individual Gaussian noise distributions within the pixels results in a Gaussian mixture model that is symmetric about the $x = 0$ mean, but can have a significant non-Gaussianity in its even cumulants/moments starting from the kurtosis/4th moment. This is also consistent with the fact that we detect no odd moments in the noise field.

We also see in Figure \ref{fig:NonGaussSNR} that for DES Y3-like data the cosmological signal exists only in the 3rd and 4th moments. At the 5th moment, the measurement is already consistent with no signal. The noise field has a 3rd moment that is consistent with 0 across the full range of scales. For the fourth moment, however, the noise has a larger fourth moment than the cosmological signal. We can infer this by seeing that the fourth moment of the observed field is very similar to that of the noise-only field. 

The significance of the 4th moment in Figure \ref{fig:NonGaussSNR} highlights the need to accurately model the noise field, since almost all the non-Gaussianity on small scales is coming from the shape noise field rather than the convergence field. Note that some previous works have also shown a strong detection of the 5th moment in the convergence field from data \citep{VanWaerbeke2013CFHTLens}, but they analyze the total fifth moment $\langle \kappa^5 \rangle$, whereas here we only consider the connected component, which is obtained as $\langle \kappa^5 \rangle - 10\langle \kappa^2 \rangle\langle \kappa^3 \rangle$, where $\langle \kappa^2 \rangle\langle \kappa^3 \rangle$ is the disconnected component.\footnote{The factor of 10 can be seen by writing all unique combinations of $\langle\kappa_i \kappa_j\rangle\langle\kappa_k \kappa_l\kappa_m\rangle$, which is the disconnected fifth moment, with $i, j, k, l, m \in \{0, 1, 2, 3\}$. There are 10 unique combinations.} Accounting for this disconnected component is important when isolating the signal in the higher orders. For example, Gaussian distributions have a non-zero fourth moment that must be accounted for --- by subtracting out this ``disconnected'' piece --- when measuring non-Gaussian features via the 4th moment. A similar scenario occurs for the fifth moment, where we subtract contributions from lower orders, namely the product of the 2nd and 3rd moments.

\subsection{PSF contributions}\label{sec:RoweStats}

\begin{figure*}
    \centering
    \includegraphics[width = 2\columnwidth]{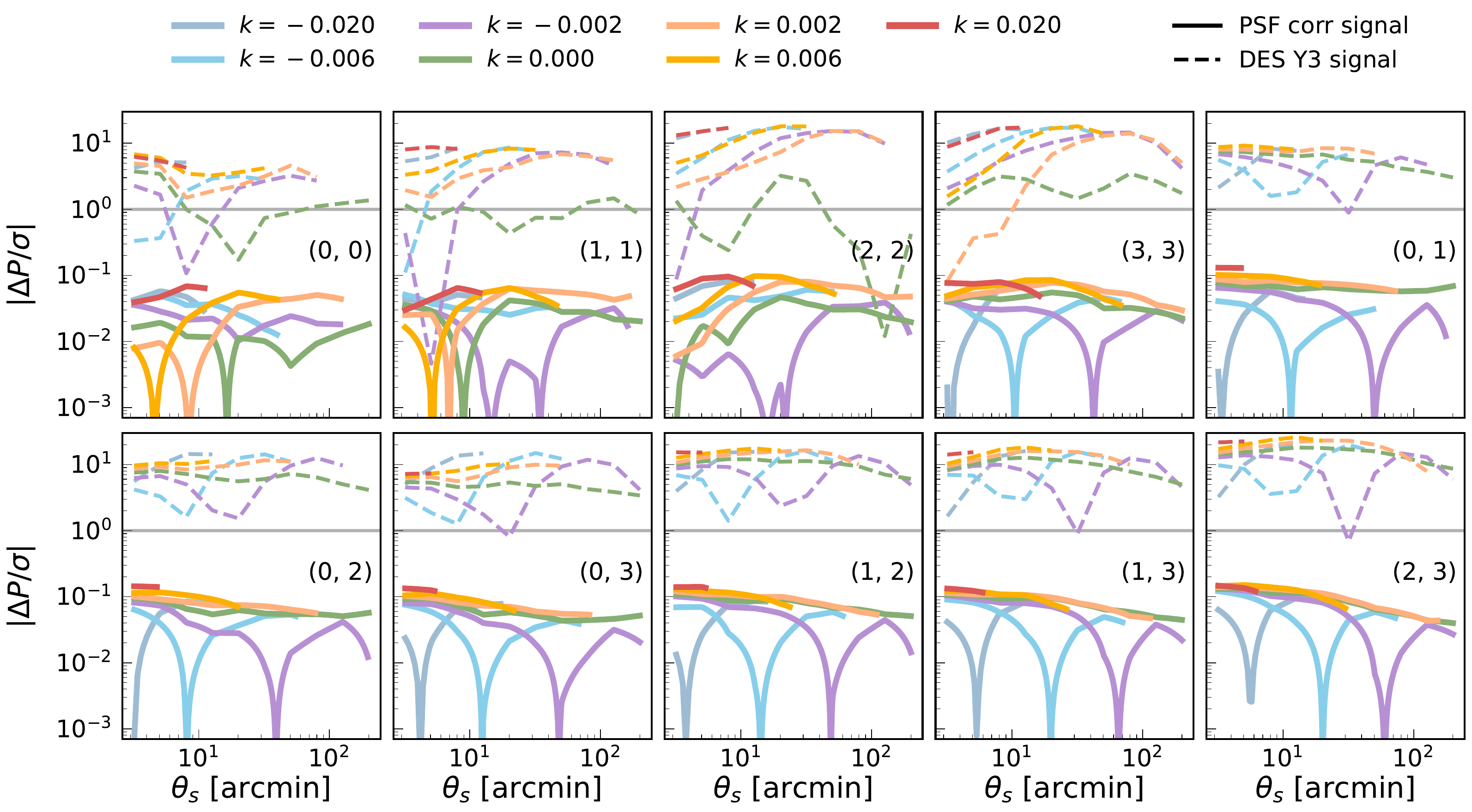}
    \caption{The difference in CDFs measured on two DES Y3-like simulated maps. One contains the Y3 PSF mass map, and the other contains a PSF mass map obtained after rotating all the PSF-based ellipticities. The contribution of any correlations from the PSF (solid lines) is below $<0.1\sigma$ and is statistically negligible for all thresholds (different colors). It is also 2-3 orders of magnitude below the cosmological signal in DES Y3 (dotted lines). The total signal-to-noise of PSF-induced correlations is $0.3\sigma$.}
    \label{fig:RhoStats}
\end{figure*}

So far we have assumed that spatial correlations between the measured galaxy shapes are a purely cosmological signal. However, this is not guaranteed to be the case as the ellipticities from the PSF can have spatial correlations as well. These correlations have been studied extensively for the 2-point functions \citep{Jarvis2021PIFF}, and the work from \citet*{Gatti2021ShearCatalog, Amon2022Shear} have explicitly shown their contributions to the cosmological signal/constraints from 2-point functions are negligible. This test has also been done at the 3-point function level \citep{Secco2022MassAp, Gatti2022MomentsDESY3} and found the contributions continue to be negligible. We now replicate this test at the CDF level, which will test the contribution of the PSFs to \textit{all higher-order moments}.

First, we detail the different PSF contributors to the galaxy shapes. The lensing convergence is obtained from the lensing shear maps, which in turn are obtained from individual galaxy ellipticities. The measured ellipticity of a single galaxy can be separated into distinct components,

\begin{equation}\label{eqn:ellipticities}
    \textbf{e}^{\rm obs} = \textbf{e}^{\rm gal} + \textbf{e}^{\rm shear} + \alpha\textbf{e}^{\rm psf, true} + \beta \Delta \textbf{e}^{\rm psf, err} + \gamma \Delta T \textbf{e}^{\rm psf, true},
\end{equation}
where $\textbf{e}^{\rm gal}$ is the intrinsic ellipticity of a given galaxy, $\textbf{e}^{\rm shear}$ is the ellipticity modification due to weak lensing from foreground structure, $\textbf{e}^{\rm psf, true}$ is the PSF ellipticity, $\Delta \textbf{e}^{\rm psf, err}$ is the PSF ellipticity error \footnote{This is defined as $\textbf{e}^{\rm psf, true} - \textbf{e}^{\rm psf}$, which is the difference between the ellipticity of a star (the ``true'' PSF) and that of the PSF model evaluated at the star's position.}, and $\Delta T \textbf{e}^{\rm psf, true}$ is the PSF size error \footnote{This is defined as $\Delta T = (T^{\rm psf, true} - T^{\rm psf})/T^{\rm psf}$, the fractional difference between the size of a star (the ``true'' PSF size) and the size of the PSF model evaluated at the location of the star.}. The first quantity of Equation \eqref{eqn:ellipticities} is assumed to average to zero, $\langle \textbf{e}^{\rm gal} \rangle = 0$, while the PSF components can still make a non-zero mean contribution. The coefficients, $\alpha, \beta, \gamma$ connect the PSF components to their effective contributions on the measured shear. The values of these coefficients can be measured directly from the data, and we use the values reported in \citet*[][see their Table 2]{Gatti2021ShearCatalog} of $\alpha = 0.001$, $\beta = 1.09$ and $\gamma = -0.5$. These PSF-based ellipticities can then be used to make a ``PSF mass map'' in the same way galaxy ellipticities are used to make the DES Y3 mass map. In practice, we make three PSF maps for each of the three PSF components in equation \eqref{eqn:ellipticities} and sum them together in the end.

We test the impact of PSFs on the CDFs by comparing measurements between two types of maps. The first type of map is the sum of the cosmological signal from the A23 simulations, the Y3-like shape noise field, and a PSF mass map for each of the three individual PSF terms of equation \eqref{eqn:ellipticities}. The second type of map contains the same signal and noise fields as the first, but the PSF mass map is now created after rotating all the PSF-related ellipticities in random directions. Thus the first map preserves any PSF-based spatial correlation signals, whereas the second map removes such correlations. Therefore, the residuals between the CDF measurements on these two maps quantify the significance of the PSF ellipticities being spatially correlated, which in turn quantifies how much this non-cosmological spatial correlation will contaminate our signal.\footnote{One could also compare maps with and without the PSF mass map. However, this would simply show that the PSF shapes are elliptical, which is already a well-established fact \citep{Jarvis2021PIFF}.} Note that we add the \textit{same} PSF mass map to all tomographic bins.

We show in Figure \ref{fig:RhoStats} the significance of the residuals between these two maps as measured by the CDFs, averaged over 8000 realizations. The results show that the significance of the PSF contribution is below $0.1 \sigma$ for all bins, scales, and thresholds. More importantly, we also show the cosmology signal seen by the CDFs --- the same results from Figure \ref{fig:SNR} --- and find the PSF contribution is multiple orders of magnitude below the cosmological signal, which has a significance of $3-10\sigma$. This also confirms that the PSF contributions at the DES Y3 data quality are negligible even beyond the 3-point information.

Note that there are some dipping/valley features in both the dashed and solid lines, which are locations where the residuals switched between positive and negative values.\footnote{Such a feature is expected if the noise-only measurement has a certain shape to it. Other higher-order statistics, such as weak lensing peaks, also find nodes in their data vector where $\texttt{signal} - \texttt{noise} = 0$ \citep[][see their Figure 5]{Zurcher2022WLPeaks}. This does not imply a lack of any cosmological signal, and is simply a consequence of the different shapes of the observed data vector and noise-only data vector.} This crossing implies there are scales where the residuals from the cosmological signal, at a given convergence threshold, are zero. This does not coincide with the scales where the same zero crossing occurs for the PSFs. So in principle, for a given threshold, there can be certain scales where the PSFs contribute more than the cosmological signal. However, this contribution would still be between $1-10\%$ of the measurement uncertainty and thus is not a concern for cosmological constraints.

\subsection{Source Clustering} \label{sec:SourceClustering}

\begin{figure*}
    \centering
    \includegraphics[width = 2\columnwidth]{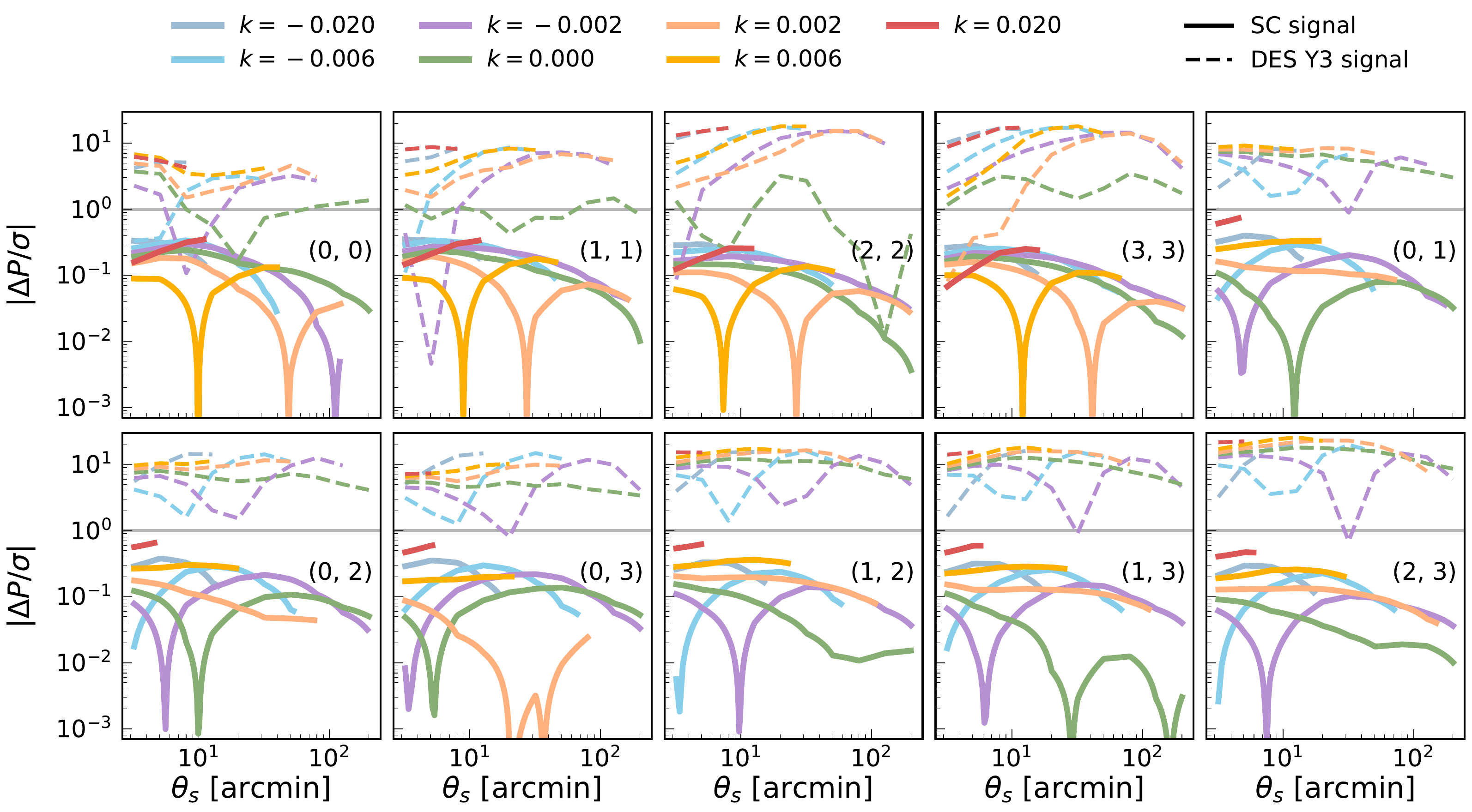}
    \caption{The difference in CDFs measured on two DES Y3-like simulated maps, where one map contains source clustering and the other does not. The signal from source clustering (solid lines)  is at $0.1-0.5\sigma$ and generally contributes $\approx 5-10\%$ to the total signal. The total signal-to-noise of source clustering-induced residuals is $1.3\sigma$.}
    \label{fig:SourceClusteringSNR}
\end{figure*}

Surveys observe the lensing field sampled at the location of source galaxies, and the ellipticities of these source galaxies are then used to infer the original lensing and convergence fields. The standard prediction for the convergence correlations has an additional correction because the source galaxies do not uniformly sample the lensing field and are themselves clustered given they trace the underlying, clustered density field. 

This clustering of source galaxies impacts the observed convergence as follows: the $n(z)$ of a survey details the weighting of the convergence field at different redshifts, and is computed across the whole survey footprint. However, the precise $n(z)$ varies across the sky. For example, at redshift $X$ in direction $\hat{a}$, we can have a significant overdensity of structure. This means the $n(z)$ in the $\hat{a}$ direction has more galaxies at redshift $X$, and the $n(z)$ must be reweighted accordingly. We will refer to this effect henceforth as source clustering (SC), as was first denoted in \citet{Bernardeau1998SC}, though this effect has also been called source-lens clustering \citep{Hamana2002SC}. The effect of source clustering is not present in the fiducial postprocessing technique described in Section \ref{sec:ForwardModelProcess}. However, it can be included through the prescription detailed in \citet[][see their Equation 5]{Gatti2023SC} and previously used in \citet{Gatti2020Moments},
\begin{equation}\label{eqn:SC}
\gamma_{\rm SC}(\nhat) = \frac{\int n(z) (1 + b_g \delta(\nhat, z)) \gamma(\nhat, z)dz}{\int n(z) (1 + b_g \delta(\nhat, z))dz},
\end{equation}
where $n(z)$ is the source redshift distribution of the tomographic bin, averaged across the survey footprint, $\delta(\nhat, z)$ and $\gamma(\nhat, z)$ are the density and true shear maps at a given direction/pixel and redshift, and $b_g$ is the source galaxy bias. In simple terms, equation \eqref{eqn:SC} modulates the $n(z)$ across the survey footprint by reweighting it in a direction-dependent way using the density fields. Note that \citet{Gatti2023SC} take $b_g = 1$, which we follow in this work as well, and this is a fair approximation for source galaxies which tend to be mostly blue galaxies.

In Figure \ref{fig:SourceClusteringSNR}, we show the difference in the CDF datavector measured on a convergence field with/without source clustering. Both sets of simulations have the same noise field, which is described in Section \ref{sec:ForwardModelProcess}. Thus, Figure \ref{fig:SourceClusteringSNR} presents the impact of source clustering on the cosmological signal. We find here that the impact on the CDFs is at most $0.1-0.5\sigma$, and it is generally 1-10\% that of the cosmological signal. \citet{Gatti2020Moments, Gatti2023SC} show the impact of source clustering on the 2nd and 3rd moments is at the $1-10\%$ level as well. \citet{Krause2021Methods} show that the source clustering effect on cosmic shear 2-point functions leads to negligible bias ($<0.15 \sigma$) in cosmological parameter constraints, but this result is obtained after performing fiducial scale cuts which remove scales where the impact of source clustering is most prominent. Thus these findings are still consistent with our statement above that source clustering is a $0.5\sigma$ effect on small scales.

We have thus far checked the impact of source clustering on the convergence field. However, source clustering will also induce a correlation between the true convergence field and the shape noise field. Both the convergence field and the source galaxy number density field depend on the density field, and are thus correlated with one another. Given the noise depends inversely on the source galaxy number density as $\sigma_\kappa \propto 1/\sqrt{n_{\rm gal}}$, the convergence field is correlated with the noise field. For example, consider two redshift bins A and B, with $z_A > z_B$. If there is an overdensity in bin B, it would simultaneously induce a large convergence in bin A and a suppressed noise in bin B, causing an anti-correlation between the convergence field of bin A and the noise field of bin B.

\begin{figure*}
    \centering
    \includegraphics[width = 2\columnwidth]{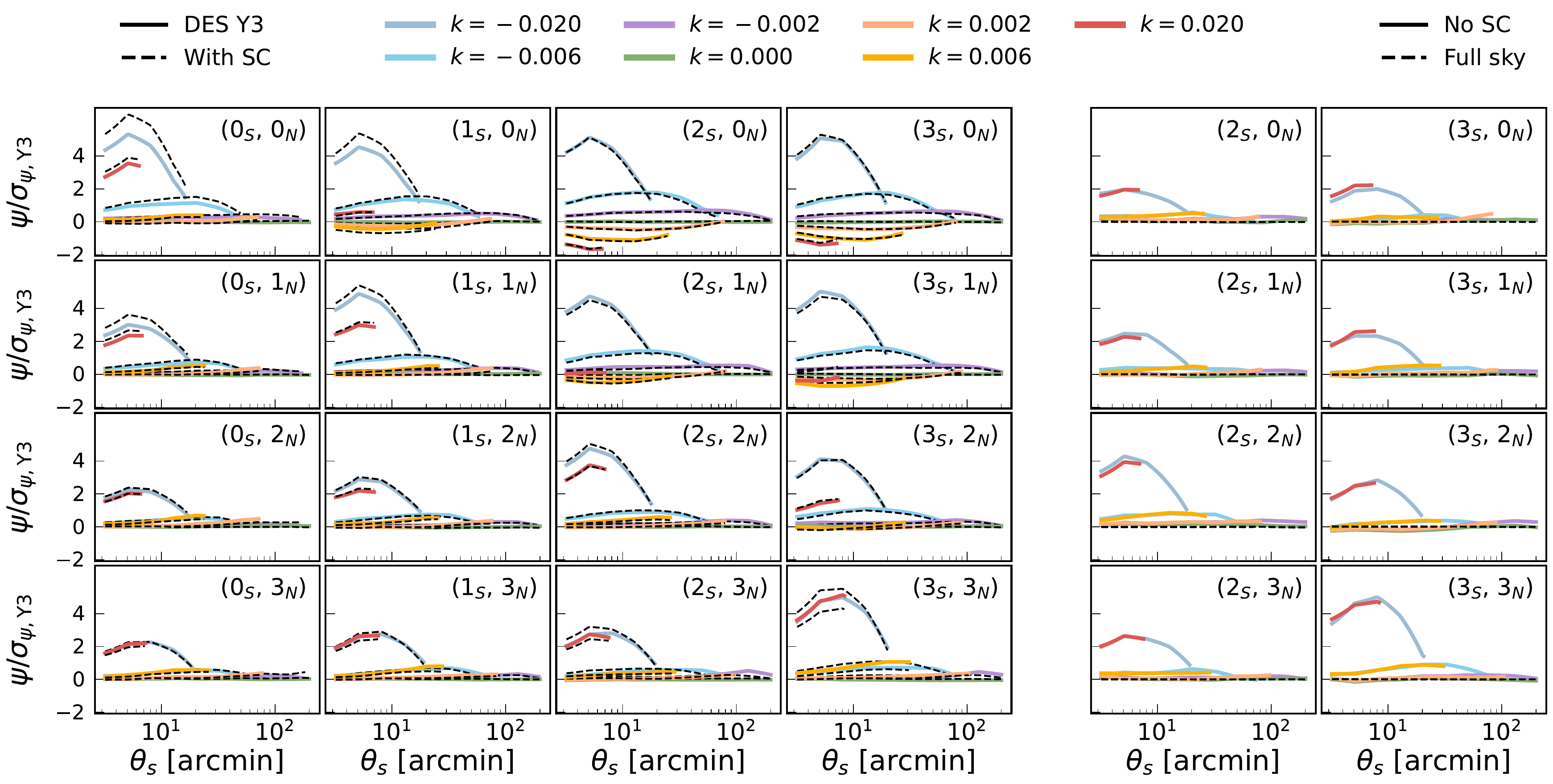}
    \caption{The correlation between two fields, which are the observed convergence field --- either from DES Y3 data or forward modelled from simulations --- and the simulated Y3-like shape noise fields. We find a significant detection of correlation. The panels show the index of the tomographic bin for the observed field (S) and the shape noise field (N). The left panels show the DES Y3 data and the A23 simulations with source clustering. The right panels show a subset of correlations for two other types of simulations --- one with no source clustering, and one with Gaussian noise and no survey mask. The simulations with no source clustering show a clear difference from those with it included. However, even without source clustering, the observed field is correlated with the noise field, and this is due to performing KS reconstruction with a survey mask. We also measure the CDFs on full sky maps that use Gaussian noise and no survey mask. In this regime, the signal and noise fields are completely uncorrelated as expected. The total signal-to-noise of the convergence--shape noise correlation, computed as the difference between the ``With SC'' and ``No SC'' models, is $13 \sigma$. The ``With SC'' model is within $3.5\sigma$ of the Y3 measurements.}
    \label{fig:NoiseSignalCorr}
\end{figure*}

\citet{Gatti2023SC} describe a simple modification of the noise field that models this correlation,
\begin{equation}\label{eqn:SCNoise}
    \gamma_{\rm SC,\, noise}(\nhat) = F(\nhat)\left(\frac{\int n(z)dz}{\int n(z) (1 + b_g \delta(\nhat, z))dz}\right)^{1/2} \gamma_{\rm noise}(\nhat),
\end{equation}
where the definitions are the same as equation \eqref{eqn:SC}, with $\gamma_{\rm noise}(\nhat)$ as the shape noise field, which is obtained as described in Section \ref{sec:ForwardModelProcess}; by using the DES Y3 galaxy shape catalog, and randomly rotating the galaxy orientations. The density factor in Equation \eqref{eqn:SCNoise} varies the number counts of source galaxies across the sky according to the underlying density field. This is the same source clustering effect discussed above but we now consider its effect on the shear noise field, $\gamma_{\rm noise}$, rather than the true shear field, $\gamma$. As a consequence of the density-based reweighting, the even moments (variance, kurtosis etc.) of the modified noise field, $\gamma_{\rm SC,\, noise}(\nhat)$, are slightly inconsistent with those of the original noise field $\gamma_{\rm noise}(\nhat)$. The factor $F(\nhat)$ is implemented as a correction for this inconsistency (see Section 3 of \citet{Gatti2023SC} for a more detailed discussion), and is modelled as
\begin{equation}\label{eqn:Fhat}
    F(\nhat) = A \sqrt{1 - B\sigma^2(\nhat)},
\end{equation}
where $\sigma^2(\nhat) = \gamma^2_{\rm noise, 1}(\nhat) + \gamma^2_{\rm noise, 2}(\nhat)$ is the shear variance, summed over both components, in a given direction/pixel and for a given noise realization. The coefficients $A$ and $B$ are calibrated in \citet{Gatti2023SC} for the four DES Y3 bins using the \textsc{Cosmogrid} simulations, with values $A \in \{0.97, 0.985, 0.990, 0.995\}$ and $B \in \{0.1, 0.05, 0.035, 0.035\}$. We have verified that the results of Figure \ref{fig:NoiseSignalCorr} below are insensitive to the inclusion/exclusion of $F(\nhat)$ in Equation \eqref{eqn:SCNoise}, which is expected as they focus on the \textit{correlations} between fields, rather than the \textit{covariance} between them.

The correction to the noise field in Equation \eqref{eqn:SCNoise} is known to improve the modelling of the 3rd moments, which are sensitive to such convergence--shape noise correlations \citep{Gatti2023SC}. We post-process our simulations using Equations \eqref{eqn:SC} and \eqref{eqn:SCNoise} to obtain convergence maps with such correlations. We then quantify the statistical significance of these correlations, as determined by the CDFs measured on these maps. The CDFs are a useful tool here as they inherit the properties of the kNN distributions, which are the discrete-field version of the CDFs and are a higher signal-to-noise estimator than the 2-point function for determining whether two fields are correlated \citep[][see their Figure 5]{Banerjee2021CrossCorr}.

Figure \ref{fig:NoiseSignalCorr} shows the convergence--shape noise correlation as seen in the CDFs. Instead of the 2-field CDFs, we show the cross-component defined in Equation \eqref{eqn:CrossCorr} and normalize it by the uncertainty in these correlations, estimated across 1000 DES Y3 realizations. Thus, the presented quantity can be interpreted as a significance of correlation. In the left panels are the results from DES Y3 and from the A23 simulations with source clustering. The DES Y3 result is the mean data vector from correlating the same DES Y3 mass map with 1000 different noise maps. The right panels show A23 simulations without source clustering, and finally the A23 simulations with purely Gaussian noise and no survey mask.


The exclusion of source clustering leads to a simulated model that is clearly different from what is observed in the data, and including source clustering brings the model and data into good agreement. The right panels of Figure \ref{fig:NoiseSignalCorr} show that even if we do not include source clustering, there are correlations between the simulated mock maps. Such correlations are expected due to the survey observing properties. The first such cause is survey depth variations, which modulate the source galaxy number density across the sky in the same way for all noise realizations and tomographic bins. The second is the presence of a common survey mask when we perform the KS reconstruction, which induces features in the map that are correlated across independent noise realizations given they all share the same mask. The black dashed lines in the right panels of Figure \ref{fig:NoiseSignalCorr} confirm that a full-sky analysis with Gaussian shape noise and no survey mask --- which by construction has removed the survey property-based effects discussed above --- has no convergence--shape noise correlations.

Figure \ref{fig:NoiseSignalCorr} also shows that convergence--shape noise correlations are statistically significant in the data vector and so are a necessary component in forward-modelling the CDFs. This is also true of other higher-order statistics. The analysis of \citet{Gatti2022MomentsDESY3} found correlations between the signal and noise field but was able to denoise the measurements to remove this effect. This was possible as they used the third moments of the field as their statistic, $\langle\kappa_{\rm obs}^3\rangle = \langle(\kappa_{\rm signal} + \kappa_{\rm noise})^3\rangle$, and so the noise-dependent terms --- such as $\langle\kappa_{\rm signal}\kappa_{\rm noise}^2\rangle$ --- that contributed to the measured moments, $\langle\kappa_{\rm obs}^3\rangle$, could be subtracted exactly. This can be done for moments of any order. For statistics like the CDFs, however, the data vectors depend on the noise in a nonlinear way, and a simple subtraction will not remove all convergence--shape noise correlations. In this case, we are reliant on an accurate forward model of the shape noise field.\footnote{It may still be possible to approximately denoise the CDFs, but we have not explored this possibility in this work.}

\subsection{Higher-order shear effects}\label{sec:HigherOrderShear}

\begin{figure*}
    \centering
    \includegraphics[width = 2\columnwidth]{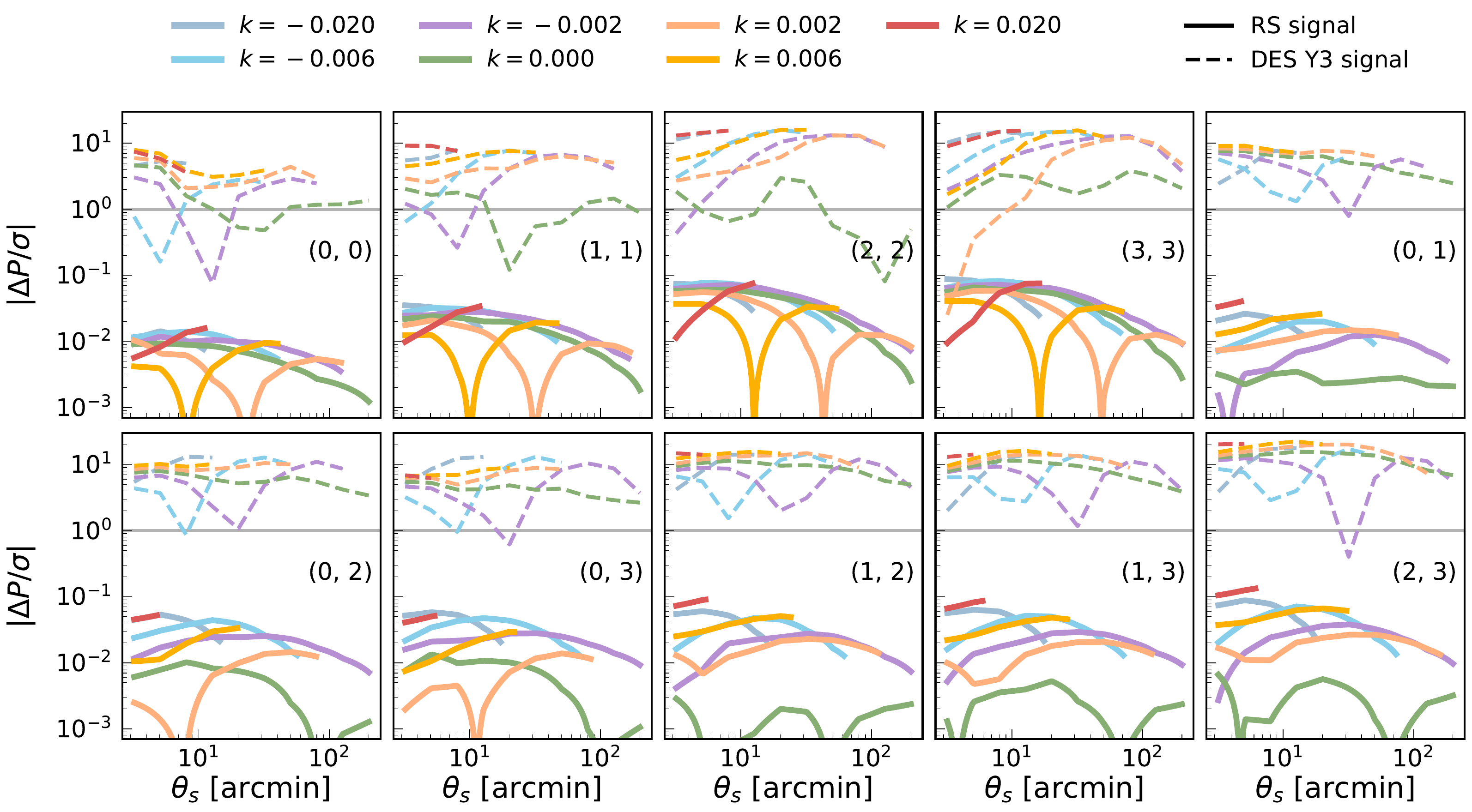}
    \caption{The difference in CDFs depending on whether or not we account for reduced shear effects, $\Delta P = P^{\rm RS} - P^{\rm fid}$. The high redshift bins, especially when looking at the 2-field CDFs, see the largest impact given source planes at high redshift have larger values of $\kappa$ and thus the $1/(1 - \kappa)$ term for the reduced shear is larger. The deviations are still within $\lesssim 0.1\sigma$ in all cases and are 2-3 orders of magnitude below the cosmological signal. The total signal-to-noise of reduced shear-induced residuals is $0.3\sigma$.}
    \label{fig:ReducedShear}
\end{figure*}

In equation \eqref{eqn:ellipticities}, the contribution to the measured ellipticity from the cosmological component is written as $\textbf{e}^{\rm shear}$. This is then connected to the shear field, $\gamma$, as $\textbf{e}^{\rm shear} = \mathbf{\gamma}/(1 - \kappa)$. In the limit of $\kappa \ll 1$, this is approximated to leading order as $\mathbf{\gamma}/(1 - \kappa) \approx \gamma$. Thus, the measured ellipticities are assumed to directly trace the shear $\gamma$, and we ignore higher-order terms, the first of which is $\gamma\kappa$.\footnote{This can be seen by expanding the reduced shear expression as a Taylor series around $\kappa = 0$, which gives $\gamma/(1 - \kappa) \sim \gamma(1 + \kappa + \kappa^2/2 + \ldots)$.} The effect of this approximation is generally known to be subdominant to the cosmological signal \citep{Krause2010ReducedShear}. The specific impact on the 2nd and 3rd moments measured in DES Y3 is also known to be negligible,  especially when compared to the uncertainties in the Y3 measurements and to other effects such as baryon imprints  \citep[][see their Figure 4]{Gatti2020Moments}.

In Figure \ref{fig:ReducedShear} we show the residuals between CDF measurements made on a mass map where the input true shear field is just $\gamma$ and a map where the input field is actually $\gamma/(1 - \kappa)$. Note that by using $\gamma/(1 - \kappa)$ rather than the approximation $\gamma(1 + \kappa + \ldots)$ we test the impact of ignoring all higher-order terms in the reduced shear approximation, rather than just the leading order correction, $\gamma\kappa$. We then perform the full postprocessing pipeline with both map versions. The differences at the data vector level are within $<0.1\sigma$ and are subdominant to the signal by multiple orders of magnitude. The impact of this approximation increases with redshift, which is expected as the variance of the $\kappa$ field increases for source galaxies at higher redshift, and so ignoring the $1/(1 - \kappa)$ factor has a larger significance.

This result also provides a validation for magnification effects, which at leading order in $\kappa$ modify the shear as $\gamma \rightarrow \gamma (1 + q\kappa)$, where $q$ is some $\mathcal{O}(1)$ constant. As was the case with the PSF contributions, these effects have been quantified up to the 3-point function for DES Y3 \citep{Gatti2020Moments}, and we have now implicitly extended it to include higher-order moments through our focus on the CDFs. 

\subsection{Baryon imprints}\label{sec:Baryons}

Finally, we check the impact of baryon modeling on this statistic. Over the past decades, it has been well-established that galaxy formation processes like gas cooling and AGN (Active Galactic Nuclei) feedback can alter the distribution of total matter within and around halos \citep{Blumenthal1986AdiabaticContraction, Gnedin2004AdiabaticContraction, Duffy2010BaryonDmProfileDensity}, which consequently will impact the weak lensing signal \citep{Chisari2018BaryonsPk}. These baryonic imprints have a strong mass/redshift-dependence \citep{Lovell2018ConcentrationBaryonImprints, Beltz-Mohrmann2021BaryonImpactTNG, Anbajagane2022Baryons} and this mass/redshift-dependent impact on the halo potential can vary across simulation prescriptions \citep[\eg][]{Anbajagane2022GalaxyVelBias, Shao2022Baryons}. 

Recently, \citet{Schneider2019Baryonification} implemented a halo-based model that can alter N-body simulations --- which are cheaper to run than full hydrodynamic simulations with galaxy formation --- to then model the baryon imprints on the density/convergence field. This technique provides a higher-level, approximate galaxy formation model that depends only on ``macro'' properties like the halo baryon fraction, the baryon density profiles, dark matter density profile etc. and the flexibility manifesting from the method's approximate nature is particularly useful for matching the range of halo property scaling relations found in the latest hydro simulations \citep[\eg][]{Anbajagane2020StellarProp, Lim2021GasProp, Lee2022rSZ, Cui2022GIZMO, Stiskalek2022TNGHorizon, Anbajagane2022Baryons}.

In this section, we once again compute residuals between CDFs measured on maps from N-body simulations and maps that have been ``baryonified''. Both sets of maps used in this section come from the \textsc{Cosmogrid} suite, and the baryonification was performed with the same model as \citet{Schneider2019Baryonification}. The parameters of the baryonification model were all given their default values, except for some of the gas model parameters which we given values of $M_c = 13.82$ and $\nu = 0$. These parameters are part of a reparameterization done in \citet{Fluri2022wCDMKIDS} and control the gas density profiles' slopes. We take the true convergence fields from \textsc{Cosmogrid} and postprocess them using the same pipeline described in Section \ref{sec:ForwardModelProcess}.

\begin{figure*}
    \centering
    \includegraphics[width = 2\columnwidth]{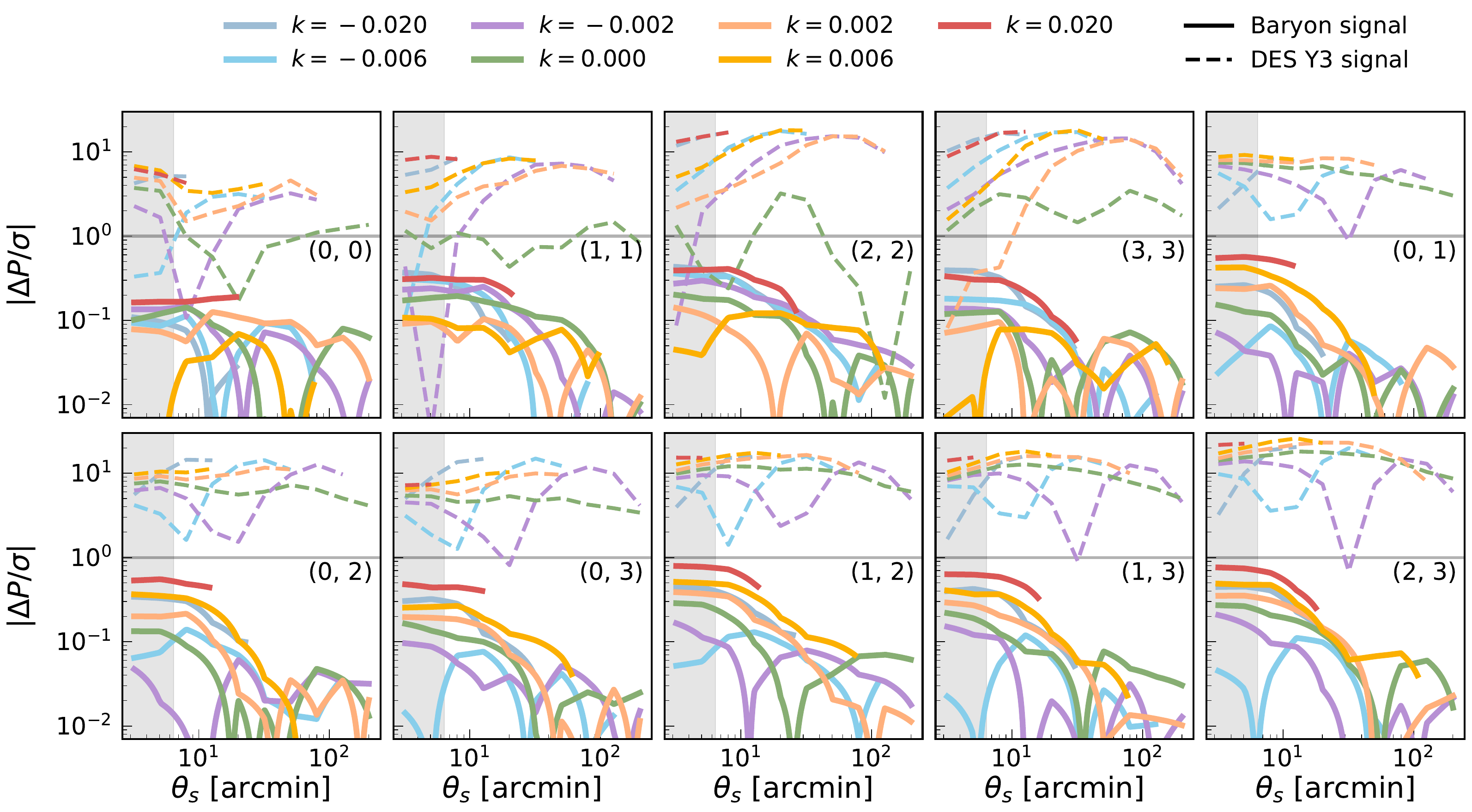}
    \caption{The difference in CDFs measured on dark matter-only (DMO) simulations and ``baryonified'' DMO simulations. As expected, baryon imprints are a significant effect on the data vector. The gray band shows the scales below $\theta < 6.4\arcmin$, which is the pixel resolution of the \textsc{Cosmogrid} DES Y3 maps, and is a factor of 2 larger than the other maps we consider in this work. Thus, the baryon effects we estimate below that scale are an underestimate of the true effect given the pixel resolution will suppress these effects. The total signal-to-noise of baryon imprints is $3.5\sigma$, though this is a lower bound given the suppression due to map resolution.}
    \label{fig:BaryonImprints}
\end{figure*}

Figure \ref{fig:BaryonImprints} shows the residuals due to baryonic imprints on DES Y3-like mock maps. In all cases, the baryon impacts are below $1\sigma$. However, note that the maps from \textsc{Cosmogrid} have a resolution of $\texttt{NSIDE} = 512$, and thus the pixel resolution is $6.4^\prime$ arcmin, instead of the $3.2^\prime$ arcmin minimum scale used in this work. Since the baryons' dominant contribution is on smaller scales, it is likely that the \textit{true} residuals at $3^\prime < \theta < 6.4^\prime$ are actually larger than what is presented in Figure \ref{fig:BaryonImprints} but are currently suppressed due to the pixel resolution of the \textsc{Cosmogrid} maps. Nevertheless, we can state that the baryon imprints for $\theta > 10\arcmin$ have a significance that is approximately 1-2 orders of magnitude below the cosmological signal.

The impact is also highest for the extreme thresholds in the CDF --- the $k = -0.006$ and $k = -0.020$ thresholds --- and this has been seen in previous, theoretical works. \citet{Osato2021KappaTNG} compared hydrodynamic simulations with a dark matter-only counterpart and showed the lensing PDF can be impacted by more than 10\% at the tails of the distribution (see their Figure 5). \citet{Sunseri2023Baryons} used the same set of simulations to show that the impact of baryons on halos, filaments, and voids affects different parts of the matter PDF.

\subsection{Scale cuts}\label{sec:ScaleCuts}

\begin{figure*}
    \centering
    \includegraphics[width = \columnwidth]{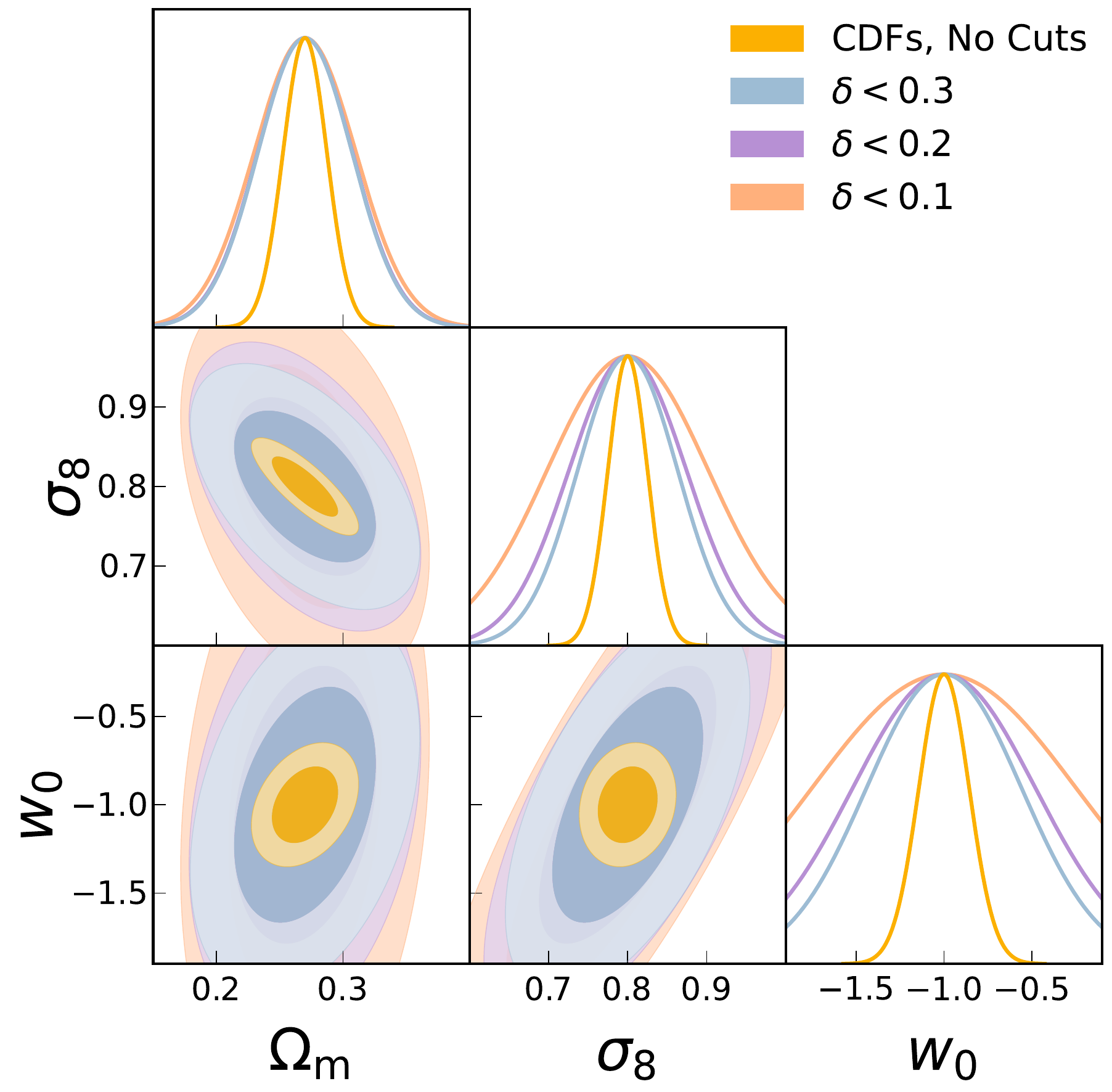}
    \includegraphics[width = \columnwidth]{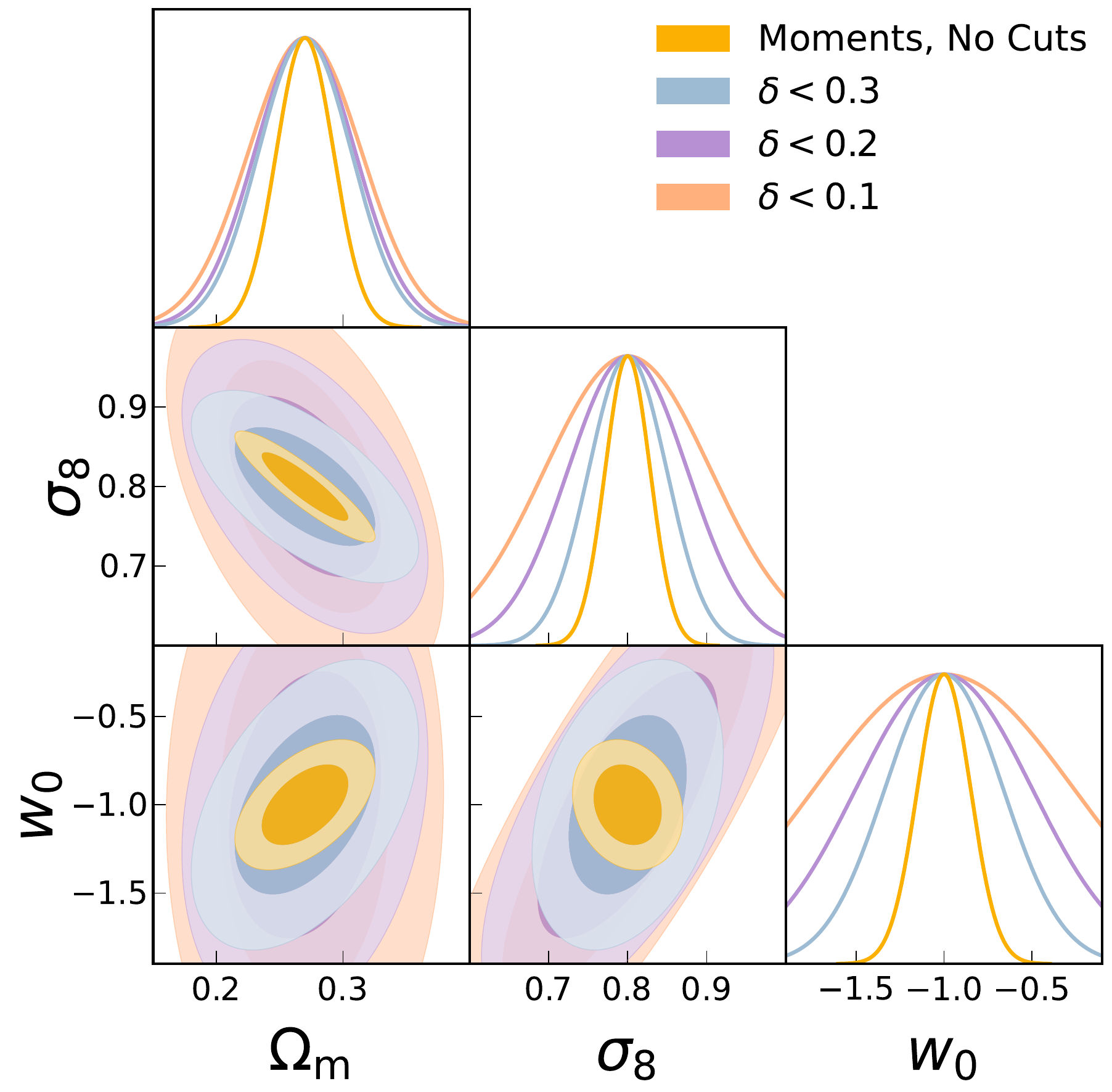}
    \caption{The Fisher constraints from CDFs (left) and 2nd + 3rd Moments (right) measured on simulations. We present four cases, where we either have no scale cuts, or cut the data vector so the parameter bias in the $\Omega_{\rm m}-\sigma_8-w_0$ contour is below a certain value; see Equation \eqref{eqn:NDbias}. The CDFs and the moments have comparable constraints, which are denoted in Table \ref{tab:ScaleCutsCombined}.}
    \label{fig:Scalecuts}
\end{figure*}

In the above sections, we have determined the impact of different systematics and modelling approximations on the CDF data vector. Some systematics are negligible for the whole data vector, such as the PSFs (Section \ref{sec:RoweStats}) and the reduced shear approximation (Section \ref{sec:HigherOrderShear}), while others are prominent at a subset of scales, such as baryon imprints (Section \ref{sec:Baryons}). Thus, using the CDFs to robustly infer cosmological constraints will require us to discard some parts of the fiducial datavector --- namely the parts where the amplitude of the systematics is high --- and obtain constraints using the remaining fraction of the data vector.

Amongst all the systematic effects considered in this work, the most significant are the baryon imprints (Figure \ref{fig:BaryonImprints}) and the source clustering effect (Figure \ref{fig:SourceClusteringSNR}). These will determine how the data vector is truncated. Our scale cuts are determined by requiring that the parameter bias due to unmodeled systematic effects is below a certain threshold. We compute this bias using the extended Fisher formalism of \citet{Amara2008SysBiasFisher,Asgari2021SysBiasFisher},
\begin{equation}\label{eqn:FisherBias}
    \Delta^{\rm bias}_p = \sum_{q} (F^{-1})_{pq}  \frac{d\widetilde{X}_{\rm fid}}{dp} \mathcal{C}^{-1} (\widetilde{X}_{\rm biased} - \widetilde{X}_{\rm fid}),
\end{equation}
where both $p$ and $q$ are indexes over the cosmological parameters of interest. The average bias in the datavector, $\widetilde{X}_{\rm biased} - \widetilde{X}_{\rm fid}$, is a quantity we have already computed and presented in the above subsections. We then summarize this bias-per-parameter, $\Delta^{\rm bias}_p$, into a bias for the full N-D posterior as
\begin{equation}\label{eqn:NDbias}
    \delta = \sqrt{\sum_{p, q}\Delta^{\rm bias}_p (C^{-1})_{pq} \Delta_q^{\rm bias}},
\end{equation}
where $C$ is the covariance of the parameters, and so $C^{-1}$ is just the Fisher matrix, $F$. Our procedure for scale cuts is simply removing datapoints until $\delta < X$, where $X$ is some chosen threshold. We will use $X \in \{0.3, 0.2, 0.1\}$. The choice $X = 0.3$ matches the tests done in the main methodology pipeline for DES Y3 \citep*[\eg][]{Krause2021Methods, Secco2022Shear, Amon2022Shear} while the other values are chosen to explore more stringent cuts that could be reflective of Stage IV surveys. Note that this threshold, $X$, is somewhat arbitrary, but that is not a concern as our goal is to see how the scale-cuts for the CDFs compare to those for the moments; as long as the same choices are applied across both statistics, the arbitrariness of the choices is not relevant.

The other component we must decide is how to determine and discard data points to achieve the condition $\delta < X$, as there is significant freedom in doing so. We could throw away all datapoints for every bin/threshold corresponding to aperture scales below a certain chosen value. However, the choice of a fixed scale threshold is suboptimal as the impact of systematics at a chosen scale varies across bins and thresholds (as seen in any of the Figures above). Thus, our choice here is a scale cut done bin-by-bin (and threshold-by-threshold, in the case of CDFs) and follows the approach of \citet*{Secco2022Shear, Amon2022Shear}. We compute the chi-squared of a given effect in a specific tomographic bin combination (and also specific threshold, in the case of CDFs), and remove the datapoints corresponding to the smallest scales until we satisfy the relation,
\begin{equation}\label{eqn:ScaleCuts}
    (\widetilde{X}_{\rm sub, biased} - \widetilde{X}_{\rm sub, fid})\,\, \mathcal{C}_{\rm sub}^{-1}\,\,(\widetilde{X}_{\rm sub, biased} - \widetilde{X}_{\rm sub, fid})^T < \Delta\chi_{\rm thresh}^2,
\end{equation}
where $\widetilde{X}_{\rm sub, biased}$ and $\widetilde{X}_{\rm sub, fid}$ are subsets of the datavectors used in Equation \eqref{eqn:FisherBias}, where the subsets correspond to specific tomographic bin combination (and threshold, when using CDFs), $\mathcal{C}_{\rm sub}$ is the covariance matrix of the subset, and $\Delta\chi_{\rm thresh}^2$ is the maximum change in $\chi^2$ we allow for the full datavector. In practice, we vary $\Delta\chi_{\rm thresh}^2$ until the parameter bias goes below our required threshold. The datapoints that have been removed to achieve this condition define the scale cuts.

Once the scale-cuts have been defined, we recompute the Fisher constraints using the truncated data vector; the results for the CDFs are tabulated in Table \ref{tab:ScaleCuts}. The table also shows constraints from generic scale cuts, where we set a fixed minimum angular scale for all tomographic bins and all thresholds. Cutting all scales below $20\arcmin$ causes a fractional change of $\approx 30\%$ in the constraints. At these scale cuts, the CDFs are comparable to combining 2nd and 3rd Moments, and we have verified that combining the CDFs with the moments still leads to a 30\% improvement in the constraints. The PSFs and reduced shear effect have no impact on scale cuts so we do not show them here. Note that, as was discussed in Section \ref{sec:Baryons}, the impact of baryonic effects is an underestimate given the baryonified \textsc{Cosmogrid} maps used to estimate the effect have a $6.4\arcmin$ minimum resolution scale. Baryon effects are more impactful at smaller scales and will be more than $10\%$ of the signal if the resolution limit is corrected. However, for our goal of consistently comparing the impacts on CDFs and Moments, this suppression is not a limiting factor. 

Table \ref{tab:ScaleCuts} shows that baryon imprints and source clustering both cause notable differences in the parameter constraints, especially in $\sigma_8$ and $w_0$. The FoM in the 3D parameter space drops by a factor of nearly 10 after implementing these scale cuts, which highlights the growing need to improve modelling of these effects instead of robustly trimming the data vector to be insensitive to the effects. Note that while the impact of source clustering on determining the scale cuts is larger than that of the baryonic imprints --- which is counter to the standard expectation --- this is once again because of the suppression of baryon effects on the small scales due to the resolution scale of the \textsc{Cosmogrid} data products.

Figure \ref{fig:Scalecuts} and Table \ref{tab:ScaleCutsCombined} also show the results from defining scale cuts using both baryon imprints and source clustering, and doing so for CDFs \textit{and} for the 2nd and 3rd moments. This provides a self-consistent reference to compare the two data vectors. The combination of scale cuts is done by looking at both baryonic effects and source clustering, and at each datapoint we pick the amplitude of the effect that is highest, i.e. $E = \max|{\rm Baryons, SC}|$ for each datapoint. We find that the moments' constraints are comparable to the CDFs' after these scale cuts.  Once we remove $w_0$ from the analysis the scale cuts cause only a factor of 3 degradation of the FoM as opposed to the factor of 10 if we include $w_0$.

Generally, one may expect the CDFs to be less sensitive to these effects than the moments; reduced shear, source clustering, and baryon imprints are all effects that grow with the amplitude of the density field and/or the convergence field. This means they impact the tails of the density/lensing distribution the most and leave the ``bulk'' of the PDF --- roughly the 68\% or the 95\% region centered around the median --- relatively unaffected. The moments are defined as an integral over the whole distribution and so cannot isolate just parts of it. The CDFs on the other hand \textit{can} perform such an isolation. They fundamentally only probe whether or not a pixel's convergence is above a given threshold; thus, if the convergence is well above/below the threshold, the measurement of the CDFs is unaffected by that pixel value shifting around due to various effects. For example, the negative thresholds $k < 0$ will be unaffected by the baryon imprints in massive halos, as massive halos exist in $\kappa > 0$ regions and baryon imprints reduce the $\kappa$ value but always keep it positive, and so the convergence around halos will always be above the $k < 0$. Of course, if the $\kappa$ values of interest are near a threshold, then any shifts will have a stronger impact on the CDF measurements at that threshold. This argument also suggests there are a particular choice of thresholds that balance constraining power while alleviating such systematics. We have not explored such an optimal selection. In Table \ref{tab:ScaleCuts} we also redo the scale cuts but now leave out $w_0$ when computing the total parameter bias, as this is a closer match to the procedures used in Stage III surveys \citep[\eg][]{Krause2021Methods}. Our qualitative findings remain the same even in this case.

\begin{table}
    \centering
    \begin{tabular}{c|c|c|c|c|c}
       \hline
       \hline
       Scale Cut & $\sigma(\Omega_{\rm m})$ & $\sigma(\sigma_8)$ & $\sigma(w_0)$ & $\rm FoM$ & $N_{\rm dof}$ \\
       \hline
       \multicolumn{6}{c}{\textit{Fixed angular cuts}}\\
       \hline
       $\theta > 3\arcmin$ & \textbf{0.018} & \textbf{0.025} & \textbf{0.15} & \textbf{1.0} & \textbf{460}\\
       $\theta > 10\arcmin$ & 0.022 & 0.032 & 0.18 & 0.85 &  410\\
       $\theta > 20\arcmin$ & 0.025 & 0.035 & 0.21 & 0.59 &  270\\
       \hline
       \multicolumn{6}{c}{\textit{Baryonic imprints cuts}}\\ 
       \hline
       $\delta < 0.3$ & 0.033 & 0.053 & 0.33  & 0.14 & 129\\
       $\delta < 0.2$ & 0.035 & 0.061 & 0.39  & 0.10 & 100\\
       $\delta < 0.1$ & 0.036 & 0.062 & 0.42  & 0.08 & 92\\
       \hline
       \multicolumn{6}{c}{\textit{Source clustering cuts}}\\ 
       \hline
       $\delta < 0.3$ & 0.038 & 0.063 & 0.44  & 0.07 & 84\\
       $\delta < 0.2$ & 0.038 & 0.078 & 0.57  & 0.05 & 71\\
       $\delta < 0.1$ & 0.042 & 0.109 & 0.82  & 0.03 & 34\\
       \hline
    \end{tabular}
    \caption{The Fisher information constraints presented in this work for CDFs measured on simulations and for a joint analysis of $\Omega_{\rm m}$, $\sigma_8$, and $w_0$, but after implementing various types of scale cuts. From top to bottom, we do (i) simple, fixed angular scale cuts, and then cuts based on (ii) Baryonic imprints, and (iii) Source Clustering. The cuts are made by removing data points until $\delta < X$, where $\delta$ --- defined in equation \eqref{eqn:NDbias} --- is the total parameter bias in a full N-D parameter space. We show the size of the modified data vector in the rightmost column. The Figure of Merit (FoM) is quoted relative to the FoM of the CDFs constraints with no scale cuts.}
    \label{tab:ScaleCuts}
\end{table}

\begin{table}
    \centering
    \begin{tabular}{c|c|c|c|c|c}
       \hline
       \hline
       Scale Cut & $\sigma(\Omega_{\rm m})$ & $\sigma(\sigma_8)$ & $\sigma(w_0)$ & FOM & $N_{\rm dof}$ \\
       \hline
       \multicolumn{6}{c}{\textit{CDFs, All cuts ($\Omega_m$, $\sigma_8$, $w_0$)}}\\ 
       \hline
       $\delta{\rm \CDF} < 0.3$ & \textbf{0.037} & \textbf{0.063} & \textbf{0.44}  & 0.07 & 84\\
       $\delta{\rm \CDF} < 0.2$ & 0.037 & 0.074 & 0.52  & 0.05 & 75\\
       $\delta{\rm \CDF} < 0.1$ & 0.040 & 0.102 & 0.75  & 0.03  & 49\\
       \hline
       \multicolumn{6}{c}{\textit{2nd \& 3rd Moments, All cuts ($\Omega_m$, $\sigma_8$, $w_0$)}}\\ 
       \hline
       $\delta{\rm Moments} < 0.3$ & 0.037 & 0.050 & 0.34 &  0.13 & 109\\
       $\delta{\rm Moments} < 0.2$ & 0.040 & 0.076 & 0.50  & 0.06 & 86\\
       $\delta{\rm Moments} < 0.1$ & 0.045 & 0.105 & 0.74 & 0.03 & 58\\
       \hline
       \hline
       \multicolumn{6}{c}{\textit{CDFs, All cuts ($\Omega_m$, $\sigma_8$)}}\\
       \hline
       $\delta{\rm \CDF} < 0.3$ & 0.033 & 0.050 & ---  & 0.27 & 99\\
       $\delta{\rm \CDF} < 0.2$ & 0.035 & 0.053 & ---  & 0.20 & 79\\
       $\delta{\rm \CDF} < 0.1$ & 0.038 & 0.057 & ---  & 0.16 & 52\\
       \hline
       \multicolumn{6}{c}{\textit{2nd \& 3rd Moments, All cuts ($\Omega_m$, $\sigma_8$)}}\\ 
       \hline
       $\delta{\rm Moments} < 0.3$ & 0.031 & 0.044 & --- & 0.34 & 118\\
       $\delta{\rm Moments} < 0.2$ & 0.036 & 0.051 & --- & 0.22 & 91\\
       $\delta{\rm Moments} < 0.1$ & 0.044 & 0.058 & --- & 0.12 & 4\\
       \hline
    \end{tabular}
    \caption{The Fisher information constraints presented in this work for a joint analysis of either $\Omega_{\rm m}$, $\sigma_8$, and $w_0$ (top two) or just $\Omega_{\rm m}$ and $\sigma_8$ (bottom two), but after implementing scale cuts to reduce the parameter bias. We show the constraints, after scale cuts, for both the CDFs and for the combination of 2nd and 3rd Moments.  We show the size of the modified data vector in the rightmost column. The Figure of Merit (FoM) is quoted relative to the FoM of the CDFs constraints with no scale cuts.}
    \label{tab:ScaleCutsCombined}
\end{table}

\section{Conclusions}\label{sec:Conclusion}

In this work, we have explored the use of the Cumulative Distribution Functions (CDFs) of the convergence field as a summary statistic for extracting cosmological information, drawing on the development of the k-Nearest Neighbor distributions for the discrete fields. The CDFs are a convenient, succinct summary of the field that approximately capture all higher moments of the field in a significantly shorter data vector that is also quicker to compute. We explore the theoretical advantages of using these CDFs and check their sensitivity to the relevant practical challenges in extracting robust cosmology constraints from Y3-like data. The conclusions of this work are as follows:

\begin{itemize}
    \item For scales of $3\arcmin < \theta < 200\arcmin$ and tomographic bins of DES Y3, the CDFs have better constraints on $\Omega_m$, $\sigma_8$ and $w$ when compared to those from the combination of both 2nd and 3rd Moments (Figure \ref{fig:Fisher}). This improvement is modest, but the CDFs still have a slightly different degeneracy direction to the moments, and combining the CDFs and moments leads to the constraints improving by $20-30\%$.
    \item The CDFs measured on a Gaussian field provide Fisher constraints that are completely consistent with the angular power spectra and 2nd Moments computed on the fully non-linear, non-Gaussian field (Figure \ref{fig:Fisher}). The CDFs and moments all have Gaussian likelihoods as well (Figure \ref{fig:CheckCov}).
    \item The DES Y3 noise field is highly non-Gaussian, with a very significant 4th moment (Figure \ref{fig:NonGaussSNR}). There is some cosmological signal at large scales in the 4th moment, but none in the 5th moment.
    \item We create a PSF ``mass map'' for testing PSF contributions at the map level, and show the signal from PSF shapes is 2-3 orders of magnitude below the cosmological signal (Figure \ref{fig:RhoStats}). This validates not only the CDFs, but also indirectly validates the minimal impact of the PSFs on information beyond the 3rd moment (existing works have already validated them at the 2nd and 3rd moment level).
    \item The presence or lack of spatial correlations in the source galaxy number counts, i.e. ``source clustering'', impacts the convergence field model at the 1-10\% level (Figure \ref{fig:SourceClusteringSNR}).
    \item The CDFs are sensitive to correlations between the convergence field and the shape noise field, induced by source clustering. We detect these correlations at $13 \sigma$, and can adequately model them in the simulated maps (Figure \ref{fig:NoiseSignalCorr}).
    \item The reduced shear approximation changes the cosmological signal at the $1-5\%$ level (Figure \ref{fig:ReducedShear}), while baryon imprints are $1-10\%$ of the cosmological signal (Figure \ref{fig:BaryonImprints}).
    \item We perform scale cuts that limit the parameter bias due to systematic effects under a certain level. The cut CDF data vector has comparable constraining power to the cut data vector of the 2nd and 3rd Moments (Table \ref{tab:ScaleCuts} and Figure \ref{fig:Scalecuts}).
\end{itemize}

Optimizing the summary of fields is a rich area of study, with a variety of approaches and outcomes. The CDFs, through their sensitivity to all moments of the field, probe both the cosmological signal at all these orders as well as any potential modeling challenges that surface at these orders (eg. the high kurtosis of the noise field that does not impact 2-point and 3-point functions). This sensitivity to all orders becomes a more relevant trait as we extend our analyses to smaller scales, which are more non-linear and thus more non-Gaussian. It may also become relevant in constraining --- and/or marginalizing over --- the impact of baryons on the density field; these effects happen predominantly within halos, and so are localized around the most non-linear regions of the density field and thus will have non-Gaussian signatures. The CDFs might also be one of the few ways to probe the highest orders of information in the field. They are more robust given they can isolate specific parts of the distribution, and this is in contrast to the higher order moments which will be increasingly sensitive to noise/outliers in the tails of the distribution. Thus, if there is significant, usable higher-order information in the cosmological field (for example, in future surveys with different noise levels and sensitivities), the CDF may be one of the only ways to robustly access it.

While efforts have already been made to obtain cosmology from up to the 3rd moment, we show there remains some information beyond the 3rd moment that can likely be accessed in a robust manner, i.e. without worrying about systematics. Effects like reduced shear, source clustering, and baryons have some impact that is at the $0.1\% - 10\%$ level depending on the effect and the angular scale. After enacting scale cuts to reduce the bias on cosmological constraints to be within $0.3\sigma$, the CDF data vector still provides constraints better than those of the  2nd and 3rd moment data vector. We have identified that accurate modelling of the noise field at higher orders is the current limiting factor in robustly inferring cosmology from statistics like the CDFs. Alternatively, an accurate way of denoising the CDFs --- which effectively bypasses requirements in modeling the noise field by removing its contribution from the data vector --- would enable robust cosmology constraints with the CDFs.

Finally, we note that even though this work has specifically focused on validating the CDF as a summary statistic, the validation results have significant implications for the broader range of lensing convergence statistics discussed in the literature. The key underlying information is the distribution of convergence as a function of scale, $P(\kappa_\theta)$, and the CDFs are a convenient and compact way of summarizing this distribution/information. Other statistics summarize this distribution in different ways, such as lensing-in-cells\footnote{This is the lensing-focused analog of counts-in-cells, where the latter is the distribution of tracer counts within a given volume, $P(k_{\rm tr}\,|\,V)$. If we replace trace counts with lensing convergence, then we obtain lensing-in-cells.} and Minkowski Functionals.\footnote{The CDFs are the same as the zeroth-order Minkowski functional, though in our formalism we also introduce a cross-correlation method --- inspired by the formalism for kNNs in \citet{Banerjee2021CrossCorr} --- which is traditionally not used/defined for the Minkowski Functionals.} As has been discussed above, another closely connected statistic is the moments of the field, $\langle \kappa_\theta^n \rangle$, which are a further summary of the distribution, $P(\kappa_\theta \,|\,\thetasmooth)$, and computing moments to an arbitrarily high order is equivalent to computing the CDF to arbitrarily many thresholds. 

As we move towards Stage IV surveys with wider survey areas and deeper observations --- both leading to higher precision measurements --- other systematics could become relevant. As a rough example, the LSST Year 10 dataset will have $\sim 3$ times the survey area as DES Y3 and $\sim 5$ times the source galaxy number density as DES Y3 \citep{LSST2018SRD}, which leads to a factor of 4 increase in precision of the data vector and in the significance of any systematic we discuss in this work. The reduced shear effect (Figure \ref{fig:ReducedShear}) --- which can be safely ignored in Stage III surveys --- will likely need to be included in the model for Stage IV, especially for LSST's highest redshift bins as the amplitude of the effect grows with redshift. However, this component can be trivially included via simulation-based modeling using the same approach we used to include its effects in our simulations (Section \ref{sec:HigherOrderShear}). Source clustering will also be a necessary modelling ingredient for Stage IV surveys as its signal-to-noise will exceed 1 for LSST. While this modeling can also be done through simulation-based modeling, it requires some galaxy bias prescription (Equation \ref{eqn:SC}) which would introduce a modeling uncertainty that has yet to be quantified. Additionally, we discussed that the Born approximation is adequate for modelling the weak lensing field under DES-like uncertainties. However, previous works have shown that for Stage IV data quality we will require ray tracing when using higher-order statistics \citep{Petri2017Born}.

These effects above --- reduced shear, source clustering, and Born approximation --- impact all statistical summaries of the lensing field, including the standard 2pt and 3pt functions. Systematics that will uniquely impact the CDFs are then effects that generate a fourth moment and beyond. We have already found in this work that the fourth moment of the noise field is a highly relevant modeling component for the CDFs. In DES Y3, this was primarily sourced by the survey depth fluctuations as well as the intrinsic, cosmological clustering of source galaxies. In general, however, any process that spatially modifies the shape noise per galaxy or the number of galaxies per pixel will generate the fourth moment. For Stage IV surveys, the precision will be high enough that effects such as spatially varying multiplicative bias --- which impacts the measured variance of the shape distribution --- could also be a required modeling component, but we must first quantify how much this bias will actually vary across the sky.

The validation steps performed in this work have implications for the statistics mentioned above --- lensing-in-cells, Minkowski Functionals, field moments etc. For example, it is likely that PSF ellipticity correlations will be a few orders of magnitude below the cosmological signal for all of these statistics. A similar case can be made for the impact of source clustering and the reduced shear approximation. Of course, it is still ideal to perform a separate validation for those statistics to explicitly verify their robustness to these effects, but the results of this work indicate --- given the statistics all summarize the same underlying distribution, $P(\kappa_\theta \,|\,\thetasmooth)$ --- that it is likely these other statistics will also be robust to these. By using the CDFs, which are approximately summarizing all higher order moments, we have tested these systematics at the map level and beyond the 3rd moment. We hope the methodologies for map-level tests that we employed and/or introduced in this work enable more checks of the large library of higher-order statistics that are being developed for the convergence field, and thus enhance the trustworthiness of these newer statistics.


\section*{Acknowledgements}

DA is supported by NSF grant No. 2108168. CC is supported by the Henry Luce Foundation and DOE grant DE-SC0021949.

Funding for the DES Projects has been provided by the U.S. Department of Energy, the U.S. National Science Foundation, the Ministry of Science and Education of Spain, 
the Science and Technology Facilities Council of the United Kingdom, the Higher Education Funding Council for England, the National Center for Supercomputing 
Applications at the University of Illinois at Urbana-Champaign, the Kavli Institute of Cosmological Physics at the University of Chicago, 
the Center for Cosmology and Astro-Particle Physics at the Ohio State University,
the Mitchell Institute for Fundamental Physics and Astronomy at Texas A\&M University, Financiadora de Estudos e Projetos, 
Funda{\c c}{\~a}o Carlos Chagas Filho de Amparo {\`a} Pesquisa do Estado do Rio de Janeiro, Conselho Nacional de Desenvolvimento Cient{\'i}fico e Tecnol{\'o}gico and 
the Minist{\'e}rio da Ci{\^e}ncia, Tecnologia e Inova{\c c}{\~a}o, the Deutsche Forschungsgemeinschaft and the Collaborating Institutions in the Dark Energy Survey. 

The Collaborating Institutions are Argonne National Laboratory, the University of California at Santa Cruz, the University of Cambridge, Centro de Investigaciones Energ{\'e}ticas, 
Medioambientales y Tecnol{\'o}gicas-Madrid, the University of Chicago, University College London, the DES-Brazil Consortium, the University of Edinburgh, 
the Eidgen{\"o}ssische Technische Hochschule (ETH) Z{\"u}rich, 
Fermi National Accelerator Laboratory, the University of Illinois at Urbana-Champaign, the Institut de Ci{\`e}ncies de l'Espai (IEEC/CSIC), 
the Institut de F{\'i}sica d'Altes Energies, Lawrence Berkeley National Laboratory, the Ludwig-Maximilians Universit{\"a}t M{\"u}nchen and the associated Excellence Cluster Universe, 
the University of Michigan, NSF's NOIRLab, the University of Nottingham, The Ohio State University, the University of Pennsylvania, the University of Portsmouth, 
SLAC National Accelerator Laboratory, Stanford University, the University of Sussex, Texas A\&M University, and the OzDES Membership Consortium.

Based in part on observations at Cerro Tololo Inter-American Observatory at NSF's NOIRLab (NOIRLab Prop. ID 2012B-0001; PI: J. Frieman), which is managed by the Association of Universities for Research in Astronomy (AURA) under a cooperative agreement with the National Science Foundation.

The DES data management system is supported by the National Science Foundation under Grant Numbers AST-1138766 and AST-1536171.
The DES participants from Spanish institutions are partially supported by MICINN under grants ESP2017-89838, PGC2018-094773, PGC2018-102021, SEV-2016-0588, SEV-2016-0597, and MDM-2015-0509, some of which include ERDF funds from the European Union. IFAE is partially funded by the CERCA program of the Generalitat de Catalunya.
Research leading to these results has received funding from the European Research
Council under the European Union's Seventh Framework Program (FP7/2007-2013) including ERC grant agreements 240672, 291329, and 306478.
We  acknowledge support from the Brazilian Instituto Nacional de Ci\^encia
e Tecnologia (INCT) do e-Universo (CNPq grant 465376/2014-2).

This manuscript has been authored by Fermi Research Alliance, LLC under Contract No. DE-AC02-07CH11359 with the U.S. Department of Energy, Office of Science, Office of High Energy Physics.



\bibliographystyle{mnras_2author}
\bibliography{References} 

\begin{thebibliography}{}
\makeatletter
\relax
\def\mn@urlcharsother{\let\do\@makeother \do\$\do\&\do\#\do\^\do\_\do\%\do\~}
\def\mn@doi{\begingroup\mn@urlcharsother \@ifnextchar [ {\mn@doi@}
  {\mn@doi@[]}}
\def\mn@doi@[#1]#2{\def\@tempa{#1}\ifx\@tempa\@empty \href
  {http://dx.doi.org/#2} {doi:#2}\else \href {http://dx.doi.org/#2} {#1}\fi
  \endgroup}
\def\mn@eprint#1#2{\mn@eprint@#1:#2::\@nil}
\def\mn@eprint@arXiv#1{\href {http://arxiv.org/abs/#1} {{\tt arXiv:#1}}}
\def\mn@eprint@dblp#1{\href {http://dblp.uni-trier.de/rec/bibtex/#1.xml}
  {dblp:#1}}
\def\mn@eprint@#1:#2:#3:#4\@nil{\def\@tempa {#1}\def\@tempb {#2}\def\@tempc
  {#3}\ifx \@tempc \@empty \let \@tempc \@tempb \let \@tempb \@tempa \fi \ifx
  \@tempb \@empty \def\@tempb {arXiv}\fi \@ifundefined
  {mn@eprint@\@tempb}{\@tempb:\@tempc}{\expandafter \expandafter \csname
  mn@eprint@\@tempb\endcsname \expandafter{\@tempc}}}

\bibitem[\protect\citeauthoryear{{Adelberger} \& {Steidel}
  et~al.,}{{Adelberger} et~al.}{1998}]{Adelberger1998CIC}
{Adelberger} K.~L.,  et~al. 1998, \mn@doi [\apj] {10.1086/306162}, \href
  {https://ui.adsabs.harvard.edu/abs/1998ApJ...505...18A} {505, 18}

\bibitem[\protect\citeauthoryear{{Allys} \& {Marchand} et~al.,}{{Allys}
  et~al.}{2020}]{Allys2020WPHandLSS}
{Allys} E.,  et~al. 2020, \mn@doi [\prd] {10.1103/PhysRevD.102.103506}, \href
  {https://ui.adsabs.harvard.edu/abs/2020PhRvD.102j3506A} {102, 103506}

\bibitem[\protect\citeauthoryear{{Amara} \& {R{\'e}fr{\'e}gier}}{{Amara} \&
  {R{\'e}fr{\'e}gier}}{2008}]{Amara2008SysBiasFisher}
{Amara} A.,  {R{\'e}fr{\'e}gier} A.,  2008, \mn@doi [\mnras]
  {10.1111/j.1365-2966.2008.13880.x}, \href
  {https://ui.adsabs.harvard.edu/abs/2008MNRAS.391..228A} {391, 228}

\bibitem[\protect\citeauthoryear{{Amon} \& {Gruen} et~al.,}{{Amon}
  et~al.}{2022}]{Amon2022Shear}
{Amon} A.,  et~al. 2022, \mn@doi [\prd] {10.1103/PhysRevD.105.023514}, \href
  {https://ui.adsabs.harvard.edu/abs/2022PhRvD.105b3514A} {105, 023514}

\bibitem[\protect\citeauthoryear{{Anbajagane} \& {Evrard} et~al.,}{{Anbajagane}
  et~al.}{2020}]{Anbajagane2020StellarProp}
{Anbajagane} D.,  et~al. 2020, \mn@doi [\mnras] {10.1093/mnras/staa1147}, \href
  {https://ui.adsabs.harvard.edu/abs/2020MNRAS.495..686A} {495, 686}

\bibitem[\protect\citeauthoryear{{Anbajagane}, {Evrard}  \&
  {Farahi}}{{Anbajagane} et~al.}{2022a}]{Anbajagane2022Baryons}
{Anbajagane} D.,  {Evrard} A.~E.,   {Farahi} A.,  2022a, \mn@doi [\mnras]
  {10.1093/mnras/stab3177}, \href
  {https://ui.adsabs.harvard.edu/abs/2022MNRAS.509.3441A} {509, 3441}

\bibitem[\protect\citeauthoryear{{Anbajagane} \& {Aung} et~al.,}{{Anbajagane}
  et~al.}{2022b}]{Anbajagane2022GalaxyVelBias}
{Anbajagane} D.,  et~al. 2022b, \mn@doi [\mnras] {10.1093/mnras/stab3587},
  \href {https://ui.adsabs.harvard.edu/abs/2022MNRAS.510.2980A} {510, 2980}

\bibitem[\protect\citeauthoryear{{Asgari} \& {Friswell} et~al.,}{{Asgari}
  et~al.}{2021a}]{Asgari2021SysBiasFisher}
{Asgari} M.,  et~al. 2021a, \mn@doi [\mnras] {10.1093/mnras/staa3810}, \href
  {https://ui.adsabs.harvard.edu/abs/2021MNRAS.501.3003A} {501, 3003}

\bibitem[\protect\citeauthoryear{{Asgari} \& {Lin} et~al.,}{{Asgari}
  et~al.}{2021b}]{Asgari2021ShearKIDS}
{Asgari} M.,  et~al. 2021b, \mn@doi [\aap] {10.1051/0004-6361/202039070}, \href
  {https://ui.adsabs.harvard.edu/abs/2021A&A...645A.104A} {645, A104}

\bibitem[\protect\citeauthoryear{{Banerjee} \& {Abel}}{{Banerjee} \&
  {Abel}}{2021a}]{Banerjee2021kNN}
{Banerjee} A.,  {Abel} T.,  2021a, \mn@doi [\mnras] {10.1093/mnras/staa3604},
  \href {https://ui.adsabs.harvard.edu/abs/2021MNRAS.500.5479B} {500, 5479}

\bibitem[\protect\citeauthoryear{{Banerjee} \& {Abel}}{{Banerjee} \&
  {Abel}}{2021b}]{Banerjee2021CrossCorr}
{Banerjee} A.,  {Abel} T.,  2021b, \mn@doi [\mnras] {10.1093/mnras/stab961},
  \href {https://ui.adsabs.harvard.edu/abs/2021MNRAS.504.2911B} {504, 2911}

\bibitem[\protect\citeauthoryear{{Banerjee} \& {Abel}}{{Banerjee} \&
  {Abel}}{2023}]{Banerjee2023TracerFieldkNN}
{Banerjee} A.,  {Abel} T.,  2023, \mn@doi [\mnras] {10.1093/mnras/stac3813},
  \href {https://ui.adsabs.harvard.edu/abs/2023MNRAS.519.4856B} {519, 4856}

\bibitem[\protect\citeauthoryear{{Barthelemy} \& {Halder} et~al.,}{{Barthelemy}
  et~al.}{2023}]{Barthelemy2023PDF}
{Barthelemy} A.,  et~al. 2023, \mn@doi [arXiv e-prints]
  {10.48550/arXiv.2307.09468}, \href
  {https://ui.adsabs.harvard.edu/abs/2023arXiv230709468B} {p. arXiv:2307.09468}

\bibitem[\protect\citeauthoryear{{Baugh}, {Gaztanaga}  \& {Efstathiou}}{{Baugh}
  et~al.}{1995}]{Baugh1995CIC}
{Baugh} C.~M.,  {Gaztanaga} E.,   {Efstathiou} G.,  1995, \mn@doi [\mnras]
  {10.1093/mnras/274.4.1049}, \href
  {https://ui.adsabs.harvard.edu/abs/1995MNRAS.274.1049B} {274, 1049}

\bibitem[\protect\citeauthoryear{{Beltz-Mohrmann} \&
  {Berlind}}{{Beltz-Mohrmann} \&
  {Berlind}}{2021}]{Beltz-Mohrmann2021BaryonImpactTNG}
{Beltz-Mohrmann} G.~D.,  {Berlind} A.~A.,  2021, arXiv e-prints, \href
  {https://ui.adsabs.harvard.edu/abs/2021arXiv210305076B} {p. arXiv:2103.05076}

\bibitem[\protect\citeauthoryear{{Bernardeau}}{{Bernardeau}}{1998}]{Bernardeau1998SC}
{Bernardeau} F.,  1998, \mn@doi [\aap] {10.48550/arXiv.astro-ph/9712115}, \href
  {https://ui.adsabs.harvard.edu/abs/1998A&A...338..375B} {338, 375}

\bibitem[\protect\citeauthoryear{{Blake}, {James}  \& {Poole}}{{Blake}
  et~al.}{2014}]{Blake2014Minkowski}
{Blake} C.,  {James} J.~B.,   {Poole} G.~B.,  2014, \mn@doi [\mnras]
  {10.1093/mnras/stt2062}, \href
  {https://ui.adsabs.harvard.edu/abs/2014MNRAS.437.2488B} {437, 2488}

\bibitem[\protect\citeauthoryear{{Blumenthal} \& {Faber} et~al.,}{{Blumenthal}
  et~al.}{1986}]{Blumenthal1986AdiabaticContraction}
{Blumenthal} G.~R.,  et~al. 1986, \mn@doi [\apj] {10.1086/163867}, \href
  {https://ui.adsabs.harvard.edu/abs/1986ApJ...301...27B} {301, 27}

\bibitem[\protect\citeauthoryear{{Boyle} \& {Uhlemann} et~al.,}{{Boyle}
  et~al.}{2021}]{Boyle2021MatterPDF}
{Boyle} A.,  et~al. 2021, \mn@doi [\mnras] {10.1093/mnras/stab1381}, \href
  {https://ui.adsabs.harvard.edu/abs/2021MNRAS.505.2886B} {505, 2886}

\bibitem[\protect\citeauthoryear{{Carlstrom}, {Holder}  \& {Reese}}{{Carlstrom}
  et~al.}{2002}]{Carlstrom2002SZReview}
{Carlstrom} J.~E.,  {Holder} G.~P.,   {Reese} E.~D.,  2002, \mn@doi [\araa]
  {10.1146/annurev.astro.40.060401.093803}, \href
  {https://ui.adsabs.harvard.edu/abs/2002ARA&A..40..643C} {40, 643}

\bibitem[\protect\citeauthoryear{{Cataneo} \& {Uhlemann} et~al.,}{{Cataneo}
  et~al.}{2022}]{Cataneo2022MatterPDFMG}
{Cataneo} M.,  et~al. 2022, \mn@doi [\mnras] {10.1093/mnras/stac904}, \href
  {https://ui.adsabs.harvard.edu/abs/2022MNRAS.513.1623C} {513, 1623}

\bibitem[\protect\citeauthoryear{{Chang} \& {Pujol} et~al.,}{{Chang}
  et~al.}{2018}]{Chang2018MassMap}
{Chang} C.,  et~al. 2018, \mn@doi [\mnras] {10.1093/mnras/stx3363}, \href
  {https://ui.adsabs.harvard.edu/abs/2018MNRAS.475.3165C} {475, 3165}

\bibitem[\protect\citeauthoryear{{Cheng} \& {M{\'e}nard}}{{Cheng} \&
  {M{\'e}nard}}{2021}]{Cheng2021WeakLensingST}
{Cheng} S.,  {M{\'e}nard} B.,  2021, \mn@doi [\mnras] {10.1093/mnras/stab2102},
  \href {https://ui.adsabs.harvard.edu/abs/2021MNRAS.507.1012C} {507, 1012}

\bibitem[\protect\citeauthoryear{{Chisari} \& {Richardson} et~al.,}{{Chisari}
  et~al.}{2018}]{Chisari2018BaryonsPk}
{Chisari} N.~E.,  et~al. 2018, \mn@doi [\mnras] {10.1093/mnras/sty2093}, \href
  {https://ui.adsabs.harvard.edu/abs/2018MNRAS.480.3962C} {480, 3962}

\bibitem[\protect\citeauthoryear{{Cui} \& {Dave} et~al.,}{{Cui}
  et~al.}{2022}]{Cui2022GIZMO}
{Cui} W.,  et~al. 2022, \mn@doi [\mnras] {10.1093/mnras/stac1402}, \href
  {https://ui.adsabs.harvard.edu/abs/2022MNRAS.514..977C} {514, 977}

\bibitem[\protect\citeauthoryear{{Davies} \& {Cautun} et~al.,}{{Davies}
  et~al.}{2021}]{Davies2021WLVoids}
{Davies} C.~T.,  et~al. 2021, \mn@doi [\mnras] {10.1093/mnras/stab2251}, \href
  {https://ui.adsabs.harvard.edu/abs/2021MNRAS.507.2267D} {507, 2267}

\bibitem[\protect\citeauthoryear{{Doux} \& {Jain} et~al.,}{{Doux}
  et~al.}{2022}]{Doux2022HarmomicShearY3}
{Doux} C.,  et~al. 2022, \mn@doi [\mnras] {10.1093/mnras/stac1826}, \href
  {https://ui.adsabs.harvard.edu/abs/2022MNRAS.515.1942D} {515, 1942}

\bibitem[\protect\citeauthoryear{{Duffy} \& {Schaye} et~al.,}{{Duffy}
  et~al.}{2010}]{Duffy2010BaryonDmProfileDensity}
{Duffy} A.~R.,  et~al. 2010, \mn@doi [\mnras]
  {10.1111/j.1365-2966.2010.16613.x}, \href
  {https://ui.adsabs.harvard.edu/abs/2010MNRAS.405.2161D} {405, 2161}

\bibitem[\protect\citeauthoryear{{Euclid Collaboration} \& {Ajani}
  et~al.,}{{Euclid Collaboration}}{2023}]{Euclid2023NGCov}
{Euclid Collaboration} 2023, \mn@doi [arXiv e-prints]
  {10.48550/arXiv.2301.12890}, \href
  {https://ui.adsabs.harvard.edu/abs/2023arXiv230112890E} {p. arXiv:2301.12890}

\bibitem[\protect\citeauthoryear{{Fluri} \& {Kacprzak} et~al.,}{{Fluri}
  et~al.}{2018}]{Fluri2018DeepLearning}
{Fluri} J.,  et~al. 2018, \mn@doi [\prd] {10.1103/PhysRevD.98.123518}, \href
  {https://ui.adsabs.harvard.edu/abs/2018PhRvD..98l3518F} {98, 123518}

\bibitem[\protect\citeauthoryear{{Fluri} \& {Kacprzak} et~al.,}{{Fluri}
  et~al.}{2019}]{Fluri2019DeepLearningKIDS}
{Fluri} J.,  et~al. 2019, \mn@doi [\prd] {10.1103/PhysRevD.100.063514}, \href
  {https://ui.adsabs.harvard.edu/abs/2019PhRvD.100f3514F} {100, 063514}

\bibitem[\protect\citeauthoryear{{Fluri} \& {Kacprzak} et~al.,}{{Fluri}
  et~al.}{2022}]{Fluri2022wCDMKIDS}
{Fluri} J.,  et~al. 2022, \mn@doi [\prd] {10.1103/PhysRevD.105.083518}, \href
  {https://ui.adsabs.harvard.edu/abs/2022PhRvD.105h3518F} {105, 083518}

\bibitem[\protect\citeauthoryear{{Friedrich} \& {Gruen} et~al.,}{{Friedrich}
  et~al.}{2018}]{Friedrich2018DensitySplit}
{Friedrich} O.,  et~al. 2018, \mn@doi [\prd] {10.1103/PhysRevD.98.023508},
  \href {https://ui.adsabs.harvard.edu/abs/2018PhRvD..98b3508F} {98, 023508}

\bibitem[\protect\citeauthoryear{{Friedrich} \& {Uhlemann} et~al.,}{{Friedrich}
  et~al.}{2020}]{Friedrich2020PDFBulkfNL}
{Friedrich} O.,  et~al. 2020, \mn@doi [\mnras] {10.1093/mnras/staa2160}, \href
  {https://ui.adsabs.harvard.edu/abs/2020MNRAS.498..464F} {498, 464}

\bibitem[\protect\citeauthoryear{{Gatti} \& {Chang} et~al.,}{{Gatti}
  et~al.}{2020}]{Gatti2020Moments}
{Gatti} M.,  et~al. 2020, \mn@doi [\mnras] {10.1093/mnras/staa2680}, \href
  {https://ui.adsabs.harvard.edu/abs/2020MNRAS.498.4060G} {498, 4060}

\bibitem[\protect\citeauthoryear{{Gatti} \& {Sheldon} et~al.,}{{Gatti}
  et~al.}{2021}]{Gatti2021ShearCatalog}
{Gatti} M.,  et~al. 2021, \mn@doi [\mnras] {10.1093/mnras/stab918}, \href
  {https://ui.adsabs.harvard.edu/abs/2021MNRAS.504.4312G} {504, 4312}

\bibitem[\protect\citeauthoryear{{Gatti} \& {Jain} et~al.,}{{Gatti}
  et~al.}{2022}]{Gatti2022MomentsDESY3}
{Gatti} M.,  et~al. 2022, \mn@doi [\prd] {10.1103/PhysRevD.106.083509}, \href
  {https://ui.adsabs.harvard.edu/abs/2022PhRvD.106h3509G} {106, 083509}

\bibitem[\protect\citeauthoryear{{Gatti} \& {Jeffrey} et~al.,}{{Gatti}
  et~al.}{2023}]{Gatti2023SC}
{Gatti} M.,  et~al. 2023, arXiv e-prints, \href
  {https://ui.adsabs.harvard.edu/abs/2023arXiv230713860G} {p. arXiv:2307.13860}

\bibitem[\protect\citeauthoryear{{Giannantonio} \& {Scranton}
  et~al.,}{{Giannantonio} et~al.}{2008}]{Giannantonio2008Gaussianize}
{Giannantonio} T.,  et~al. 2008, \mn@doi [\prd] {10.1103/PhysRevD.77.123520},
  \href {https://ui.adsabs.harvard.edu/abs/2008PhRvD..77l3520G} {77, 123520}

\bibitem[\protect\citeauthoryear{{Giblin}, {Cai}  \&
  {Harnois-D{\'e}raps}}{{Giblin} et~al.}{2023}]{Giblin2023LensingPDF}
{Giblin} B.,  {Cai} Y.-C.,   {Harnois-D{\'e}raps} J.,  2023, \mn@doi [\mnras]
  {10.1093/mnras/stad230}, \href
  {https://ui.adsabs.harvard.edu/abs/2023MNRAS.520.1721G} {520, 1721}

\bibitem[\protect\citeauthoryear{{Gnedin} \& {Kravtsov} et~al.,}{{Gnedin}
  et~al.}{2004}]{Gnedin2004AdiabaticContraction}
{Gnedin} O.~Y.,  et~al. 2004, \mn@doi [\apj] {10.1086/424914}, \href
  {https://ui.adsabs.harvard.edu/abs/2004ApJ...616...16G} {616, 16}

\bibitem[\protect\citeauthoryear{{Gong} \& {Halder} et~al.,}{{Gong}
  et~al.}{2023}]{Gong2023Integrated3pt}
{Gong} Z.,  et~al. 2023, \mn@doi [arXiv e-prints] {10.48550/arXiv.2304.01187},
  \href {https://ui.adsabs.harvard.edu/abs/2023arXiv230401187G} {p.
  arXiv:2304.01187}

\bibitem[\protect\citeauthoryear{{Gough} \& {Uhlemann}}{{Gough} \&
  {Uhlemann}}{2022}]{Gough2022MatterPDFMG}
{Gough} A.,  {Uhlemann} C.,  2022, \mn@doi [Universe]
  {10.3390/universe8010055}, \href
  {https://ui.adsabs.harvard.edu/abs/2022Univ....8...55G} {8, 55}

\bibitem[\protect\citeauthoryear{{Gruen} \& {Friedrich} et~al.,}{{Gruen}
  et~al.}{2018}]{Gruen2018DensitySplitY1}
{Gruen} D.,  et~al. 2018, \mn@doi [\prd] {10.1103/PhysRevD.98.023507}, \href
  {https://ui.adsabs.harvard.edu/abs/2018PhRvD..98b3507G} {98, 023507}

\bibitem[\protect\citeauthoryear{{Halder} \& {Friedrich} et~al.,}{{Halder}
  et~al.}{2021}]{Halder2021Integrated3ptShear}
{Halder} A.,  et~al. 2021, \mn@doi [\mnras] {10.1093/mnras/stab1801}, \href
  {https://ui.adsabs.harvard.edu/abs/2021MNRAS.506.2780H} {506, 2780}

\bibitem[\protect\citeauthoryear{{Hamana} \& {Colombi} et~al.,}{{Hamana}
  et~al.}{2002}]{Hamana2002SC}
{Hamana} T.,  et~al. 2002, \mn@doi [\mnras] {10.1046/j.1365-8711.2002.05103.x},
  \href {https://ui.adsabs.harvard.edu/abs/2002MNRAS.330..365H} {330, 365}

\bibitem[\protect\citeauthoryear{{Hartlap}, {Simon}  \& {Schneider}}{{Hartlap}
  et~al.}{2007}]{Hartlap2007}
{Hartlap} J.,  {Simon} P.,   {Schneider} P.,  2007, \mn@doi [\aap]
  {10.1051/0004-6361:20066170}, \href
  {https://ui.adsabs.harvard.edu/abs/2007A&A...464..399H} {464, 399}

\bibitem[\protect\citeauthoryear{{Heydenreich}, {Br{\"u}ck}  \&
  {Harnois-D{\'e}raps}}{{Heydenreich}
  et~al.}{2021}]{Heydenreich2021BettiNumbersWL}
{Heydenreich} S.,  {Br{\"u}ck} B.,   {Harnois-D{\'e}raps} J.,  2021, \mn@doi
  [\aap] {10.1051/0004-6361/202039048}, \href
  {https://ui.adsabs.harvard.edu/abs/2021A&A...648A..74H} {648, A74}

\bibitem[\protect\citeauthoryear{{Heydenreich} \& {Br{\"u}ck}
  et~al.,}{{Heydenreich} et~al.}{2022}]{Heydenreich2022Y1BettiNumberWL}
{Heydenreich} S.,  et~al. 2022, \mn@doi [\aap] {10.1051/0004-6361/202243868},
  \href {https://ui.adsabs.harvard.edu/abs/2022A&A...667A.125H} {667, A125}

\bibitem[\protect\citeauthoryear{{Heydenreich} \& {Linke}
  et~al.,}{{Heydenreich} et~al.}{2023}]{Heydenreich2023ThirdOrder}
{Heydenreich} S.,  et~al. 2023, \mn@doi [\aap] {10.1051/0004-6361/202244820},
  \href {https://ui.adsabs.harvard.edu/abs/2023A&A...672A..44H} {672, A44}

\bibitem[\protect\citeauthoryear{{Hinshaw} \& {Larson} et~al.,}{{Hinshaw}
  et~al.}{2013}]{Hinshaw2013WMAP9}
{Hinshaw} G.,  et~al. 2013, \mn@doi [\apjs] {10.1088/0067-0049/208/2/19}, \href
  {https://ui.adsabs.harvard.edu/abs/2013ApJS..208...19H} {208, 19}

\bibitem[\protect\citeauthoryear{{Jain}, {Seljak}  \& {White}}{{Jain}
  et~al.}{1998}]{Jain1998LensingPDF}
{Jain} B.,  {Seljak} U.,   {White} S.,  1998, arXiv e-prints, \href
  {https://ui.adsabs.harvard.edu/abs/1998astro.ph..4238J} {pp
  astro--ph/9804238}

\bibitem[\protect\citeauthoryear{{Jarvis} \& {Bernstein} et~al.,}{{Jarvis}
  et~al.}{2021}]{Jarvis2021PIFF}
{Jarvis} M.,  et~al. 2021, \mn@doi [\mnras] {10.1093/mnras/staa3679}, \href
  {https://ui.adsabs.harvard.edu/abs/2021MNRAS.501.1282J} {501, 1282}

\bibitem[\protect\citeauthoryear{{Jeffrey} \& {Lanusse} et~al.,}{{Jeffrey}
  et~al.}{2020}]{Jeffrey2020DLReconstruction}
{Jeffrey} N.,  et~al. 2020, \mn@doi [\mnras] {10.1093/mnras/staa127}, \href
  {https://ui.adsabs.harvard.edu/abs/2020MNRAS.492.5023J} {492, 5023}

\bibitem[\protect\citeauthoryear{{Jeffrey} \& {Gatti} et~al.,}{{Jeffrey}
  et~al.}{2021}]{Niall2021MassMap}
{Jeffrey} N.,  et~al. 2021, \mn@doi [\mnras] {10.1093/mnras/stab1495}, \href
  {https://ui.adsabs.harvard.edu/abs/2021MNRAS.505.4626J} {505, 4626}

\bibitem[\protect\citeauthoryear{{Kacprzak} \& {Fluri} et~al.,}{{Kacprzak}
  et~al.}{2023}]{Kacprzak2023Cosmogrid}
{Kacprzak} T.,  et~al. 2023, \mn@doi [\jcap] {10.1088/1475-7516/2023/02/050},
  \href {https://ui.adsabs.harvard.edu/abs/2023JCAP...02..050K} {2023, 050}

\bibitem[\protect\citeauthoryear{{Kaiser} \& {Squires}}{{Kaiser} \&
  {Squires}}{1993}]{Kaiser1993KS}
{Kaiser} N.,  {Squires} G.,  1993, \mn@doi [\apj] {10.1086/172297}, \href
  {https://ui.adsabs.harvard.edu/abs/1993ApJ...404..441K} {404, 441}

\bibitem[\protect\citeauthoryear{{Kratochvil}, {Haiman}  \& {May}}{{Kratochvil}
  et~al.}{2010}]{Kratochvil2010WLPeaks}
{Kratochvil} J.~M.,  {Haiman} Z.,   {May} M.,  2010, \mn@doi [\prd]
  {10.1103/PhysRevD.81.043519}, \href
  {https://ui.adsabs.harvard.edu/abs/2010PhRvD..81d3519K} {81, 043519}

\bibitem[\protect\citeauthoryear{{Krause} \& {Hirata}}{{Krause} \&
  {Hirata}}{2010}]{Krause2010ReducedShear}
{Krause} E.,  {Hirata} C.~M.,  2010, \mn@doi [\aap]
  {10.1051/0004-6361/200913524}, \href
  {https://ui.adsabs.harvard.edu/abs/2010A&A...523A..28K} {523, A28}

\bibitem[\protect\citeauthoryear{{Krause} \& {Fang} et~al.,}{{Krause}
  et~al.}{2021}]{Krause2021Methods}
{Krause} E.,  et~al. 2021, \mn@doi [arXiv e-prints]
  {10.48550/arXiv.2105.13548}, \href
  {https://ui.adsabs.harvard.edu/abs/2021arXiv210513548K} {p. arXiv:2105.13548}

\bibitem[\protect\citeauthoryear{{Kruse} \& {Schneider}}{{Kruse} \&
  {Schneider}}{2000}]{Kruse2000NGTail}
{Kruse} G.,  {Schneider} P.,  2000, \mn@doi [\mnras]
  {10.1046/j.1365-8711.2000.03389.x}, \href
  {https://ui.adsabs.harvard.edu/abs/2000MNRAS.318..321K} {318, 321}

\bibitem[\protect\citeauthoryear{{Lanzieri} \& {Lanusse} et~al.,}{{Lanzieri}
  et~al.}{2023}]{Lanzieri2023HOS}
{Lanzieri} D.,  et~al. 2023, \mn@doi [arXiv e-prints]
  {10.48550/arXiv.2305.07531}, \href
  {https://ui.adsabs.harvard.edu/abs/2023arXiv230507531L} {p. arXiv:2305.07531}

\bibitem[\protect\citeauthoryear{{Lee} \& {Anbajagane} et~al.,}{{Lee}
  et~al.}{2022}]{Lee2022rSZ}
{Lee} E.,  et~al. 2022, \mn@doi [\mnras] {10.1093/mnras/stac2781}, \href
  {https://ui.adsabs.harvard.edu/abs/2022MNRAS.517.5303L} {517, 5303}

\bibitem[\protect\citeauthoryear{{Lim} \& {Barnes} et~al.,}{{Lim}
  et~al.}{2021}]{Lim2021GasProp}
{Lim} S.~H.,  et~al. 2021, \mn@doi [\mnras] {10.1093/mnras/stab1172}, \href
  {https://ui.adsabs.harvard.edu/abs/2021MNRAS.504.5131L} {504, 5131}

\bibitem[\protect\citeauthoryear{{Lovell} \& {Pillepich} et~al.,}{{Lovell}
  et~al.}{2018}]{Lovell2018ConcentrationBaryonImprints}
{Lovell} M.~R.,  et~al. 2018, \mn@doi [\mnras] {10.1093/mnras/sty2339}, \href
  {https://ui.adsabs.harvard.edu/abs/2018MNRAS.481.1950L} {481, 1950}

\bibitem[\protect\citeauthoryear{{MacCrann} \& {Becker} et~al.,}{{MacCrann}
  et~al.}{2022}]{MacCrann2022ImsimsY3}
{MacCrann} N.,  et~al. 2022, \mn@doi [\mnras] {10.1093/mnras/stab2870}, \href
  {https://ui.adsabs.harvard.edu/abs/2022MNRAS.509.3371M} {509, 3371}

\bibitem[\protect\citeauthoryear{{Martinet} \& {Schneider} et~al.,}{{Martinet}
  et~al.}{2018}]{Martinet2018HOSKiDS}
{Martinet} N.,  et~al. 2018, \mn@doi [\mnras] {10.1093/mnras/stx2793}, \href
  {https://ui.adsabs.harvard.edu/abs/2018MNRAS.474..712M} {474, 712}

\bibitem[\protect\citeauthoryear{{Mecke}, {Buchert}  \& {Wagner}}{{Mecke}
  et~al.}{1994}]{Mecke1994Minkowski}
{Mecke} K.~R.,  {Buchert} T.,   {Wagner} H.,  1994, \mn@doi [\aap]
  {10.48550/arXiv.astro-ph/9312028}, \href
  {https://ui.adsabs.harvard.edu/abs/1994A&A...288..697M} {288, 697}

\bibitem[\protect\citeauthoryear{{Munshi} \& {Jung} et~al.,}{{Munshi}
  et~al.}{2023}]{Munshi2023PosDepCorr}
{Munshi} D.,  et~al. 2023, \mn@doi [\prd] {10.1103/PhysRevD.107.043516}, \href
  {https://ui.adsabs.harvard.edu/abs/2023PhRvD.107d3516M} {107, 043516}

\bibitem[\protect\citeauthoryear{{Myles} \& {Alarcon} et~al.,}{{Myles}
  et~al.}{2021}]{Myles2021PhotoZ}
{Myles} J.,  et~al. 2021, \mn@doi [\mnras] {10.1093/mnras/stab1515}, \href
  {https://ui.adsabs.harvard.edu/abs/2021MNRAS.505.4249M} {505, 4249}

\bibitem[\protect\citeauthoryear{{Omori}}{{Omori}}{2022}]{Omori2022Agora}
{Omori} Y.,  2022, \mn@doi [arXiv e-prints] {10.48550/arXiv.2212.07420}, \href
  {https://ui.adsabs.harvard.edu/abs/2022arXiv221207420O} {p. arXiv:2212.07420}

\bibitem[\protect\citeauthoryear{{Osato}, {Liu}  \& {Haiman}}{{Osato}
  et~al.}{2021}]{Osato2021KappaTNG}
{Osato} K.,  {Liu} J.,   {Haiman} Z.,  2021, \mn@doi [\mnras]
  {10.1093/mnras/stab395}, \href
  {https://ui.adsabs.harvard.edu/abs/2021MNRAS.502.5593O} {502, 5593}

\bibitem[\protect\citeauthoryear{{Park} \& {Allys} et~al.,}{{Park}
  et~al.}{2022}]{Park2022NGCov}
{Park} C.~F.,  et~al. 2022, \mn@doi [arXiv e-prints]
  {10.48550/arXiv.2204.05435}, \href
  {https://ui.adsabs.harvard.edu/abs/2022arXiv220405435P} {p. arXiv:2204.05435}

\bibitem[\protect\citeauthoryear{{Parroni} \& {Cardone} et~al.,}{{Parroni}
  et~al.}{2020}]{Parroni2020Minkowski}
{Parroni} C.,  et~al. 2020, \mn@doi [\aap] {10.1051/0004-6361/201935988}, \href
  {https://ui.adsabs.harvard.edu/abs/2020A&A...633A..71P} {633, A71}

\bibitem[\protect\citeauthoryear{{Peacock}}{{Peacock}}{1983}]{Peacock1983TwoSideTest}
{Peacock} J.~A.,  1983, \mn@doi [\mnras] {10.1093/mnras/202.3.615}, \href
  {https://ui.adsabs.harvard.edu/abs/1983MNRAS.202..615P} {202, 615}

\bibitem[\protect\citeauthoryear{{Peel} \& {Pettorino} et~al.,}{{Peel}
  et~al.}{2018}]{Peel2018Moments}
{Peel} A.,  et~al. 2018, \mn@doi [\aap] {10.1051/0004-6361/201833481}, \href
  {https://ui.adsabs.harvard.edu/abs/2018A&A...619A..38P} {619, A38}

\bibitem[\protect\citeauthoryear{{Petri} \& {Liu} et~al.,}{{Petri}
  et~al.}{2015}]{Petri2015MomentsMinkowski}
{Petri} A.,  et~al. 2015, \mn@doi [\prd] {10.1103/PhysRevD.91.103511}, \href
  {https://ui.adsabs.harvard.edu/abs/2015PhRvD..91j3511P} {91, 103511}

\bibitem[\protect\citeauthoryear{{Petri}, {Haiman}  \& {May}}{{Petri}
  et~al.}{2017}]{Petri2017Born}
{Petri} A.,  {Haiman} Z.,   {May} M.,  2017, \mn@doi [\prd]
  {10.1103/PhysRevD.95.123503}, \href
  {https://ui.adsabs.harvard.edu/abs/2017PhRvD..95l3503P} {95, 123503}

\bibitem[\protect\citeauthoryear{{Planck Collaboration} \& {Ade}
  et~al.,}{{Planck Collaboration}}{2016a}]{Planck2016CosmoParams}
{Planck Collaboration} 2016a, \mn@doi [\aap] {10.1051/0004-6361/201525830},
  \href {https://ui.adsabs.harvard.edu/abs/2016A&A...594A..13P} {594, A13}

\bibitem[\protect\citeauthoryear{{Planck Collaboration} \& {Ade}
  et~al.,}{{Planck Collaboration}}{2016b}]{Planck2016GaussianityTest}
{Planck Collaboration} 2016b, \mn@doi [\aap] {10.1051/0004-6361/201526681},
  \href {https://ui.adsabs.harvard.edu/abs/2016A&A...594A..16P} {594, A16}

\bibitem[\protect\citeauthoryear{{Planck Collaboration} \& {Akrami}
  et~al.,}{{Planck Collaboration}}{2020}]{Planck2020GaussianityTest}
{Planck Collaboration} 2020, \mn@doi [\aap] {10.1051/0004-6361/201935201},
  \href {https://ui.adsabs.harvard.edu/abs/2020A&A...641A...7P} {641, A7}

\bibitem[\protect\citeauthoryear{{Potter}, {Stadel}  \& {Teyssier}}{{Potter}
  et~al.}{2017}]{Potter2017Pkdgrav3}
{Potter} D.,  {Stadel} J.,   {Teyssier} R.,  2017, \mn@doi [Computational
  Astrophysics and Cosmology] {10.1186/s40668-017-0021-1}, \href
  {https://ui.adsabs.harvard.edu/abs/2017ComAC...4....2P} {4, 2}

\bibitem[\protect\citeauthoryear{{Schneider} \& {Teyssier} et~al.,}{{Schneider}
  et~al.}{2019}]{Schneider2019Baryonification}
{Schneider} A.,  et~al. 2019, \mn@doi [\jcap] {10.1088/1475-7516/2019/03/020},
  \href {https://ui.adsabs.harvard.edu/abs/2019JCAP...03..020S} {2019, 020}

\bibitem[\protect\citeauthoryear{{Secco} \& {Samuroff} et~al.,}{{Secco}
  et~al.}{2022a}]{Secco2022Shear}
{Secco} L.~F.,  et~al. 2022a, \mn@doi [\prd] {10.1103/PhysRevD.105.023515},
  \href {https://ui.adsabs.harvard.edu/abs/2022PhRvD.105b3515S} {105, 023515}

\bibitem[\protect\citeauthoryear{{Secco} \& {Jarvis} et~al.,}{{Secco}
  et~al.}{2022b}]{Secco2022MassAp}
{Secco} L.~F.,  et~al. 2022b, \mn@doi [\prd] {10.1103/PhysRevD.105.103537},
  \href {https://ui.adsabs.harvard.edu/abs/2022PhRvD.105j3537S} {105, 103537}

\bibitem[\protect\citeauthoryear{{Sellentin} \& {Loureiro} et~al.,}{{Sellentin}
  et~al.}{2023}]{Sellentin2023Almanac}
{Sellentin} E.,  et~al. 2023, \mn@doi [arXiv e-prints]
  {10.48550/arXiv.2305.16134}, \href
  {https://ui.adsabs.harvard.edu/abs/2023arXiv230516134S} {p. arXiv:2305.16134}

\bibitem[\protect\citeauthoryear{{Sevilla-Noarbe} \& {Bechtol}
  et~al.,}{{Sevilla-Noarbe} et~al.}{2021}]{Sevilla2021Y3Gold}
{Sevilla-Noarbe} I.,  et~al. 2021, \mn@doi [\apjs] {10.3847/1538-4365/abeb66},
  \href {https://ui.adsabs.harvard.edu/abs/2021ApJS..254...24S} {254, 24}

\bibitem[\protect\citeauthoryear{{Shan} \& {Liu} et~al.,}{{Shan}
  et~al.}{2018}]{Shan2018HOSKiDS}
{Shan} H.,  et~al. 2018, \mn@doi [\mnras] {10.1093/mnras/stx2837}, \href
  {https://ui.adsabs.harvard.edu/abs/2018MNRAS.474.1116S} {474, 1116}

\bibitem[\protect\citeauthoryear{{Shao}, {Anbajagane}  \& {Chang}}{{Shao}
  et~al.}{2022}]{Shao2022Baryons}
{Shao} M.,  {Anbajagane} D.,   {Chang} C.,  2022, \mn@doi [arXiv e-prints]
  {10.48550/arXiv.2212.05964}, \href
  {https://ui.adsabs.harvard.edu/abs/2022arXiv221205964S} {p. arXiv:2212.05964}

\bibitem[\protect\citeauthoryear{{Springel}}{{Springel}}{2005}]{Springel2005Gagdet2}
{Springel} V.,  2005, \mn@doi [\mnras] {10.1111/j.1365-2966.2005.09655.x},
  \href {https://ui.adsabs.harvard.edu/abs/2005MNRAS.364.1105S} {364, 1105}

\bibitem[\protect\citeauthoryear{{Stiskalek} \& {Bartlett} et~al.,}{{Stiskalek}
  et~al.}{2022}]{Stiskalek2022TNGHorizon}
{Stiskalek} R.,  et~al. 2022, \mn@doi [\mnras] {10.1093/mnras/stac1609}, \href
  {https://ui.adsabs.harvard.edu/abs/2022MNRAS.514.4026S} {514, 4026}

\bibitem[\protect\citeauthoryear{{Sunseri}, {Li}  \& {Liu}}{{Sunseri}
  et~al.}{2023}]{Sunseri2023Baryons}
{Sunseri} J.,  {Li} Z.,   {Liu} J.,  2023, \mn@doi [\prd]
  {10.1103/PhysRevD.107.023514}, \href
  {https://ui.adsabs.harvard.edu/abs/2023PhRvD.107b3514S} {107, 023514}

\bibitem[\protect\citeauthoryear{{Sunyaev} \& {Zeldovich}}{{Sunyaev} \&
  {Zeldovich}}{1972}]{Sunyaev1972SZEffect}
{Sunyaev} R.~A.,  {Zeldovich} Y.~B.,  1972, Comments on Astrophysics and Space
  Physics, \href {https://ui.adsabs.harvard.edu/abs/1972CoASP...4..173S} {4,
  173}

\bibitem[\protect\citeauthoryear{{Takahashi} \& {Hamana} et~al.,}{{Takahashi}
  et~al.}{2017}]{Takahashi2017Sims}
{Takahashi} R.,  et~al. 2017, \mn@doi [\apj] {10.3847/1538-4357/aa943d}, \href
  {https://ui.adsabs.harvard.edu/abs/2017ApJ...850...24T} {850, 24}

\bibitem[\protect\citeauthoryear{{The Dark Energy Survey Collaboration}}{{The
  Dark Energy Survey Collaboration}}{2005}]{DES2005}
{The Dark Energy Survey Collaboration} 2005, arXiv e-prints, \href
  {https://ui.adsabs.harvard.edu/abs/2005astro.ph.10346T} {pp
  astro--ph/0510346}

\bibitem[\protect\citeauthoryear{{The LSST Dark Energy Science Collaboration}
  \& {Mandelbaum} et~al.,}{{The LSST Dark Energy Science
  Collaboration}}{2018}]{LSST2018SRD}
{The LSST Dark Energy Science Collaboration} 2018, \mn@doi [arXiv e-prints]
  {10.48550/arXiv.1809.01669}, \href
  {https://ui.adsabs.harvard.edu/abs/2018arXiv180901669T} {p. arXiv:1809.01669}

\bibitem[\protect\citeauthoryear{{Uhlemann} \& {Friedrich} et~al.,}{{Uhlemann}
  et~al.}{2020}]{Uhlemann2020PDFNeutrino}
{Uhlemann} C.,  et~al. 2020, \mn@doi [\mnras] {10.1093/mnras/staa1155}, \href
  {https://ui.adsabs.harvard.edu/abs/2020MNRAS.495.4006U} {495, 4006}

\bibitem[\protect\citeauthoryear{{Van Waerbeke} \& {Benjamin} et~al.,}{{Van
  Waerbeke} et~al.}{2013}]{VanWaerbeke2013CFHTLens}
{Van Waerbeke} L.,  et~al. 2013, \mn@doi [\mnras] {10.1093/mnras/stt971}, \href
  {https://ui.adsabs.harvard.edu/abs/2013MNRAS.433.3373V} {433, 3373}

\bibitem[\protect\citeauthoryear{{Villaescusa-Navarro} \& {Hahn}
  et~al.,}{{Villaescusa-Navarro} et~al.}{2020}]{Navarro2020Quijote}
{Villaescusa-Navarro} F.,  et~al. 2020, \mn@doi [\apjs]
  {10.3847/1538-4365/ab9d82}, \href
  {https://ui.adsabs.harvard.edu/abs/2020ApJS..250....2V} {250, 2}

\bibitem[\protect\citeauthoryear{{Wang}, {Banerjee}  \& {Abel}}{{Wang}
  et~al.}{2022}]{Wang2022kNNSDSS}
{Wang} Y.,  {Banerjee} A.,   {Abel} T.,  2022, \mn@doi [\mnras]
  {10.1093/mnras/stac1551}, \href
  {https://ui.adsabs.harvard.edu/abs/2022MNRAS.514.3828W} {514, 3828}

\bibitem[\protect\citeauthoryear{{White} \& {Hu}}{{White} \&
  {Hu}}{2000}]{White2000WLsimulations}
{White} M.,  {Hu} W.,  2000, \mn@doi [\apj] {10.1086/309009}, \href
  {https://ui.adsabs.harvard.edu/abs/2000ApJ...537....1W} {537, 1}

\bibitem[\protect\citeauthoryear{{Z{\"u}rcher} \& {Fluri}
  et~al.,}{{Z{\"u}rcher} et~al.}{2021}]{Zurcher2021WLForecast}
{Z{\"u}rcher} D.,  et~al. 2021, \mn@doi [\jcap]
  {10.1088/1475-7516/2021/01/028}, \href
  {https://ui.adsabs.harvard.edu/abs/2021JCAP...01..028Z} {2021, 028}

\bibitem[\protect\citeauthoryear{{Z{\"u}rcher} \& {Fluri}
  et~al.,}{{Z{\"u}rcher} et~al.}{2022}]{Zurcher2022WLPeaks}
{Z{\"u}rcher} D.,  et~al. 2022, \mn@doi [\mnras] {10.1093/mnras/stac078}, \href
  {https://ui.adsabs.harvard.edu/abs/2022MNRAS.511.2075Z} {511, 2075}

\makeatother
\end{thebibliography}




\appendix

\section{3-field CDFs and beyond}\label{appx:Nfield_CDFs}

Formally, in the Gaussian limit, the 3-field CDF contains no new information beyond those from the 2-field CDFs, since they can also be described completely by the multivariate normal in equation \eqref{eqn:multivariate_normal}. Thus, the 3-field CDFs can be predicted exactly using the covariance of the fields as a function of smoothing scale.

\begin{figure}
    \centering
    \includegraphics[width = \columnwidth]{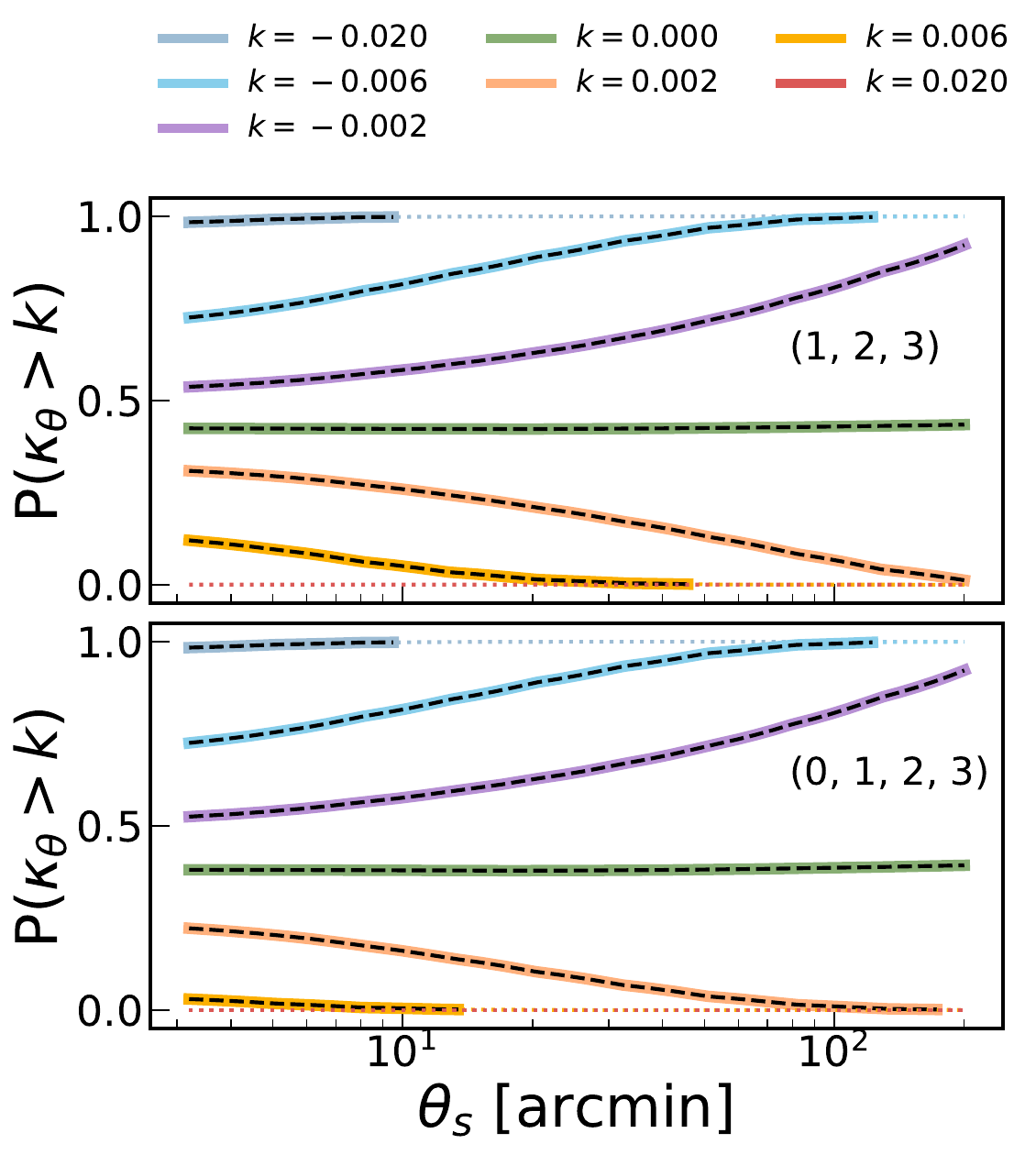}
    \caption{Measurements for the 3-field and 4-field CDFs on the noiseless DES Y3-like simulations (colored lines), and a theoretical prediction in the limit of the field being Gaussian (black, dashed lines). The latter follows the same procedure of Section \ref{sec:GaussianConsistency}. The Gaussian model fits the data well, as is expected in this limit. The bin indices show the different tomographic bins used in the measurement.}
    \label{fig:Nfield}
\end{figure}

We show this explicitly in Figure \ref{fig:Nfield}. We make measurements of the 3-field and 4-field CDFs on Gaussian fields, and then exactly predict the measurements given the covariance matrix as a function of smoothing scale. The covariance matrix is measured directly on the map. We have verified the residuals between the measured N-field CDFs and the prediction is within $0.1\sigma$, where $\sigma$ comes solely from cosmic variance. This test is an extension of Figure \ref{fig:ConsistencyRelation} for N-field CDFs of higher N.

\section{Gaussianity of Covariance Matrix}\label{appx:Cov_Matrix}

\begin{figure}
    \centering
    \includegraphics[width = \columnwidth]{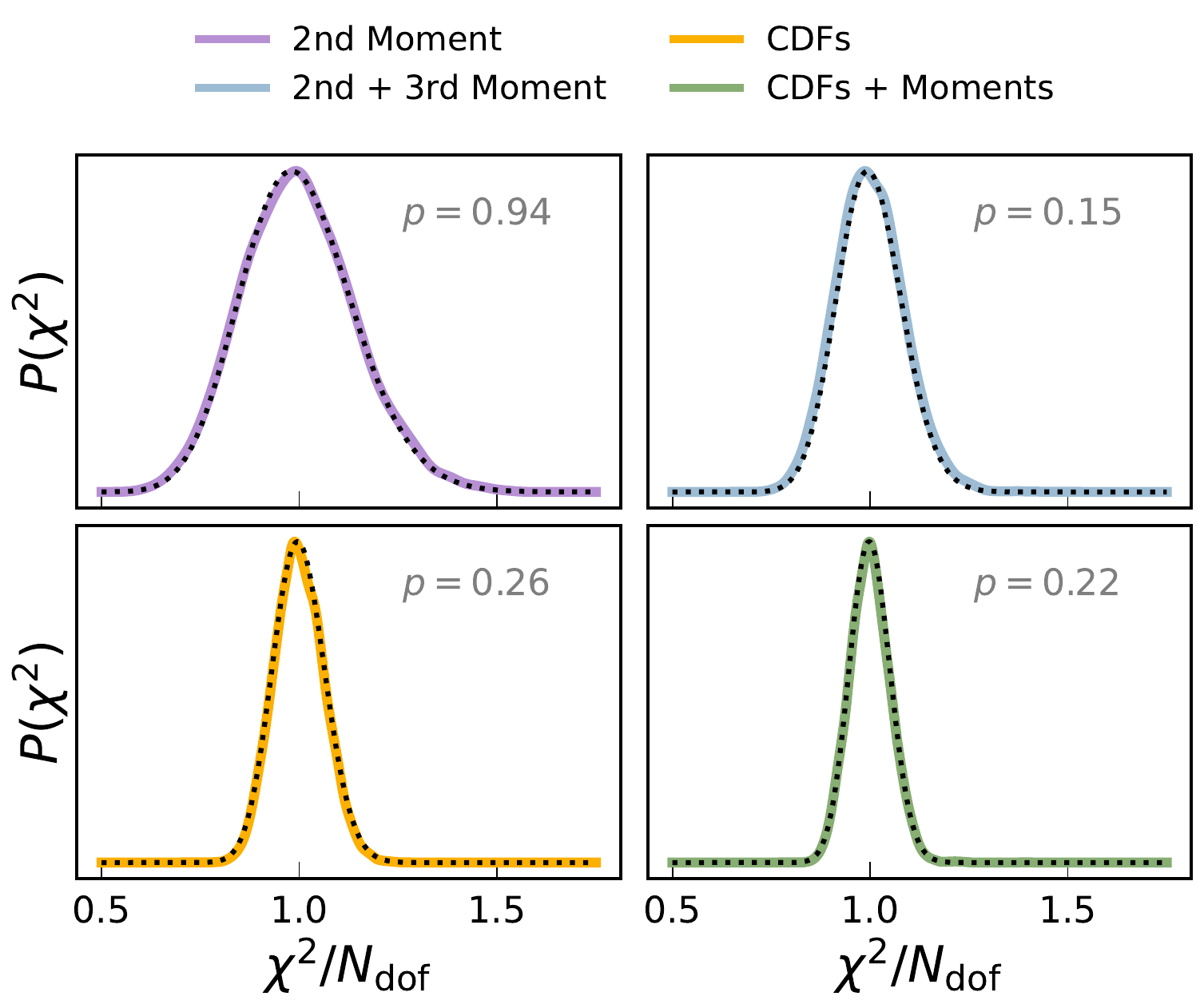}
    \caption{The chi-squared distributions of the data vectors (solid lines), compared with a theoretical chi-squared distribution (dotted black line) with $N_{\rm dof}$ given by the size of the data vector. In the Gaussian likelihood limit the theoretical distributions will match the measured distribution. A Kolmogorov-Smirnov test shows the probability that the observed and expected distributions are similar exceeds $p > 0.1$. The data vectors considered in this work have a sufficiently Gaussian likelihood.}
    \label{fig:CheckCov}
\end{figure}

The process of performing a Fisher forecast, or obtaining constraints using likelihood minimization, assumes the likelihood of the data vector is Gaussian, i.e. the measurement uncertainty in the data vector is distributed as a multivariate Gaussian. We test here the validity of that assumption. We do so by first transforming every realization $i$ of a data vector by removing its mean, $S_i = D_i - \langle D \rangle$, where the mean is computed over all $i$ realizations. We then compute $\chi^2 = S_i \mathcal{C}^{-1} S_i$, where $C$ is the covariance matrix estimated using all realizations of $D$. In the limit that the likelihood is Gaussian, the distribution of $\chi^2$ must follow a standard $\chi^2$ distribution. 

In Figure \ref{fig:CheckCov}, we show the measured and expected distributions for four different data vectors, and in all cases we find the measured distributions match the expected Gaussian-limit distributions. We also compute a Kolmogorov-Smirnov statistic to quantify the level of agreement between the measured and expected distribution \citep{Peacock1983TwoSideTest}. This validates that the Fisher formalism is an accurate way to estimate potential constraints from the statistics considered in this work. Some additional techniques can also be used to quantify this Gaussianity of the likelihood \citep{Park2022NGCov, Euclid2023NGCov}, and they are roughly similar to the approach we have taken here.


\section*{Affiliations}

$^{1}$ Department of Astronomy and Astrophysics, University of Chicago, Chicago, IL 60637, USA\\
$^{2}$ Kavli Institute for Cosmological Physics, University of Chicago, Chicago, IL 60637, USA\\
$^{3}$ Indian Institute of Science Education and Research,  Pune,  411008, India\\
$^{4}$ Department of Physics, Stanford University, 382 Via Pueblo Mall, Stanford, CA 94305, USA\\
$^{5}$ Kavli Institute for Particle Astrophysics \& Cosmology, P. O. Box 2450, Stanford University, Stanford, CA 94305, USA\\
$^{6}$ SLAC National Accelerator Laboratory, Menlo Park, CA 94025, USA\\
$^{7}$ Department of Physics and Astronomy, University of Pennsylvania, Philadelphia, PA 19104, USA\\
$^{8}$ Department of Physics, ETH Zurich, Wolfgang-Pauli-Strasse 16, CH-8093 Zurich, Switzerland\\
$^{9}$ Argonne National Laboratory, 9700 South Cass Avenue, Lemont, IL 60439, USA\\
$^{10}$ Institute of Astronomy, University of Cambridge, Madingley Road, Cambridge CB3 0HA, UK\\
$^{11}$ Kavli Institute for Cosmology, University of Cambridge, Madingley Road, Cambridge CB3 0HA, UK\\
$^{12}$ Institute for Astronomy, University of Hawai'i, 2680 Woodlawn Drive, Honolulu, HI 96822, USA\\
$^{13}$ Physics Department, 2320 Chamberlin Hall, University of Wisconsin-Madison, 1150 University Avenue Madison, WI  53706-1390\\
$^{14}$ Department of Physics, Carnegie Mellon University, Pittsburgh, Pennsylvania 15312, USA\\
$^{15}$ Instituto de Astrofisica de Canarias, E-38205 La Laguna, Tenerife, Spain\\
$^{16}$ Laborat\'orio Interinstitucional de e-Astronomia - LIneA, Rua Gal. Jos\'e Cristino 77, Rio de Janeiro, RJ - 20921-400, Brazil\\
$^{17}$ Universidad de La Laguna, Dpto. AstrofÃ­sica, E-38206 La Laguna, Tenerife, Spain\\
$^{18}$ Center for Astrophysical Surveys, National Center for Supercomputing Applications, 1205 West Clark St., Urbana, IL 61801, USA\\
$^{19}$ Department of Astronomy, University of Illinois at Urbana-Champaign, 1002 W. Green Street, Urbana, IL 61801, USA\\
$^{20}$ Department of Physics, Duke University Durham, NC 27708, USA\\
$^{21}$ NASA Goddard Space Flight Center, 8800 Greenbelt Rd, Greenbelt, MD 20771, USA\\
$^{22}$ Lawrence Berkeley National Laboratory, 1 Cyclotron Road, Berkeley, CA 94720, USA\\
$^{23}$ Fermi National Accelerator Laboratory, P. O. Box 500, Batavia, IL 60510, USA\\
$^{24}$ NSF AI Planning Institute for Physics of the Future, Carnegie Mellon University, Pittsburgh, PA 15213, USA\\
$^{25}$ Universit\'e Grenoble Alpes, CNRS, LPSC-IN2P3, 38000 Grenoble, France\\
$^{26}$ Department of Physics and Astronomy, University of Waterloo, 200 University Ave W, Waterloo, ON N2L 3G1, Canada\\
$^{27}$ Jet Propulsion Laboratory, California Institute of Technology, 4800 Oak Grove Dr., Pasadena, CA 91109, USA\\
$^{28}$ University Observatory, Faculty of Physics, Ludwig-Maximilians-Universit\"at, Scheinerstr. 1, 81679 Munich, Germany\\
$^{29}$ School of Physics and Astronomy, Cardiff University, CF24 3AA, UK\\
$^{30}$ Department of Astronomy, University of Geneva, ch. d'\'Ecogia 16, CH-1290 Versoix, Switzerland\\
$^{31}$ Department of Physics \& Astronomy, University College London, Gower Street, London, WC1E 6BT, UK\\
$^{32}$ Department of Applied Mathematics and Theoretical Physics, University of Cambridge, Cambridge CB3 0WA, UK\\
$^{33}$ Instituto de F\'isica Gleb Wataghin, Universidade Estadual de Campinas, 13083-859, Campinas, SP, Brazil\\
$^{34}$ Department of Physics, University of Genova and INFN, Via Dodecaneso 33, 16146, Genova, Italy\\
$^{35}$ Jodrell Bank Center for Astrophysics, School of Physics and Astronomy, University of Manchester, Oxford Road, Manchester, M13 9PL, UK\\
$^{36}$ Centro de Investigaciones Energ\'eticas, Medioambientales y Tecnol\'ogicas (CIEMAT), Madrid, Spain\\
$^{37}$ Brookhaven National Laboratory, Bldg 510, Upton, NY 11973, USA\\
$^{38}$ Department of Physics and Astronomy, Stony Brook University, Stony Brook, NY 11794, USA\\
$^{39}$ D\'{e}partement de Physique Th\'{e}orique and Center for Astroparticle Physics, Universit\'{e} de Gen\`{e}ve, 24 quai Ernest Ansermet, CH-1211 Geneva, Switzerland\\
$^{40}$ Institut d'Estudis Espacials de Catalunya (IEEC), 08034 Barcelona, Spain\\
$^{41}$ Institut de Recherche en Astrophysique et Plan\'etologie (IRAP), Universit\'e de Toulouse, CNRS, UPS, CNES, 14 Av. Edouard Belin, 31400 Toulouse, France\\
$^{42}$ Cerro Tololo Inter-American Observatory, NSF's National Optical-Infrared Astronomy Research Laboratory, Casilla 603, La Serena, Chile\\
$^{43}$ Department of Astronomy, University of Michigan, Ann Arbor, MI 48109, USA\\
$^{44}$ Department of Physics, University of Michigan, Ann Arbor, MI 48109, USA\\
$^{45}$ Institute of Cosmology and Gravitation, University of Portsmouth, Portsmouth, PO1 3FX, UK\\
$^{46}$ Department of Physics, Northeastern University, Boston, MA 02115, USA\\
$^{47}$ Physics Department, William Jewell College, Liberty, MO, 64068\\
$^{48}$ Hamburger Sternwarte, Universit\"{a}t Hamburg, Gojenbergsweg 112, 21029 Hamburg, Germany\\
$^{49}$ School of Mathematics and Physics, University of Queensland,  Brisbane, QLD 4072, Australia\\
$^{50}$ Department of Physics, IIT Hyderabad, Kandi, Telangana 502285, India\\
$^{51}$ Institute of Theoretical Astrophysics, University of Oslo. P.O. Box 1029 Blindern, NO-0315 Oslo, Norway\\
$^{52}$ Institut de F\'{\i}sica d'Altes Energies (IFAE), The Barcelona Institute of Science and Technology, Campus UAB, 08193 Bellaterra (Barcelona) Spain\\
$^{53}$ Santa Cruz Institute for Particle Physics, Santa Cruz, CA 95064, USA\\
$^{54}$ Center for Cosmology and Astro-Particle Physics, The Ohio State University, Columbus, OH 43210, USA\\
$^{55}$ Department of Physics, The Ohio State University, Columbus, OH 43210, USA\\
$^{56}$ Center for Astrophysics $\vert$ Harvard \& Smithsonian, 60 Garden Street, Cambridge, MA 02138, USA\\
$^{57}$ Australian Astronomical Optics, Macquarie University, North Ryde, NSW 2113, Australia\\
$^{58}$ Lowell Observatory, 1400 Mars Hill Rd, Flagstaff, AZ 86001, USA\\
$^{59}$ George P. and Cynthia Woods Mitchell Institute for Fundamental Physics and Astronomy, and Department of Physics and Astronomy, Texas A\&M University, College Station, TX 77843,  USA\\
$^{60}$ Instituci\'o Catalana de Recerca i Estudis Avan\c{c}ats, E-08010 Barcelona, Spain\\
$^{61}$ Observat\'orio Nacional, Rua Gal. Jos\'e Cristino 77, Rio de Janeiro, RJ - 20921-400, Brazil\\
$^{62}$ School of Physics and Astronomy, University of Southampton,  Southampton, SO17 1BJ, UK\\
$^{63}$ National Center for Supercomputing Applications, 1205 West Clark St., Urbana, IL 61801, USA\\

\bsp	
\label{lastpage}
\end{document}